\documentclass[sigconf,authorversion,nonacm]{acmart}

\usepackage{amsthm}
\usepackage[utf8x]{inputenc}
\usepackage{xcolor,colortbl}
\usepackage{blindtext}
\usepackage{xcolor,colortbl}
\usepackage{booktabs} 
\usepackage{subfigure}
\usepackage{multirow}				
\usepackage{fixmath} 
\usepackage{graphicx}
\usepackage{dsfont}
\usepackage{marginnote}
\usepackage{multicol}
\usepackage[ruled,linesnumbered]{algorithm2e}
\usepackage{algpseudocode}
\usepackage{mathtools}
\usepackage{bm}
\usepackage{caption} 
\usepackage{enumerate}
\usepackage{lipsum}
\usepackage[most]{tcolorbox}
\usepackage{stfloats}
\usepackage{tikz}
\usepackage{makecell}
\usepackage{enumitem}
\usepackage{cleveref}
\usepackage{listings}
\usepackage{tablefootnote}
\usepackage{threeparttable}

\newcommand{\eat}[1]{}
\newcommand{\reminder}[1]{[\vadjust{\vbox to0pt{\vss\hbox to0pt{\hss{\Large $\Longrightarrow$}}}}{{\textsf{\small #1}}}]}

\newcommand{\RA}[1]{\textcolor{red}{\reminder{RA:~#1}}}
\newcommand{\BH}[1]{\textcolor{blue}{\reminder{BH:~#1}}}
\renewcommand{\BH}[1]{}

\newcommand {\sys}{{GEqO}\xspace}
\newcommand{\rev}[1]{#1}

\newcommand {\ecpm}{\textsc{\textit{EMF}}\xspace}
\newcommand {\vpm}{\textsc{\textit{VMF}}\xspace}
\newcommand {\sfm}{\textsc{\textit{SF}}\xspace}
\newcommand {\ssfl}{\textsc{\textit{SSFL}}\xspace}

\newcommand {\cloudViews}{CloudViews}

\newcommand{\todox}[1]{\textcolor{orange}{\{\{#1\}\}}}

\newtheorem{definition}{Definition}[section]

\def\ojoin{\setbox0=\hbox{$\Join$}%
\rule[0.05ex]{.27em}{.7pt}\llap{\rule[0.97ex]{.27em}{.7pt}}}
\def\leftouterjoin{\mathbin{\ojoin\mkern-6.8mu\Join}}
\def\rightouterjoin{\mathbin{\Join\mkern-6.8mu\ojoin}}
\def\fullouterjoin{\mathbin{\ojoin\mkern-6.8mu\bowtie\mkern-6.8mu\ojoin}}

\newcommand{\ie}{i.e.,~}
\newcommand{\eg}{e.g.,~}

\newcommand{\LR}{\textbf{LR}}
\newcommand{\BRF}{\textbf{BRF}}
\newcommand{\LP}{\textbf{LP}}
\newcommand{\WR}{\textbf{WR}}

\newtheorem*{problem*}{Problem}

\newcommand*{\approxequiv}{%
  \mathrel{\vcenter{\offinterlineskip
\hbox{$\sim$}\vskip-.35ex\hbox{$\sim$}\vskip-.35ex\hbox{$\sim$}}}}

\usepackage{sourcecodepro}
\usepackage{listingsutf8}

\usepackage{framed}
\usepackage{tcolorbox}
\newenvironment{roundedframe}{%
  \MakeFramed {\advance\hsize-\width \FrameRestore}}%
 {\endMakeFramed}

\lstnewenvironment{sqllisting}
    {\lstset{language=sql,escapeinside={(*@}{@*)}}}
    {}

\lstnewenvironment{boxedlisting}
    {\lstset{language=sql,escapeinside={(*@}{@*)}}}
    {}

\usepackage{etoolbox}
\BeforeBeginEnvironment{boxedlisting}{\begin{roundedframe}}
\AfterEndEnvironment{boxedlisting}{\end{roundedframe}}

\newcommand{\squishlist}{
   \begin{list}{$\bullet$}{%
        \setlength{\itemsep}{0pt}%
        \setlength{\parsep}{0pt}%
        \setlength{\topsep}{0pt}%
        \setlength{\partopsep}{0pt}%
        \setlength{\listparindent}{-2pt}%
        \setlength{\itemindent}{-5pt}%
        \setlength{\leftmargin}{1.2em}%
        \setlength{\labelwidth}{0em}%
        \setlength{\labelsep}{0.5em}%
    }
}

\newcommand{\squishend}{
    \end{list}  }

\definecolor{codepurple}{HTML}{C42043}
\lstdefinestyle{sqlstyle}{
basicstyle=\footnotesize\ttfamily,
breaklines=true,
keywordstyle=\bfseries\color{codepurple},
language=sql
}
\usepackage[T1]{fontenc}
\usepackage{contour}
\usepackage{varwidth}

\renewcommand*\copyright{{%
  \textcopyright}}

\begin{document}

\title{\sys: ML-Accelerated Semantic Equivalence Detection}

\author{Brandon Haynes}
\email{brandon.haynes@microsoft.com}
\orcid{0000-0002-1501-9586}
\affiliation{%
  \institution{Microsoft Gray Systems Lab}
 \country{USA}
}

\author{Rana Alotaibi}
\email{ranaalotaibi@microsoft.com}
\orcid{0009-0005-0457-8429}
\affiliation{%
  \institution{Microsoft Gray Systems Lab}
 \country{USA}
}

\author{Anna Pavlenko}
\email{anna.pavlenko@microsoft.com}
\orcid{0009-0006-7442-4254}
\affiliation{%
  \institution{Microsoft Gray Systems Lab}
 \country{USA}
}

\author{Jyoti Leeka}
\email{jyoti.leeka@microsoft.com}
\orcid{0000-0003-2920-1431}
\affiliation{%
  \institution{Microsoft}
 \country{USA}
}

\author{Alekh Jindal}
\email{alekh@smart-apps.ai}
\orcid{0000-0001-8844-8165}
\authornote{Work done while at Microsoft.}
\affiliation{%
  \institution{SmartApps}
 \country{USA}
}

\author{Yuanyuan Tian}
\email{yuanyuantian@microsoft.com}
\orcid{0000-0002-6835-8434}
\affiliation{%
  \institution{Microsoft Gray Systems Lab}
 \country{USA}
}

\renewcommand{\shortauthors}{Brandon Haynes et al.}

\newcommand{\tpcdsHeatmapFigure}{
\begin{figure}[!t]
\includegraphics[width=0.35\textwidth]{Figure/HeatMap.JPG}
\caption{Heatmap of TPC-DS Subexpressions Containment and Equivalence Results}
\label{fig:tpcds-heatmap}
\end{figure}
}
\newcommand{\queryFeaturizer}{
\begin{figure*}[!htbp]
\centering
\includegraphics[width=14cm]{Figure/queries-featurization.pdf}
\caption{Query Featurizer and Convolution} 
\label{fig:query_featurizer}
\end{figure*}
}

\newcommand{\workloadEncoding}{
\begin{figure*}[t!] %
\centering
\includegraphics[width=0.8\textwidth,trim={0 3em 0 0},clip]{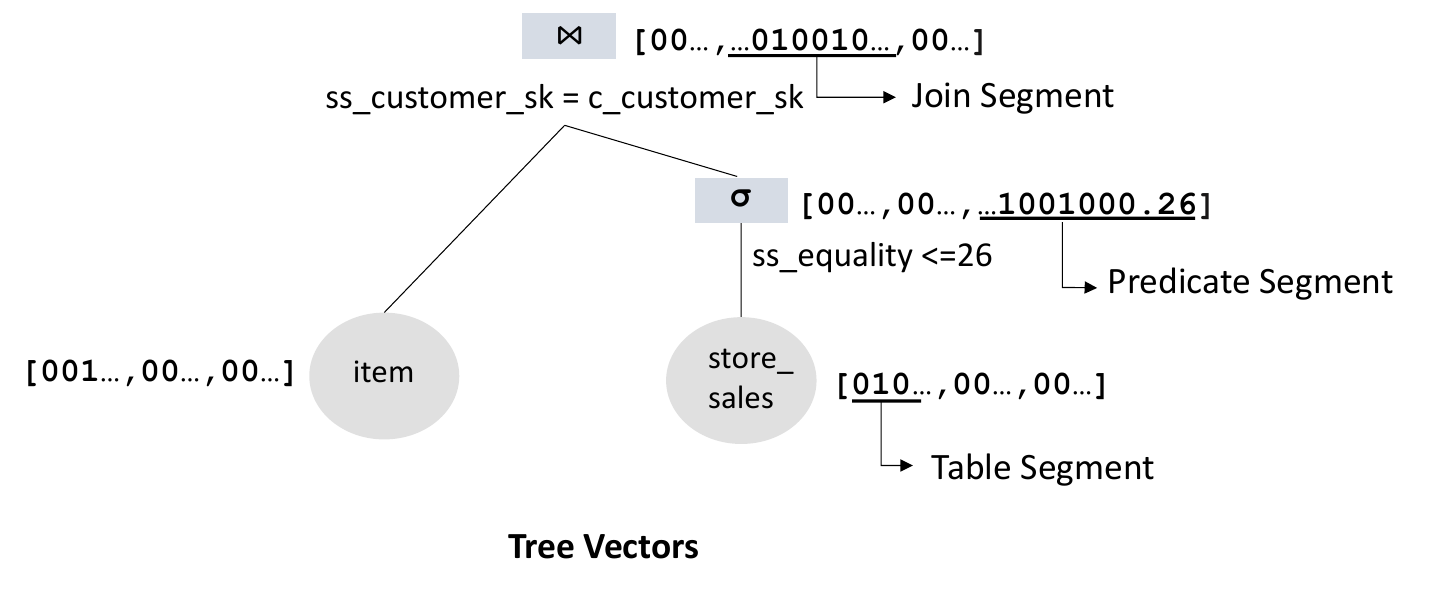}
\caption{Instance-based node vector encoding of an SPJ subexpression. %
Each operator's metadata is converted to its ``vector segment''; %
unrelated segments are set to zero.
}
\label{fig:workloadEncoding}
\end{figure*}
}

\newcommand{\encodingConverter}{
\begin{figure*}[t!]
\centering
\includegraphics[width=\textwidth, height=3.5cm]{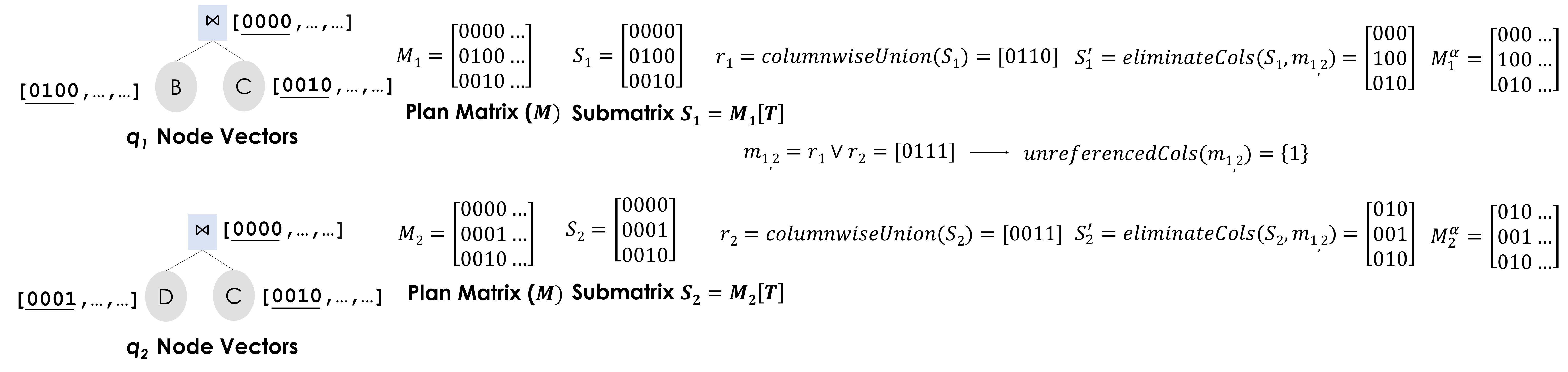} %
\caption{%
Example of converting instance-based to db-agnostic encoding (table segments only)
}
\label{fig:encoding-converter}
\end{figure*}
}

\newcommand{\workloadAnalysis}{
\begin{figure*}[!htbp]
\centering
\includegraphics[width=0.9\textwidth]{Figure/workloadanalysis}
\caption{Workload analysis \RA{Updated the Figure. This refers now to: \sfm and \vpm-based Pesudo-equivalent Sampling}}
\label{fig:workloadAnalysis}
\end{figure*}
}

\newcommand{\arch}{
\begin{figure*}[!t]
\centering
\includegraphics[width=1\textwidth]{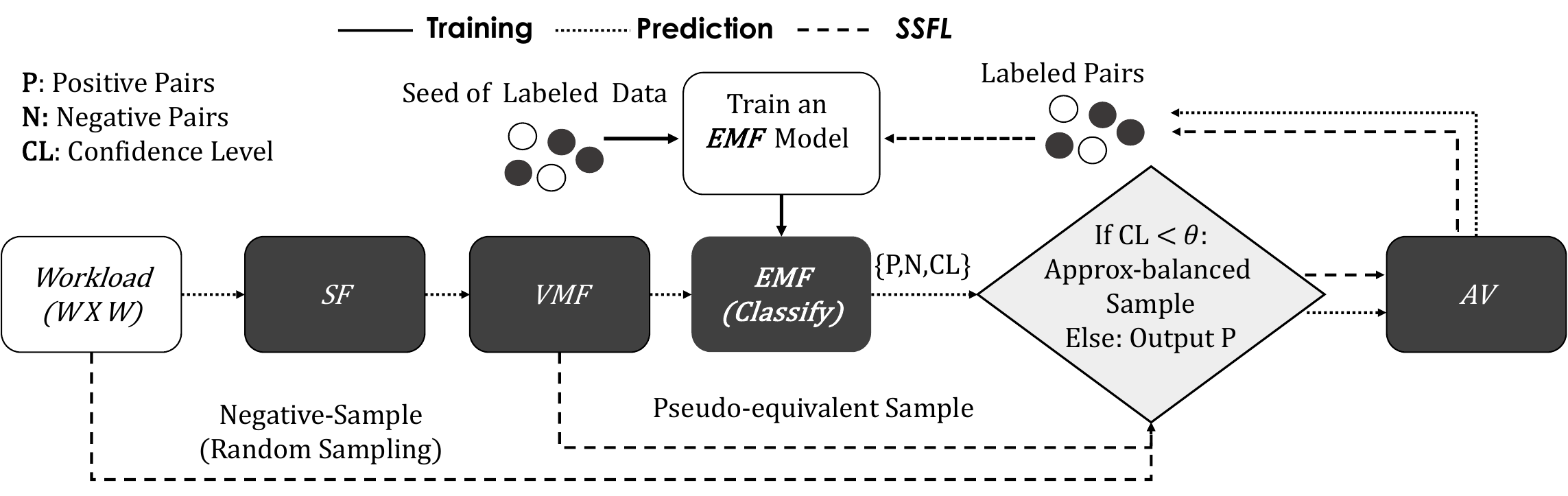}
\caption{\sys Overview 
} %
\vspace{0.5em}
\label{fig:arch}
\end{figure*}
}

\newcommand{\ITSTesting}{
\begin{figure*}[!htbp]
\centering
	\subfigure[\textbf{TPCH-CNT}]{\includegraphics[height=4cm]{Figure/ITS-CNT-TPCH.pdf}}
	\subfigure[\textbf{TPSCD-CNT}]{\includegraphics[height=4cm]{Figure/ITS-CNT-TPCDS.pdf}}
	\subfigure[\textbf{TPCH-EQ}]{\includegraphics[height=4cm]{Figure/ITS-EQ-TPCH.pdf}}
	\subfigure[\textbf{TPSCD-EQ}]{\includegraphics[height=4cm]{Figure/ITS-EQ-TPCDS.pdf}}
	\caption{Feedback Loop Validation Test - Interesting Table Subsets Sampling}
	\label{fig:ITSTesting}
\end{figure*}
}

\newcommand{\TEMPTesting}{
\begin{figure*}[!htbp]
\centering
	\subfigure[\textbf{TPCH-CNT}]{\includegraphics[height=4cm]{Figure/TEMP-CNT-TPCH.pdf}}
	\subfigure[\textbf{TPSCD-CNT}]{\includegraphics[height=4cm]{Figure/TEMP-CNT-TPCDS.pdf}}
	\subfigure[\textbf{TPCH-EQ}]{\includegraphics[height=4cm]{Figure/TEMP-EQ-TPCH.pdf}}
	\subfigure[\textbf{TPSCD-EQ}]{\includegraphics[height=4cm]{Figure/TEMP-EQ-TPCDS.pdf}}
	\caption{Feedback Loop Validation Test - Query Templates-based Sampling}
	\label{fig:TEMPTesting}
\end{figure*}
}

\newcommand{\LSTesting}{
\begin{figure*}[!htbp]
\centering
	\subfigure[\textbf{TPCH-CNT}]{\includegraphics[height=4cm]{Figure/LS-CNT-TPCH.pdf}}
	\subfigure[\textbf{TPSCD-CNT}]{\includegraphics[height=4cm]{Figure/LS-CNT-TPCDS.pdf}}
	\subfigure[\textbf{TPCH-EQ}]{\includegraphics[height=4cm]{Figure/LS-EQ-TPCH.pdf}}
	\subfigure[\textbf{TPSCD-EQ}]{\includegraphics[height=4cm]{Figure/LS-EQ-TPCDS.pdf}}
	\caption{Feedback Loop Validation Test - Latent Space-based Sampling}
	\label{fig:LSTesting}
\end{figure*}
}

\newcommand{\FREQTesting}{
\begin{figure*}[!htbp]
\centering
	\subfigure[\textbf{TPCH-CNT}]{\includegraphics[height=4cm]{Figure/FREQ-CNT-TPCH.pdf}}
	\subfigure[\textbf{TPSCD-CNT}]{\includegraphics[height=4cm]{Figure/FREQ-CNT-TPCDS.pdf}}
	\subfigure[\textbf{TPCH-EQ}]{\includegraphics[height=4cm]{Figure/FREQ-EQ-TPCH.pdf}}
	\subfigure[\textbf{TPSCD-EQ}]{\includegraphics[height=4cm]{Figure/FREQ-EQ-TPCDS.pdf}}
	\caption{Feedback Loop Validation Test - Query Frequency based Sampling}
	\label{fig:FREQTesting}
\end{figure*}
}

\newcommand{\LRTWR}{
\begin{figure*}[!htbp]
\centering
	\subfigure[\textbf{SP Queries}]{\includegraphics[height=4cm]{charts/results/EXP1_SP_LOG_Training_Error_R.pdf}}
	\subfigure[\textbf{SPJ1 Queries}]{\includegraphics[height=4cm]{charts/results/EXP1_SPJ1_LOG_Training_Error_R.pdf}}
	\subfigure[\textbf{SPJ2 Queries}]{\includegraphics[height=4cm]{charts/results/EXP1_SPJ2_LOG_Training_Error_R.pdf}}
	\caption{Logistic Regression (\LR)-Workload \WR-Training Error-Log scale}
	\label{fig:LRTWR}
\end{figure*}
}

\newcommand{\LRCWR}{
\begin{figure*}[!htbp]
\subfigure[\textbf{SP Queries}]{\includegraphics[height=4cm]{charts/results/EXP1_SP_LOG_Testing_Error_R.pdf}}
\subfigure[\textbf{SPJ1 Queries}]{\includegraphics[height=4cm]{charts/results/EXP1_SPJ1_LOG_Testing_Error_R.pdf}}
\subfigure[\textbf{SPJ2 Queries}]{\includegraphics[height=4cm]{charts/results/EXP1_SPJ2_LOG_Testing_Error_R.pdf}}
\caption{Logistic Regression (\LR)-Workload \WR-Cross Validation-Log scale}
	\label{fig:LRCWR}
\end{figure*}
}

\newcommand{\BNFCWR}{
\begin{figure*}[!htbp]
\subfigure[\textbf{SP Queries}]{\includegraphics[height=4cm]{charts/results/EXP1_SP_BNF_Testing_Error_128_R.pdf}\label{fig:BNFCWR_SP}}
\subfigure[\textbf{SPJ1 Queries}]{\includegraphics[height=4cm]{charts/results/EXP1_SPJ1_BNF_Testing_Error_128_R.pdf}\label{fig:BNFCWR_SPJ1}}
\subfigure[\textbf{SPJ2 Queries}]{\includegraphics[height=4cm]{charts/results/EXP1_SPJ2_BNF_Testing_Error_128_R.pdf}\label{fig:BNFCWR_SPJ2}}
\caption{Balanced Random Forest (\BRF)-Workload \WR-Cross Validation-Log scale}
	\label{fig:BNFCWR}
\end{figure*}
}
	
\newcommand{\LRACWRSP}{
\begin{figure}[!t]
\includegraphics[height=4cm]{charts/results/LOG_AUC_CNF_SP_R.pdf}
\caption{Logistic Regression (\LR)-Workload \WR\ (SP)-AUC and Confusion Matrix}
\label{fig:LRACWRSP}
\end{figure}
}

\newcommand{\LPTWRRS}{
\begin{figure}[!t]
\includegraphics[height=4cm]{charts/results/EXP3_LP_RandomSampling_TrainError}
\caption{ \LP-Random Sampling-Workload \WR-Train Error}
\label{fig:LPTWRRS}
\end{figure}
}

\newcommand{\LPLRWRRS}{
\begin{figure}[!t]
\includegraphics[height=4cm]{charts/results/EXP3_LP_RandomSampling_LR_TrainError}
\caption{ \LP-Random Sampling-LR-Workload \WR-Train Error}
\label{fig:LPLRWRRS}
\end{figure}
}

\newcommand{\AvgPredictionTimeNoFeaturization}{
\begin{figure}[!t]
\includegraphics[height=4cm]{charts/results/AVG_Prediction_Time_No_Feature}
\caption{\LR, \BRF, and \LP Average Prediction w/o Query Featurization}
\label{fig:AvgPredictionTimeNoFeature}
\end{figure}
}

\newcommand{\AvgPredictionTimeWithFeaturization}{
\begin{figure}[!t]
\includegraphics[height=4cm]{charts/results/AVG_Prediction_Time_with_Feature}
\caption{\LR, \BRF, and \LP Average Prediction w/ Query Featurization}
\label{fig:AvgPredictionTimeWithFeature}
\end{figure}
}

\newcommand{\containmentBaselineResults}{
\begin{figure}[!t]
\includegraphics[height=4cm]{charts/results/BASELINE_CALCITE}
\caption{\sys\ ML- Pre filter Model  VS Calcite CNT }
\label{fig:containmentBaselineResults}
\end{figure}
}

\newcommand{\equivalenceBaselineResults}{
\begin{figure}[!t]
\includegraphics[height=4cm]{charts/results/BASELINE_SPES}
\caption{\sys\ ML- Pre filter Model  VS SPES EQ }
\label{fig:equivalenceBaselineResults}
\end{figure}
}

\newcommand{\equivalencePerformance}{
\begin{figure}[!t]
\includegraphics[height=4cm]{charts/results/performance/performance-equivalences-PLACEHOLDER}
\caption{\sys equivalence performance}
\label{fig:equivalence-performance}
\end{figure}
}

\newcommand{\containmentPerformance}{
\begin{figure}[!t]
\includegraphics[height=4cm]{charts/results/performance/performance-containment-PLACEHOLDER}
\caption{\sys containment performance}
\label{fig:equivalence-performance}
\end{figure}
}

\newcommand{\eqfeedbackloopTPCDS}{
\begin{figure}[!t]
\includegraphics[height=4cm]{Figure/EQ_Feedback_Loop_Results_TPCDS.pdf}
\caption{Feedback Loop Evaluation on EQ TPCDS - MLP}
\label{fig:feedbackloop}
\end{figure}
}

\newcommand{\eqfeedbackloopTPCDSTPCH}{
\begin{figure}[!htbp]
\includegraphics[height=4cm]{Figure/Random_VS_LS_Accuracy.pdf}\label{fig:feedback_loop_smapling}
\caption{Feedback Training Loop Results}
	\label{fig:EQCMFeedbackLoop}
\end{figure}
}

\newcommand{\eqfeedbackloopPer}{
\begin{figure*}[t]
\includegraphics[width=0.55\textwidth]{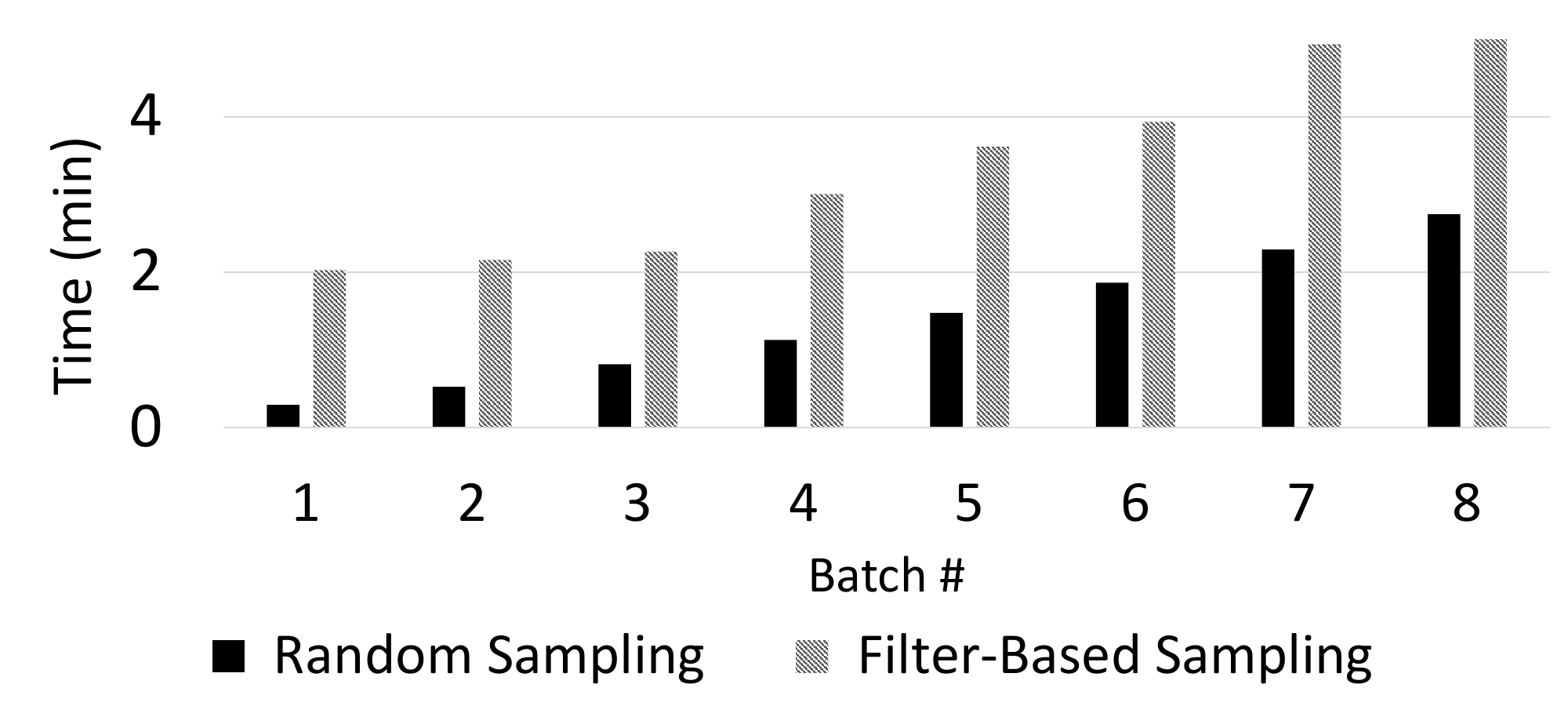}
\label{fig:feedback_loop_performance}
\caption{\ssfl training time for filter-based sampling versus random sampling.  Each iteration trains over a batch of 512 newly-labeled samples.}
\label{fig:EQCMFeedbackLoop}
\end{figure*}
}

\newcommand{\GECOFeedbackLoopDrilldown}{
\begin{figure}[t]
\includegraphics[width=1\columnwidth]{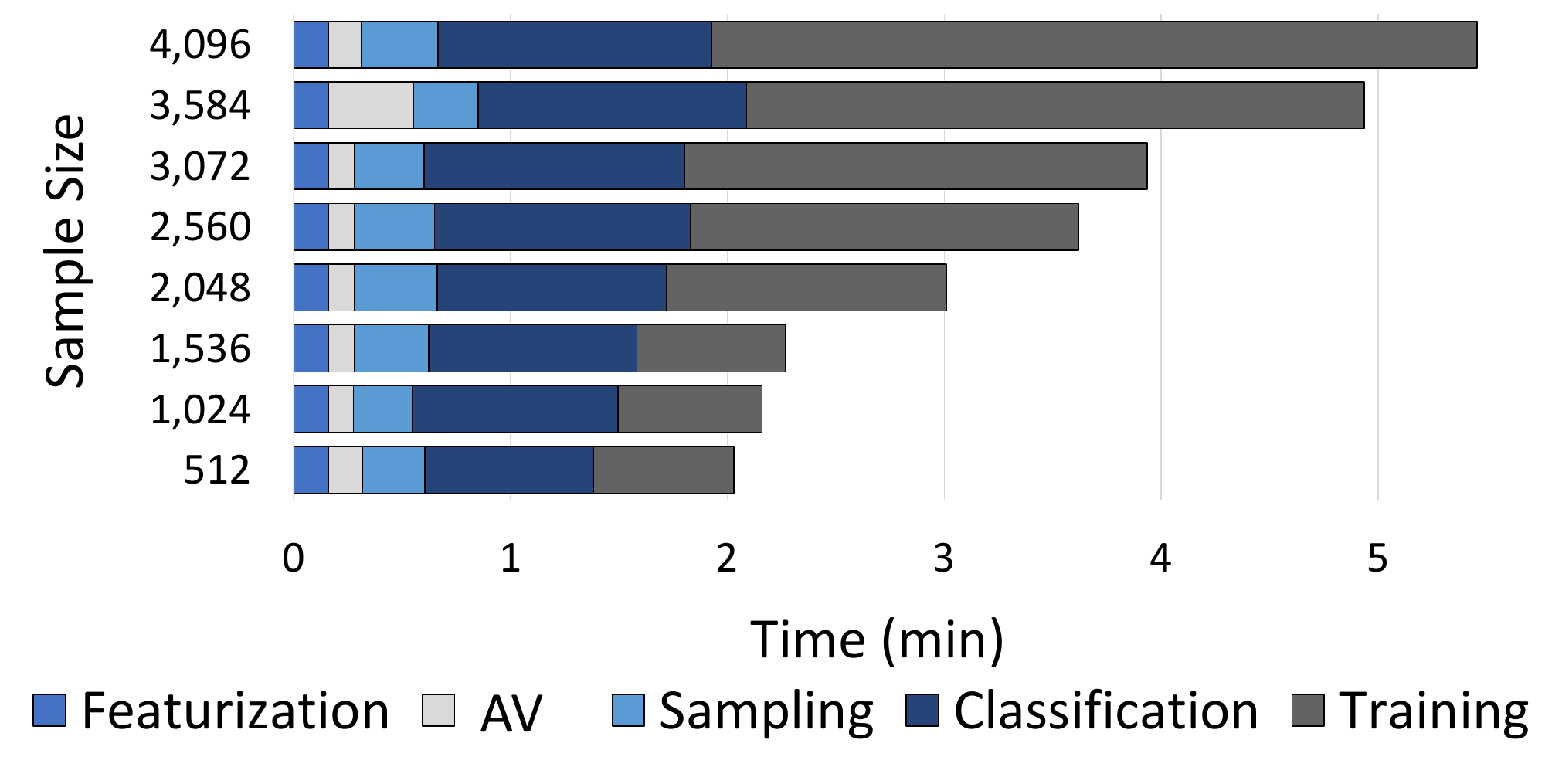}
\label{fig:feedback_loop_DrillDown}
\caption{Runtime %
of the operations applied by the \ssfl.
}
\label{fig:EQCMFeedbackLoopdrilldown}
\end{figure}
}

\newcommand{\GECOEndtoEnd}{
\begin{figure*}[t]
  \centering
  \includegraphics[trim={0 0 0 17.5cm},clip,width=0.5\textwidth]{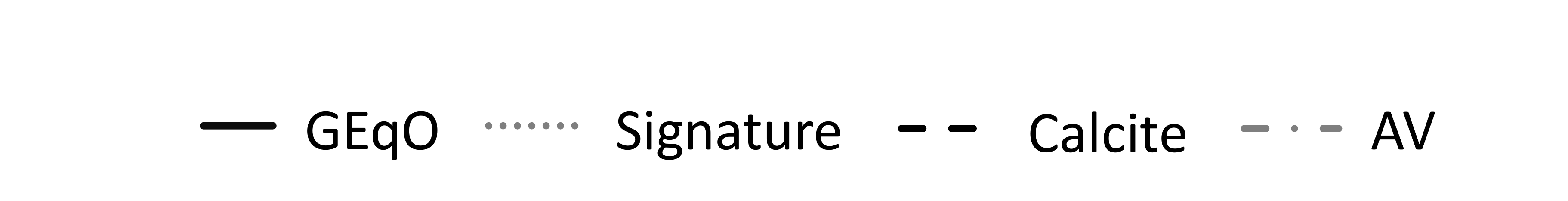} %
  
  \subfigure[\textbf{True positive rate}]{
  \includegraphics[width=0.33\textwidth]{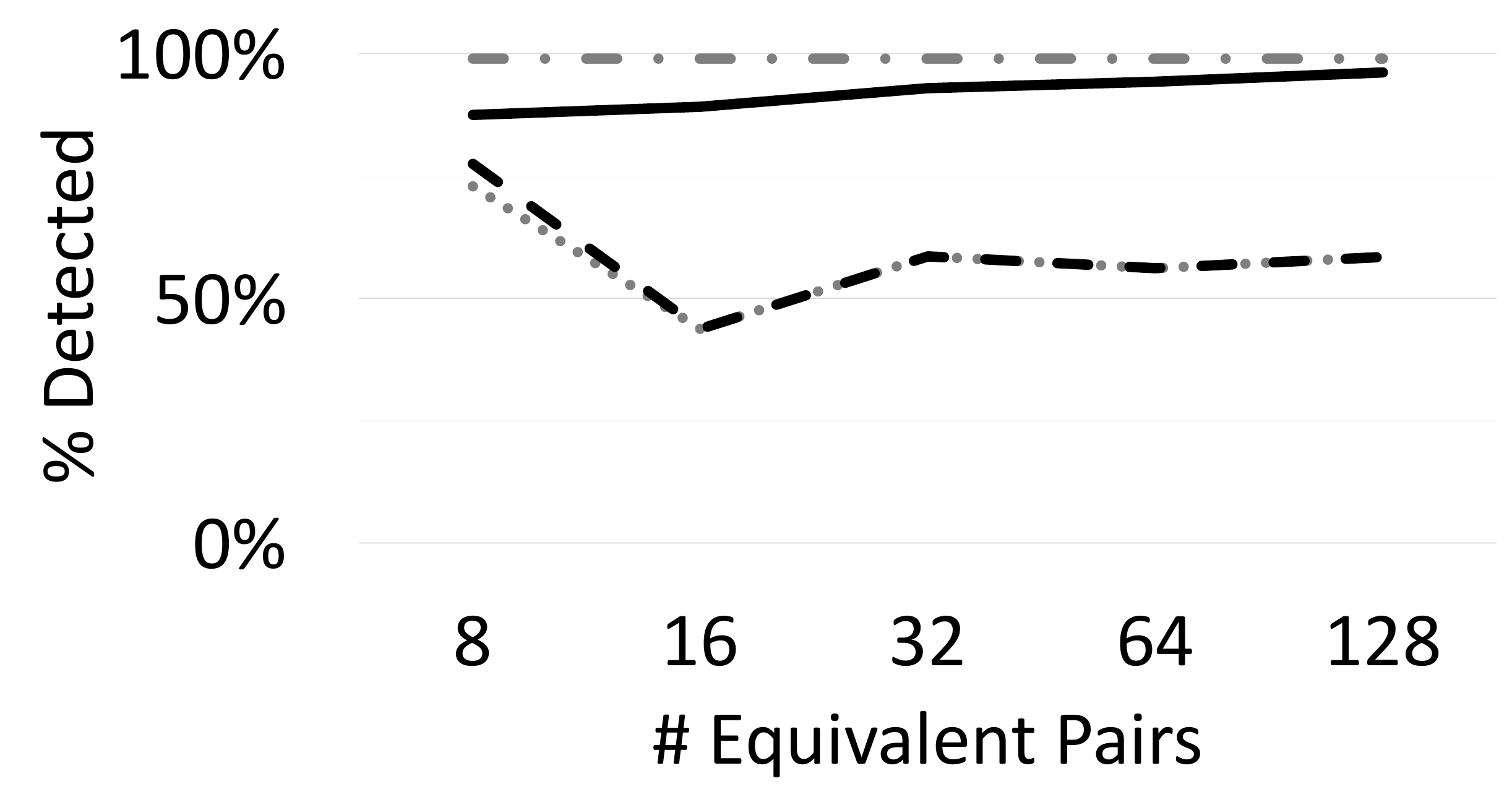}
  \label{subfig:e2e-tpr}
}%
\subfigure[\textbf{Total runtime}]{
  \includegraphics[width=0.33\textwidth]{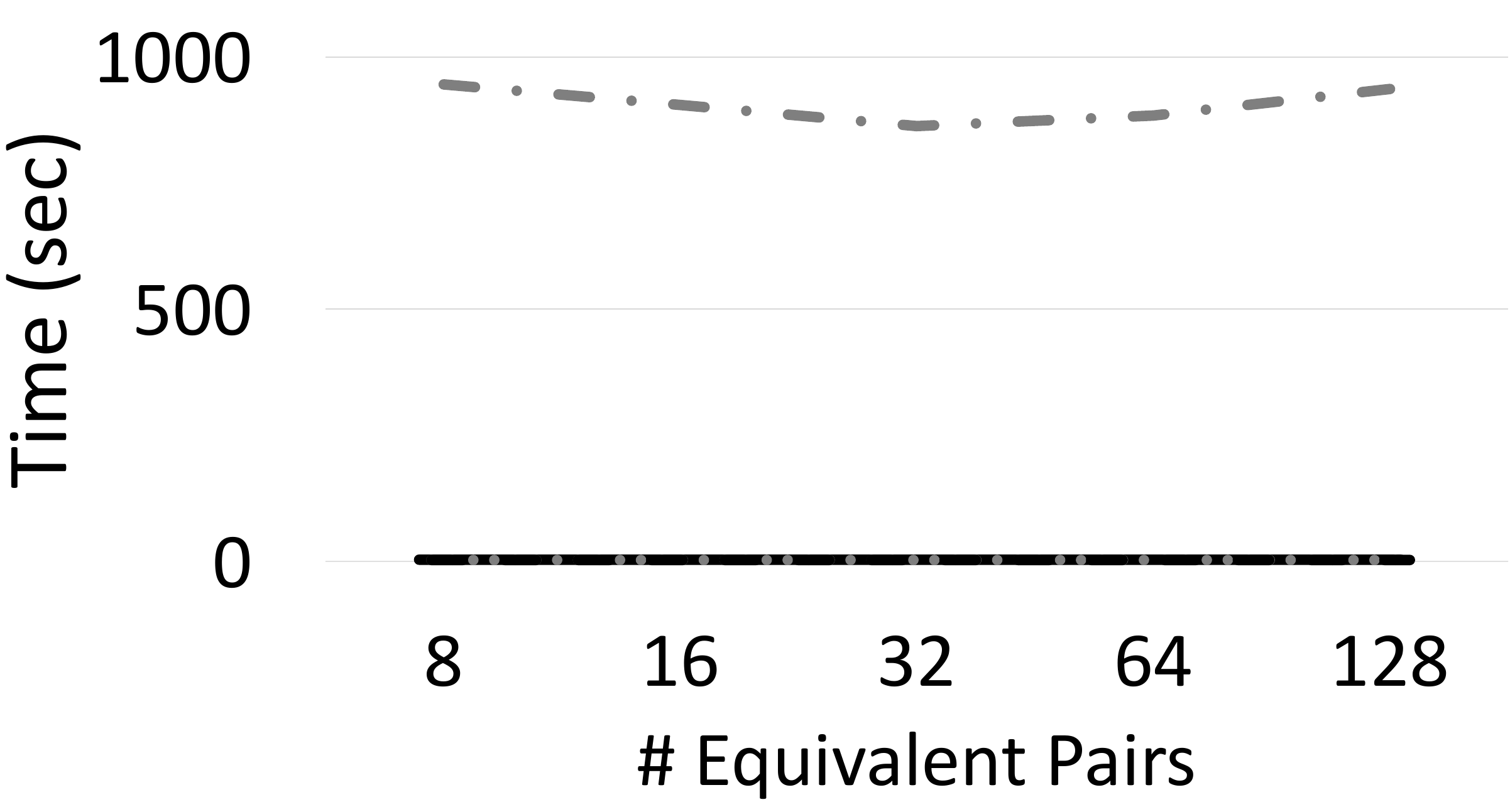}
  \label{subfig:e2e-spes}
}%
\vspace{3ex}
\subfigure[\textbf{Total runtime (w/o SPES)%
}]{
  \includegraphics[width=0.33\textwidth]{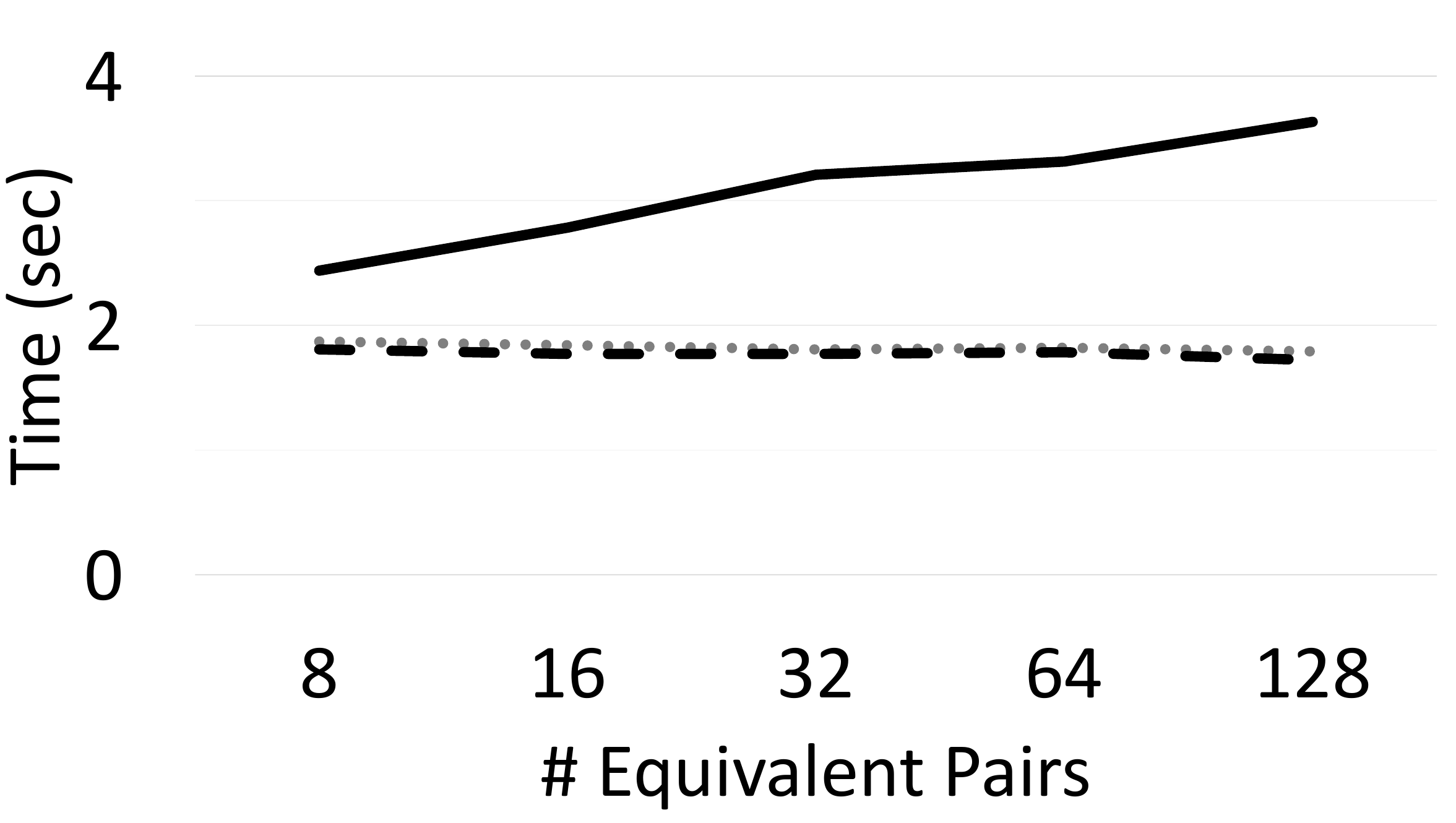}
  \label{subfig:e2e-runtime}
}%
\subfigure[\textbf{Time / equivalence (w/o SPES)%
}]{
  \includegraphics[width=0.33\textwidth]{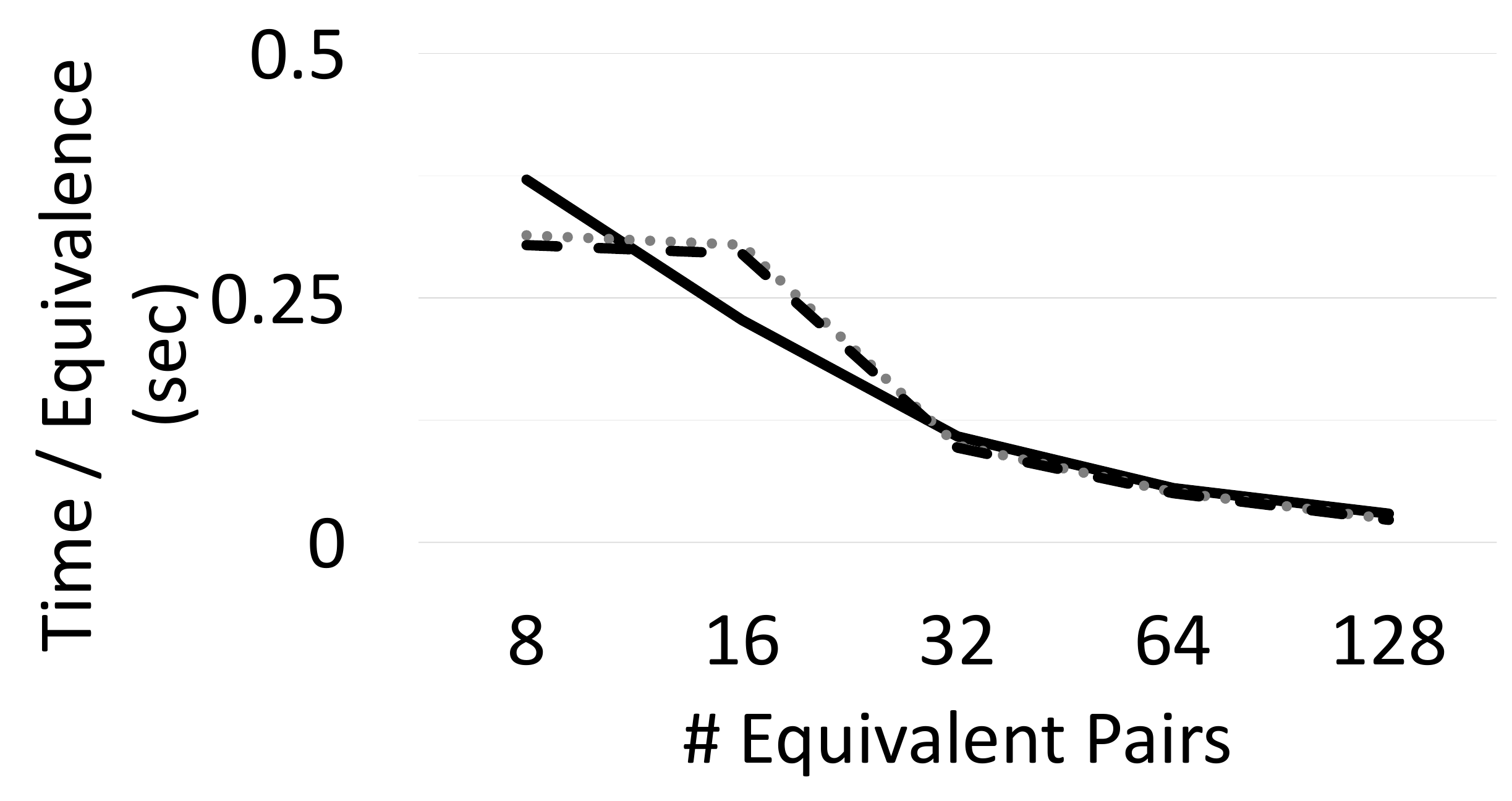}
  \label{subfig:e2e-velocity}
}%
\caption{End-to-end \sys performance vs. the Calcite query optimizer, signature-based detection, and SPES. %
}
\label{fig:EQCMFeedbackendtoend}
\end{figure*}
}

\newcommand{\eqfeedbackloopTPCDSTPCHSep}{
\begin{figure}[t!]
\includegraphics[trim={0 0 0 13.5cm},clip,width=0.5\columnwidth]{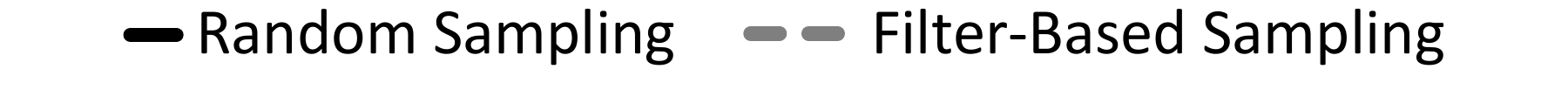}  
\subfigure[\textbf{Accuracy}]{
  \includegraphics[trim={0 2cm 0 0},clip,width=0.49\columnwidth]{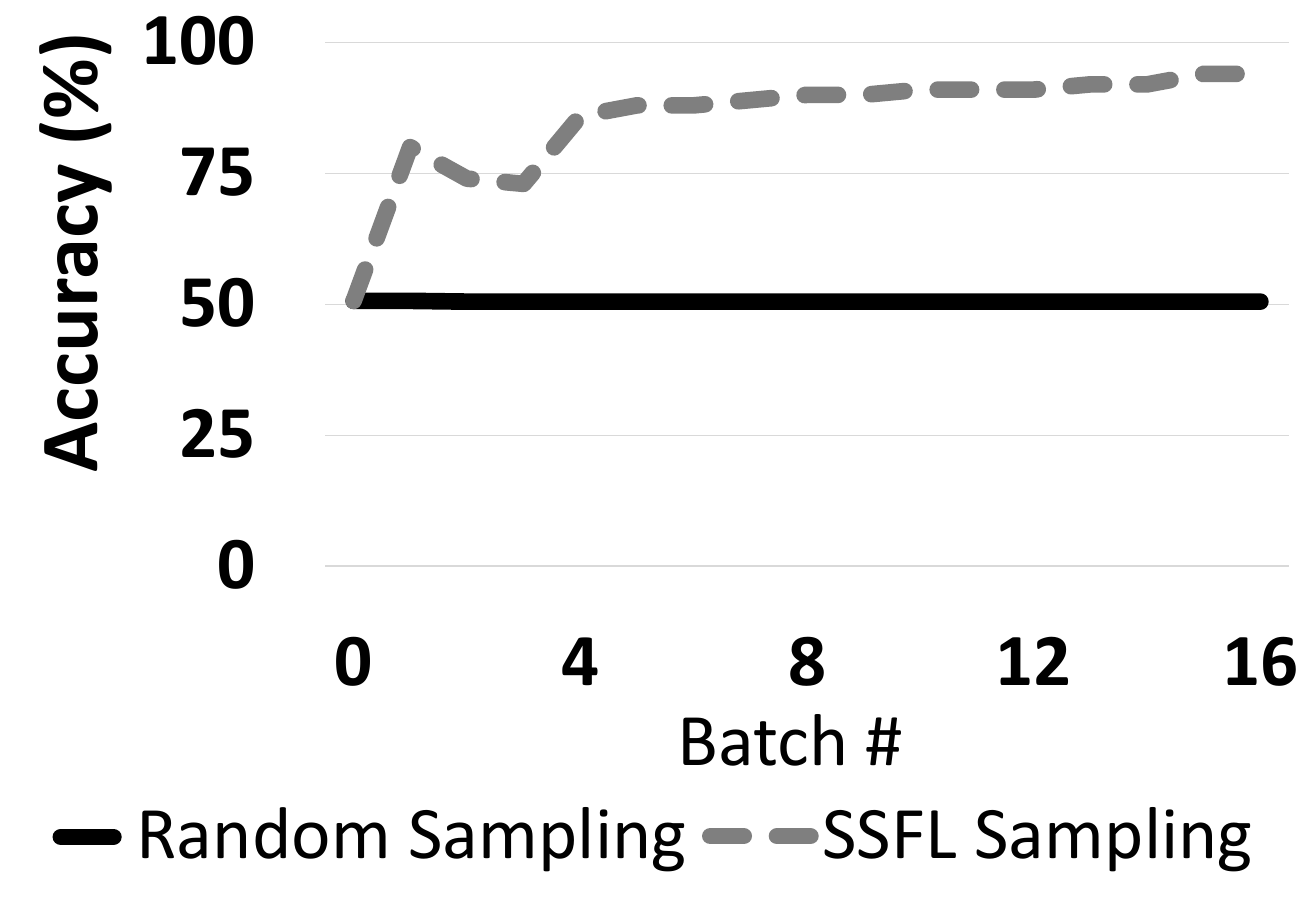}
}%
\subfigure[\textbf{F1 Score}]{
  \includegraphics[trim={0 2cm 0 0},clip,width=0.49\columnwidth]{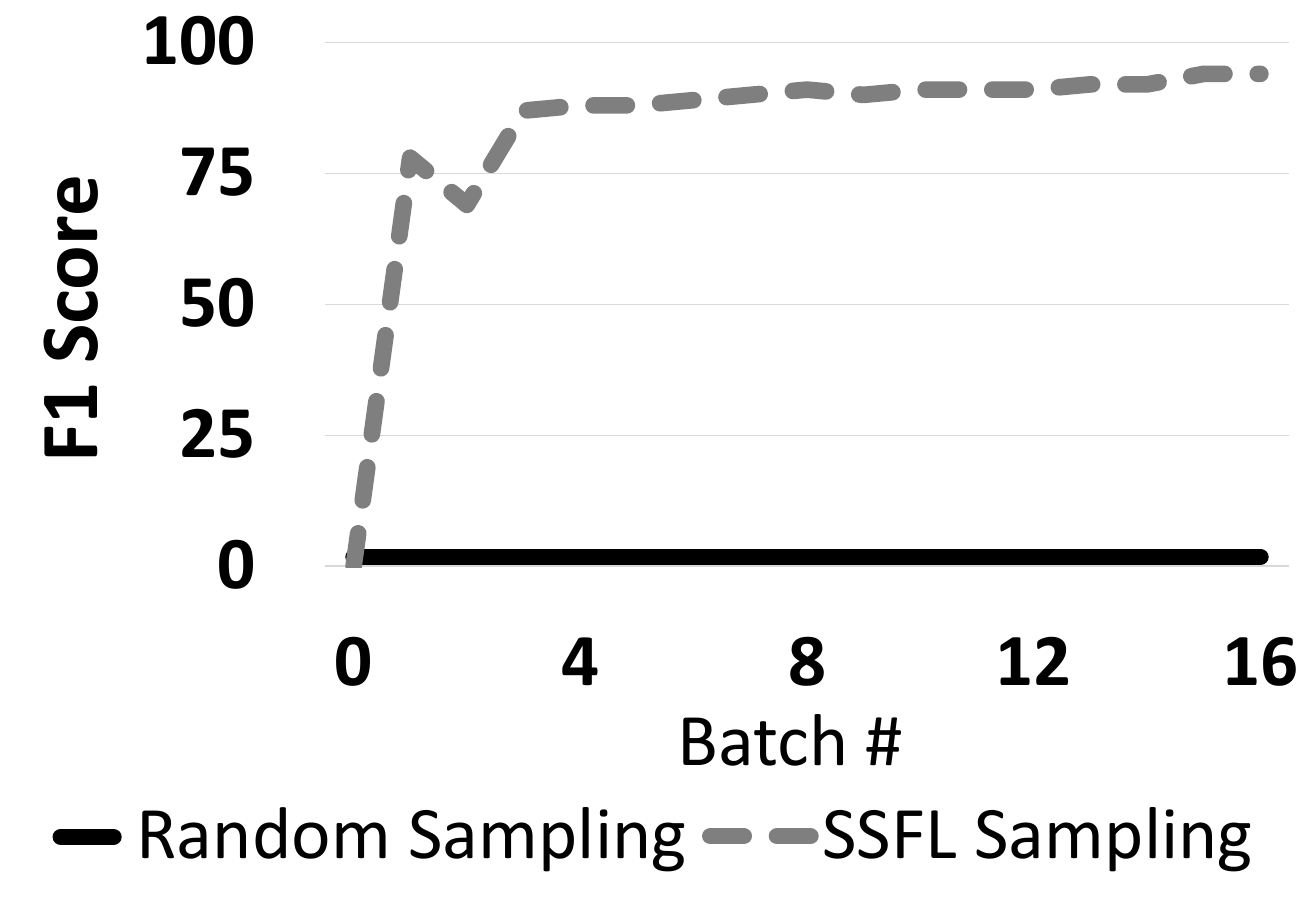}
}
\caption{\ssfl accuracy and F1 for filter-based and random samples.  Each sampling method was used to iteratively select and train over successive batches of 512 samples.}
\label{fig:EQCMFeedbackLoopAcc}
\end{figure}
}

\newcommand{\eqgmmpairsclusters}{
\begin{figure}[!t]
\includegraphics[height=4cm]{Figure/EQ_GMM_NM_Pairs_GMM_Relaxed_TPCH-TPCDS.png}
\caption{Workload Pairs Clustering TPCDS - GMM}
\label{fig:eq_clusters}
\end{figure}
}

\newcommand{\CMTransferLearning}{
\begin{figure*}[t]

    \subfigure[\textbf{MLP}]  {\includegraphics[width=0.30\textwidth]{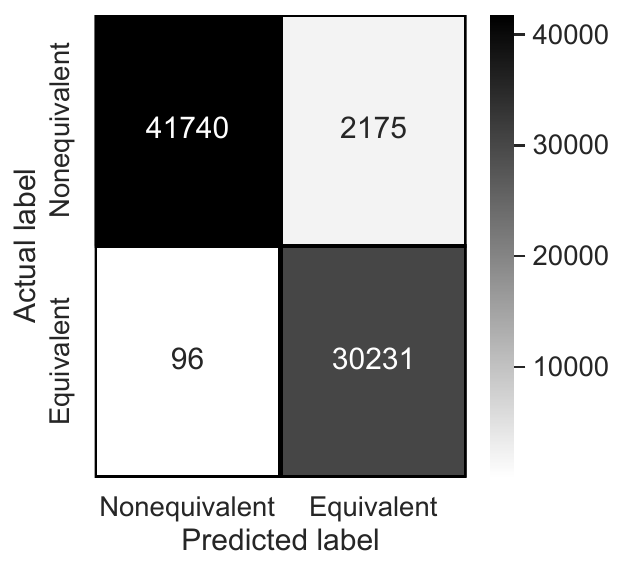}\label{fig:Netowrk_Classifier_CM}}
    \subfigure[\textbf{RF}]{\includegraphics[width=0.30\textwidth]{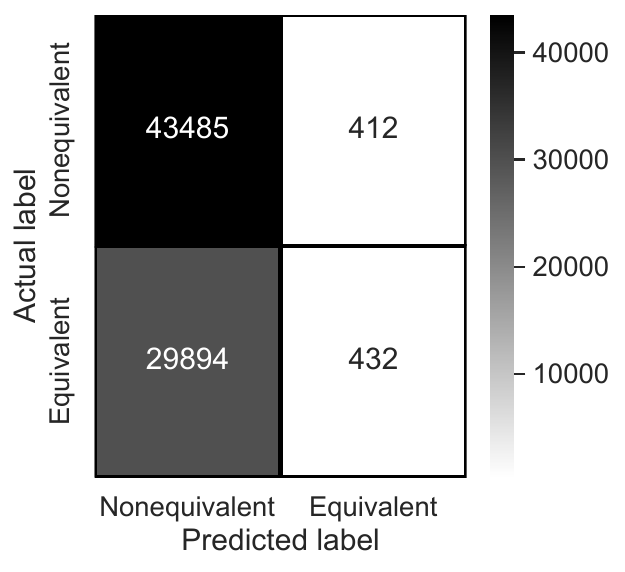}\label{fig:RF_CM}}
    \subfigure[\textbf{LR}]{\includegraphics[width=0.30\textwidth]{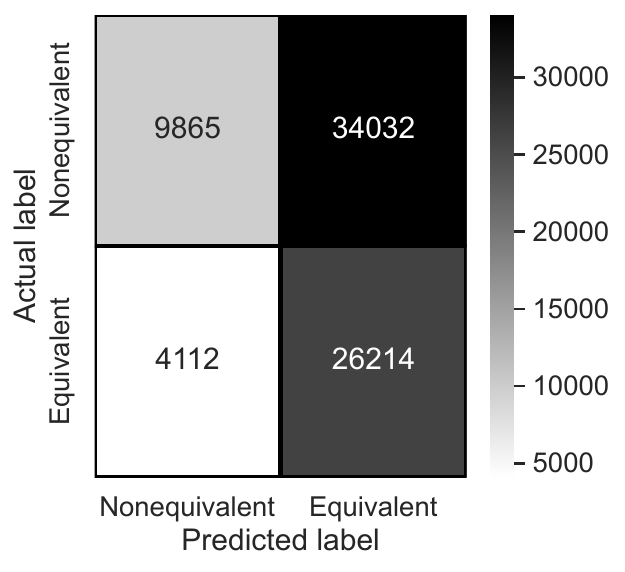}
    }%
    \label{fig:LR_CM}
    \caption{Confusion matrices of three candidate \ecpm models (trained on TPC-H and tested on TPC-DS).} 
    \label{fig:heatmap}
\end{figure*}
}

\newcommand{\CMTransferLearningOriginal}{
\begin{figure*}[t]
    \centering
    \small
    \begin{minipage}[h]{0.34\textwidth}
        \centering
        \footnotesize
        \begin{tabular}{ l c c }
            \toprule
\textbf{Model Type} & \textbf{Accuracy} & \textbf{F1} \\
            \midrule
            $MLP$ & \textbf{0.970} &  \textbf{0.964} \\     
            $RF$  & 0.592 & 0.030\\ 
            $LR$ & 0.588 & 0.486\\            
            \bottomrule
        \end{tabular}
        \captionsetup{type=table}
        \caption{Classifier performance (train TPC-H, test TPC-DS).}
        \label{table:mlTransferLearning}
        \vspace{0.5em}
        \begin{tabular}{ c c c c c }
            \toprule
\textbf{Dataset size} & \textbf{Precision} & \textbf{Recall} & \textbf{F1} \\
            \hline
            \phantom{0}1.2k & 
            0.94 & 0.99 & 0.97 \\
            \phantom{0}5.0k & 
            0.93 & 0.98 & 0.97 \\
            11.0k & 0.90 & 0.96 & 0.94 \\
            19.9k & 0.93 & 0.97 & 0.95 \\
            44.9k & 0.88 & 0.96 & 0.94 \\
            \bottomrule
        \end{tabular}
        \captionsetup{type=table}
        \caption{Transfer learning performance on randomly-generated schema.}        \label{table:mlTransferLearningAlt}
    \end{minipage}%
    \hfill%
    \begin{minipage}[h]{0.62\textwidth}
        \subfigure[\textbf{MLP}]{\includegraphics[width=0.3\textwidth,trim={0 0 14cm 0},clip]{Figure/confusion-matrix-mlp}\label{fig:Netowrk_Classifier_CM}}
        \subfigure[\textbf{RF}]{\includegraphics[width=0.3\textwidth,trim={0 0 14cm 0},clip]{Figure/confusion-matrix-rf}\label{fig:RF_CM}}
        \subfigure[\textbf{LR}]{\includegraphics[width=0.3\textwidth,trim={0 0 14cm 0},clip]{Figure/confusion-matrix-lr}
        }%
        \includegraphics[width=2.33cm,trim={20cm 0 0 0},clip]{Figure/confusion-matrix-mlp}
        \label{fig:LR_CM}
        \caption{Confusion matrices of three candidate \ecpm models (trained on TPC-H and tested on TPC-DS).} 
        \label{fig:heatmap}
        \end{minipage}
\end{figure*}
}

 \newcommand{\emfArchitecture}{
\begin{figure*}[t]
\centering
\includegraphics[width=\textwidth]{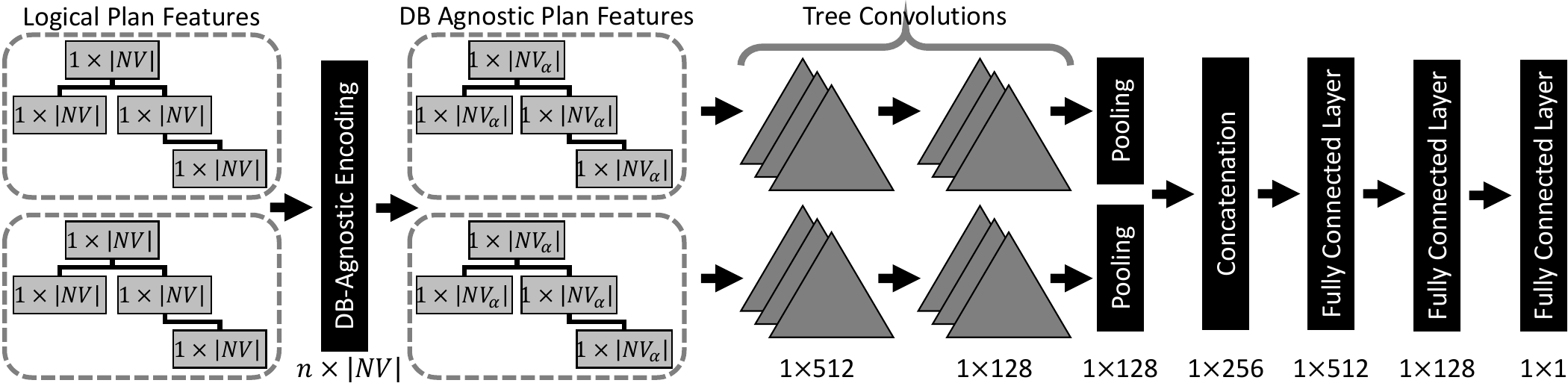}
\caption{The \ecpm architecture. %
Instance-based encoding ($NV$) %
(\S\ref{sec:instance-encoding}); db-agnostic encoding (${NV}_\alpha$) (\S\ref{sec:db-agnostic-encoding}).}
\label{fig:emf-archiecture}
\end{figure*}
}

\newcommand{\ablation}{
\begin{figure}[t!]
\includegraphics[width=0.75\columnwidth]{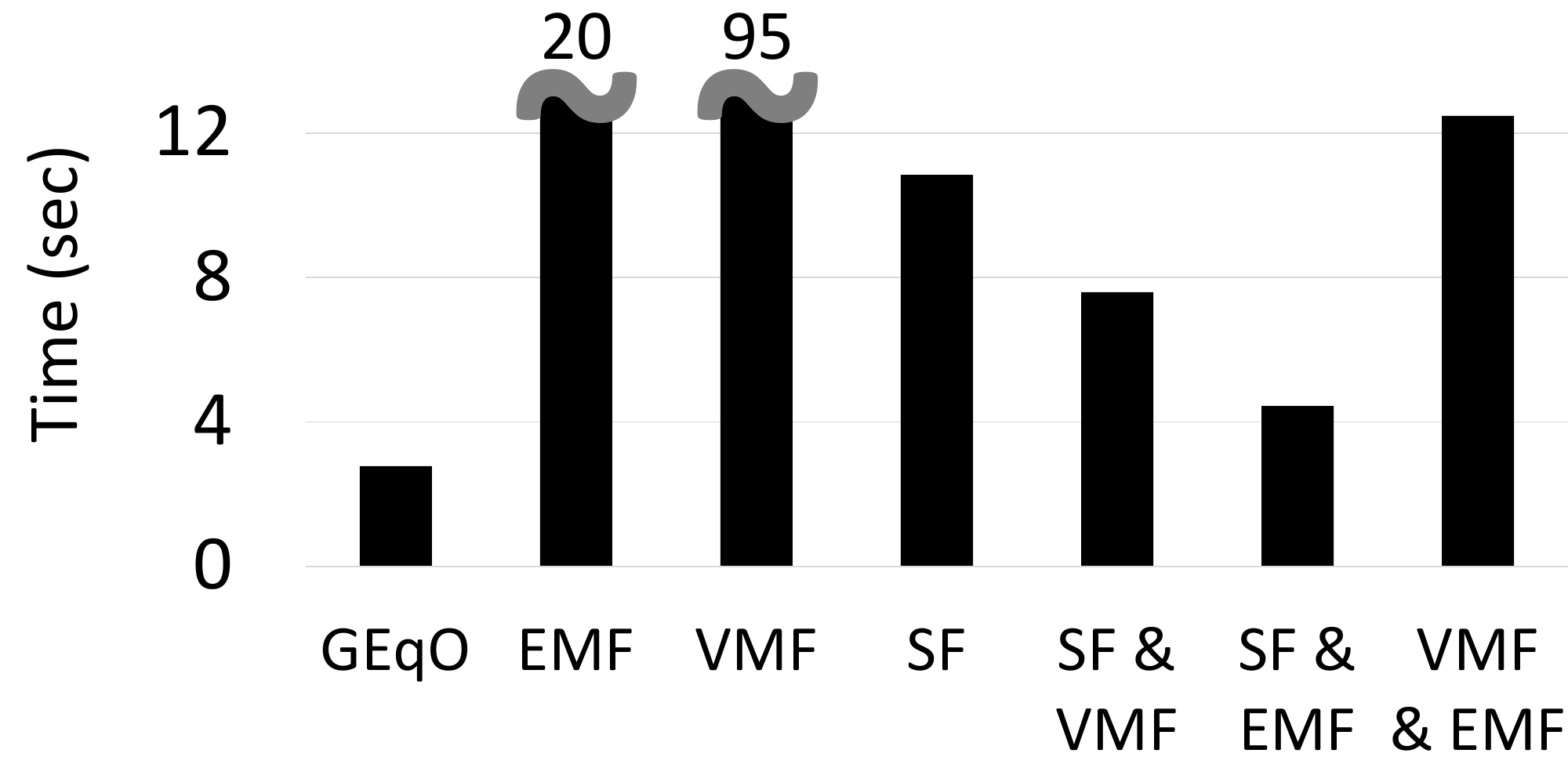}
\caption{\rev{
Total runtime under various filter combinations.
}}
\label{fig:ablation}
\end{figure}
}

\newcommand{\reuse}{
\begin{figure}[t!]
\includegraphics[width=0.750\columnwidth]{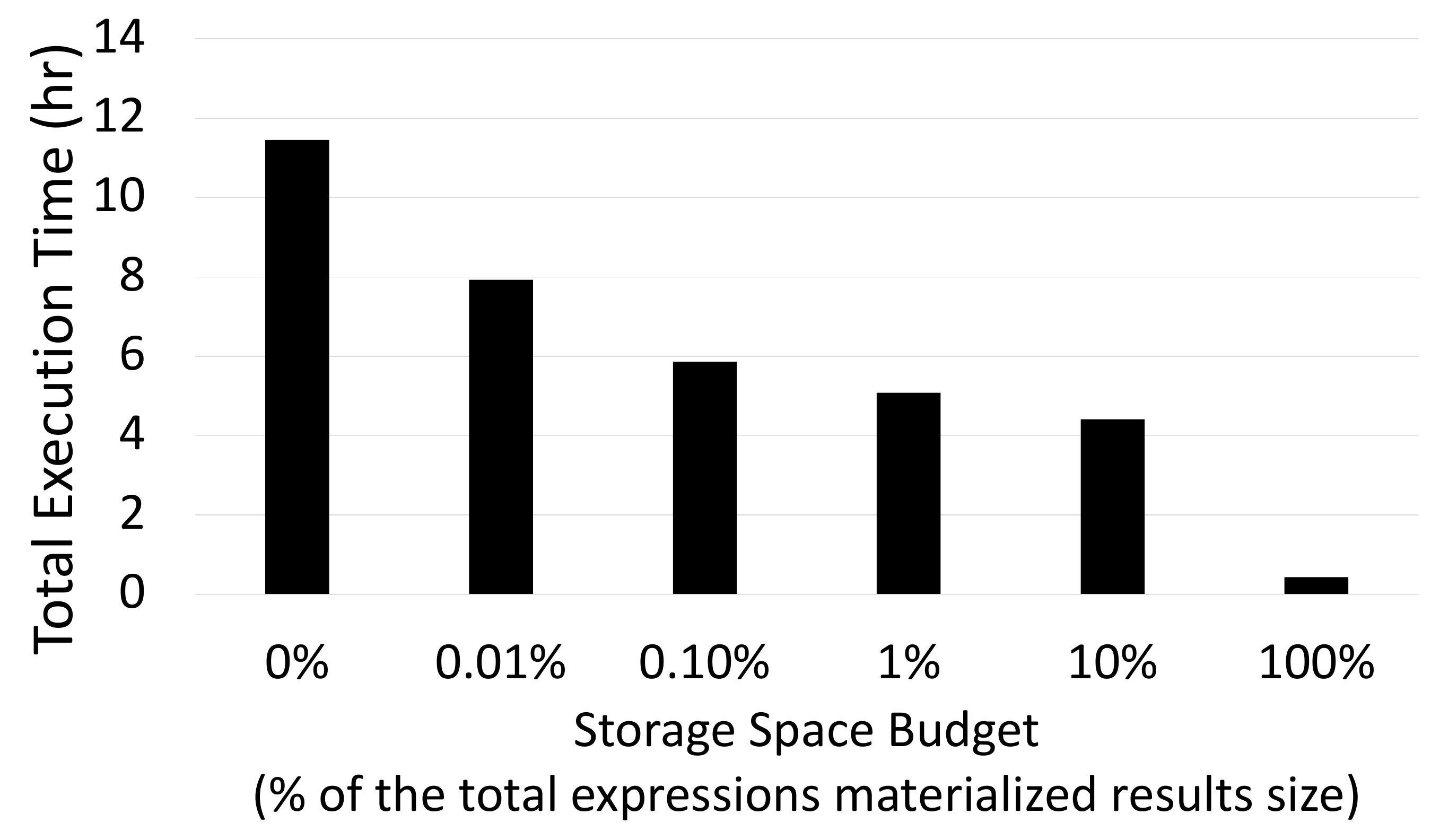}
\caption{\rev{Result Caching Performance. }}
\label{fig:resue}
\end{figure}
}

\newcommand{\filterTable}{
    \begingroup
    \begin{table}
        \centering
        \caption{\sys Filters}
        \begin{tabular}{ l r l l c c }
            \toprule
            \textbf{Filters} & \textbf{PPS} & \textbf{Precision} & \textbf{Recall} & 
            \textbf{\%Reduction} &
            \textbf{Complexity} \\
            \midrule
            \rowcolor{gray!30}
             \sfm & 204\kilo & 0.038 &  	0.986 & & $\mathcal{O}(n)$ \\
             \vpm & 96\kilo & 0.002 & 0.986 & & $\mathcal{O}(n \log n)$\\
             \ecpm & 5k & 0.004 & 0.959 & & $\mathcal{O}(n^3)$\\
            \rowcolor{gray!30}
            AV & 0.02k & 1.0 & 1.0 & & EXPTIME \\
            \sys & \todox{204\kilo} & 1.0 & 1.0 & & EXPTIME \\
            \bottomrule
        \end{tabular}
        \label{table:filter-characteristics}
    \end{table}
    \endgroup
}

\newcommand{\filterTableAlternate}{
    \begingroup
    \begin{table}
        \centering
        \caption{\sys Filters \todox{\textsuperscript {\dagger} We assume AV recall is 1.0, it's actually not.}}
        \begin{tabular}{ l r l c c }
            \toprule

            \textbf{Filters} & 
            \textbf{PPS} &    
            \textbf{Recall} & 
            \textbf{Precision} &
            \textbf{Complexity} 
            \\
            \midrule
            
            \sfm & 
            204\kilo & 
            0.986 & 0.966 & 
            $\mathcal{O}(n)$ 
            \\

            \rowcolor{gray!30}
            \vpm & 
            96\kilo & 
            0.986 & 
            0.731 & 
            $\mathcal{O}(n \log n)$
            \\

            \ecpm & 
            5\kilo & 
            0.959 & 
            0.916 & 
            $\mathcal{O}(n^3)$
            \\

            \rowcolor{gray!30}
            AV & 
            0.02\kilo & 
            $1.000^{\dagger}$ & 
            1.000 & 
            $\mathcal{O}(n \cdot 2^{\Omega(s)})$ \\
            
            $\text{Oracle}{+}\text{AV}$ & 
            0.02\kilo & 
            1.000 & 
            1.000 & 
            $opt = \mathcal{O}(\lvert E \rvert \cdot 2^{\Omega(s)})$
            \\

            \midrule

            \rowcolor{gray!30}
            \textbf{\sys} &             \textbf{90\kilo} & 
            \textbf{0.955} & 
            & 
            \textbf{$\mathcal{O}(\epsilon \cdot 2^{\Omega(s)}) + opt$}
            \\
            
            \bottomrule
        \end{tabular}
        \label{table:filter-characteristics}
    \end{table}
    \endgroup
}

\newcommand{\filterTableAlternateX}{
    \begingroup
    \begin{table}
        \centering
        \caption{\sys filter performance.  The ``Oracle+AV'' row shows the optimal case where an oracle correctly identifies all equivalent pairs which are then verified.  We assume a verifier with perfect recall.}
        \begin{tabular}{ l c c c c c }
            \toprule

            \textbf{Method} & 
            \textbf{Time (s)} &    
            \textbf{Acc.} &
            \textbf{Prec.} &
            \textbf{Rec.} & 
            \textbf{Complexity} 
            \\
            \midrule

            \sfm & 
            \phantom{00}0.3 & 
            0.97 & 
            0.03 & 
            0.99 & 
            \footnotesize$\mathcal{O}(n)$ 
            \\

            \rowcolor{gray!30}
            \vpm & 
            \phantom{00}0.5 & 
            0.98 & 
            0.06 & 
            0.98 &
            \footnotesize$\mathcal{O}(n \log n)$
            \\

            \ecpm &
            \phantom{00}9.4 & 
            0.95 &
            0.10 & 
            0.96 & 
            \footnotesize$\mathcal{O}(n^3)$
            \\

            \rowcolor{gray!30}
            AV & 
            898.5 & 
            1.00 &
            1.00 & 
            1.00 & 
            \footnotesize$\mathcal{O}(n \cdot 2^{\Omega(s)})$ \\
            
            $\text{Oracle}{+}\text{AV}$ & 
            \phantom{00}1.0 & 
            1.00 &
            1.00 & 
            1.00 & 
            \footnotesize$opt = \mathcal{O}(\lvert E \rvert \cdot 2^{\Omega(s)})$
            \\

            \midrule

            \rowcolor{gray!30}
            \textbf{\sys} &          \textbf{\phantom{00}3.1} & 
            \textbf{1.00} &
            \textbf{1.00} & 
            \textbf{0.93} & 
            \footnotesize\boldmath {$\mathcal{O}(\epsilon \cdot 2^{\Omega(s)}) + opt$}
            \\

            \bottomrule
        \end{tabular}
        \label{table:filter-characteristics}
    \end{table}
    \endgroup
}

\newcommand{\mlTransferLearningFinal}{
  \begin{table}[t]
    \centering
    \caption{Classifier performance (train TPC-H, test TPC-DS).}%
    \begin{tabular}{ l c c }
        \toprule
\textbf{Model Type} & \textbf{Accuracy} & \textbf{F1} \\
        \midrule
        $MLP$ & \textbf{0.970} &  \textbf{0.964} \\     
        $RF$  & 0.592 & 0.030\\ 
        $LR$ & 0.588 & 0.486\\            
        \bottomrule
    \end{tabular}
    \label{table:mlTransferLearning}
  \end{table}
}

\newcommand{\mlTransferLearningAltFinal}{
\begin{table}[t]
    \centering
\caption{Transfer learning performance on randomly-generated schema.}            \begin{tabular}{ c c c c c }
    \toprule
\textbf{Dataset Size} & \textbf{Precision} & \textbf{Recall} & \textbf{F1} \\
    \midrule
    \phantom{0}1.2k & 
    0.94 & 0.99 & 0.97 \\
    \phantom{0}5.0k & 
    0.93 & 0.98 & 0.97 \\
    11.0k & 0.90 & 0.96 & 0.94 \\
    19.9k & 0.93 & 0.97 & 0.95 \\
    44.9k & 0.88 & 0.96 & 0.94 \\
    \bottomrule
\end{tabular}
\label{table:mlTransferLearningAlt}
\end{table}
}

\newcommand{\filterTableAlternateXX}{
    \begingroup
    \begin{table*}[t!]
        \centering
        \caption{Performance of \sys and its filters (\sfm, \vpm, \ecpm) on ${\sim}$50k subexpression pairs and 50 equivalences generated using a TPC-DS schema; see \S\ref{sec:baseline_expr}.}
        \label{table:filter-characteristics}
        \begin{tabular}{ l c c c l }
            \toprule
            \textbf{Filter} & 
            \textbf{Time (sec)} &    
            \textbf{TPR} &
            \textbf{TNR} &
            \textbf{Complexity} 
            \\
            \midrule
            Schema Filter (\sfm) & 
            \phantom{00}0.3 & 
            0.98 & 
            0.37 & 
            $\mathcal{O}(n)$ 
            \\
            Vector Matching Filter (\vpm) & 
            \phantom{00}0.5 & 
            0.98 & 
            0.66 &
            $\mathcal{O}(n \log n)$
            \\
            Equivalence Model Filter (\ecpm) &
            \phantom{00}1.3 & %
            0.98 & 
            0.80 & 
            $\mathcal{O}(n^3)$
            \\
            Automated Verifier (\textit{AV}) & 
            898.5 & 
            1.00 & 
            1.00 & 
            \rev{$\mathcal{O}(n \cdot 2^{\Omega(\gamma)})$}
            \\
            \midrule
            \textbf{\sys} &          
            \textbf{\phantom{00}3.1} & 
            \textbf{0.93} & 
            \textbf{1.00} & 
            \rev{$\mathcal{O}(\epsilon \cdot 2^{\Omega(\gamma)}) + opt$}
            \\
            $\text{Oracle}+\text{AV}$ & 
            \phantom{00}1.0 & 
            1.00 & 
            1.00 & 
            \rev{$opt = \mathcal{O}(\lvert E \rvert \cdot 2^{\Omega(\gamma)})$}
            \\
            \bottomrule
        \end{tabular}
    \begin{tablenotes}
    \small
    \vspace{1mm}
\item We report true positive rate (TPR) and true negative rate (TNR). The ``Oracle+AV'' row shows a hypothetical optimal case where an oracle correctly identifies all equivalent pairs, which are then verified. We assume a verifier with perfect recall. \rev{$n$ is the number of subexpressions, $\gamma$ is the number of symbols in the AV's SAT formulation, and $E$ is the set of equivalent subexpression pairs.  \sys verifies $\epsilon$ more pairs than the oracle, which we empirically show to be ${\sim}5\text{--}10\%$ (see \autoref{sec:baseline_expr}).}
  \end{tablenotes}
    \end{table*}
    \endgroup
}

\newcommand{\queriesInvolvedTable}{
    \begingroup
    \begin{table}
        \centering
        \caption{Queries with a subexpression equivalent to or contained in another query}
        \begin{tabular}{ l l l }
            \toprule
Workload & Equivalence & Containment \\
            \hline

            TPC-DS & x / 99 (z\%) & y / 99 (a\%) \\            
            TPC-H & x / 99 (d\%) & y / 99 (e\%) \\            
            \bottomrule
        \end{tabular}
        \label{table:queriesInvolvedTable}
    \end{table}
    \endgroup
}

\newcommand{\mlModelsSize}{
    \begingroup
    \begin{table}[htbp!]
        \centering
        \caption{ML Models Sizes}
        \begin{tabular}{ l l }
            \toprule
\textbf{ML Model} & \textbf{Size (MB)}\\
            \hline
            $LR$ & x \\            
            $MLP$ & x\\      
            $RF$  & x\\ 
            \bottomrule
        \end{tabular}
        \label{table:mlModelSizes}
    \end{table}
    \endgroup
}

\newcommand{\mlTransferLearning}{
    \begingroup
    \begin{table}[t!]
        \centering
        \footnotesize
        \begin{tabular}{ l c c }
            \toprule
\textbf{Model Type} & \textbf{Accuracy} & \textbf{F1} \\
            \midrule
            $MLP$ & \textbf{0.970} &  \textbf{0.964} \\     
            $RF$  & 0.592 & 0.030\\ 
            $LR$ & 0.588 & 0.486\\            
            \bottomrule
        \end{tabular}
        \caption{Classifier performance (train TPC-H, test TPC-DS).}
        \label{table:mlTransferLearning}
    \end{table}
    \endgroup
}

\newcommand{\mlTransferLearningAlt}{
    \begingroup
    \begin{table}[t]
        \centering
        \caption{Transfer learning performance on random schema.}
        \begin{tabular}{ c c c c }
            \toprule
\textbf{Accuracy} & \textbf{Precision} & \textbf{Recall} & \textbf{F1} \\
            \midrule
            0.93 & 0.91 & 0.98 & 0.96 \\
            \bottomrule
        \end{tabular}
        \label{table:mlTransferLearningAlt}
    \end{table}
    \endgroup
}

\newcommand{\VMFperformance}{
    \begingroup
    \begin{table}[t!]
        \centering
        \caption{\vpm performance (train TPC-H, test TPC-DS).
        }
        \begin{tabular}{ c c c c }
            \toprule
\textbf{Accuracy} & \textbf{Precision} & \textbf{Recall} & \textbf{F1} \\
            \midrule
            0.74 & 0.42 & 0.98 & 0.60 \\
            \bottomrule
        \end{tabular}
        \label{table:vmf}
    \end{table}
    \endgroup
}

\begin{abstract}
Large scale analytics engines have become a core dependency for modern data-driven enterprises to derive business insights and drive actions. These engines support a large number of analytic jobs processing huge volumes of data on a daily basis, and workloads are often inundated with overlapping computations across multiple jobs.
Reusing common computation is crucial for efficient cluster resource utilization and reducing job execution time. Detecting common computation
is the first and key step for reducing this computational redundancy. However, detecting equivalence on large-scale analytics engines requires \textit{efficient} and \textit{scalable} solutions that are fully \textit{automated}. In addition, to maximize computation reuse, equivalence needs to be detected at the \textit{semantic} level instead of just the \textit{syntactic} level (\ie the ability to detect semantic equivalence of seemingly different-looking queries). Unfortunately, existing solutions fall short of satisfying these requirements. 

In this paper, we take a major step towards filling this gap by proposing \sys, a portable and lightweight machine-learning-based framework for efficiently identifying \textit{semantically} equivalent computations at scale. \sys introduces two machine-learning-based filters that quickly prune out nonequivalent subexpressions and employs a semi-supervised learning feedback loop to iteratively improve its model with an intelligent sampling mechanism. Further, with its novel \textit{database-agnostic} featurization method, \sys can \textit{transfer} the learning from one workload and database to another. Our extensive empirical evaluation shows that, on TPC-DS-like queries, \sys yields significant performance gains---up to $200\times$ faster than automated verifiers---and finds up to $2\times$ more equivalences than optimizer and signature-based equivalence approaches.
\end{abstract}

\maketitle

\section{Introduction}

Modern data-driven enterprises fundamentally rely on large-scale analytics engines (e.g.,  
Spark~\cite{spark}, SCOPE~\cite{scope}, Synapse~\cite{synapse}, BigQuery~\cite{bigquery}, Redshift~\cite{redshift})
to derive business insights and drive actions.
Concretely, engines such as SCOPE 
process exabytes of data and execute millions of jobs, with trillions of operators~\cite{jindal2021production} per  cluster~\cite{zhu2021kea}.
Computational redundancy within these analytics engines is strikingly common~\cite{jindal2021production,alibaba}, 
where intermediate results are duplicated across different queries (\ie they contain \textit{equivalent subexpressions}). 
According to Jindal et al. ~\cite{jindal2018computation}, 
about 40\% of the jobs in SCOPE contain equivalent subexpressions (i.e., at least one subexpression is equivalent to a subexpression in another job). %
Because of this pervasive redundancy, identifying and reusing common computation has long been recognized as a critical technique to improve query performance and reduce operational costs.
For example, a wide range of tools and approaches for leveraging materialized views have been developed, including CloudViews~\cite{jindal2018selecting}, Google Napa~\cite{napa}, and Redshift AutoMV~\cite{automv}. Common computation reuse has also been exploited for multi-query optimization~\cite{mqo22,mqo88} in the context of multiple-query-at-a-time systems. 
 
\rev{For all these tools and techniques, detecting equivalent subexpressions is the first and crucial step.
For example, view selection algorithms (e.g. \cite{agrawal2000automated}) maximize the benefit of 
materializing computation that is most redundant in cost or frequency of use, under a storage or maintenance cost constraint. %
Similarly, view matching relies on detecting and leveraging equivalent views 
to improve query performance. At the query level, identifying equivalence is also a %
crucial step in efficient rewriting (either automatically by an optimizer or manually by a DBA), where a query is transformed into an equivalent---but better-performing---variant~\cite{cascadeopt, volcanoopt}.  %
Finally, determining query equivalence is also important in generating functional or performance tests %
for database implementations~\cite{liu2022automatic,wetune}.}

In this paper, we focus on the problem of \textit{detecting subexpression equivalence at scale}.\footnotemark %
\xspace There are a number of distinct challenges in doing so. 
First, the detection process must be \textit{automatic}
 due to the sheer number of developers and jobs involved. 
Second, scalability is crucial as quadratic pairwise comparison over
trillions of subexpressions is intractable in most current solutions.
Third, to maximize computation reuse, equivalence detection needs to be sufficiently \textit{general} to identify common computation expressed in different ways by different users.
This means that a detection algorithm should 
go beyond merely
``judging a book by its cover'' (i.e., only identifying superficially- or syntactically-equivalent subexpressions)
but rather ``look beneath the surface'' to detect \textit{semantic equivalence} between %
subexpressions with dissimilar structures. 
Figure~\ref{lst:example1} shows such an example, where the highlighted subexpressions differ syntactically but are nonetheless semantically equivalent.

\footnotetext{\rev{This work does not propose a novel view selection or rewriting algorithm. Rather, it presents a framework designed to accelerate equivalence detection, which is considered a fundamental step for these and other algorithms}.}

\begin{figure}[t]
\centering
\begin{sqllisting}
        Q1: SELECT y, AVG(x) FROM (
        \end{sqllisting}
        \vspace{-0.75em}
        \begin{boxedlisting}     
SELECT A.x, B.y FROM A, B
WHERE A.joinKey = B.joinKey 
  AND A.val > B.val + 10 
  AND B.val > 10
        \end{boxedlisting}
        \vspace{-0.75em}
        \begin{sqllisting}
            ) GROUP BY B.y
    
        Q2: SELECT SUM(x), SUM(y) FROM (
        \end{sqllisting}
        \vspace{-0.75em}
        \begin{boxedlisting}
SELECT A.x, B.y FROM B, A
WHERE B.joinKey = A.joinKey 
  AND B.val + 10 < A.val 
  AND B.val + 10 > 20 
  AND A.val > 20
        \end{boxedlisting}
        \vspace{-3mm}
        \begin{sqllisting}
            )
    \end{sqllisting}
\caption{%
Two queries that contain semantically-equivalent subexpressions highlighted by shaded boxes.}
\label{lst:example1}
\end{figure}

Existing approaches to detecting subexpression equivalence
do not address all of the above challenges.
Optimizer-based approaches, which are used by
many classical materialized view selection and matching algorithms~\cite{agrawal2000automated, goldstein2001optimizing}, defer to the query optimizer to detect equivalence. 
This approach lacks generality, given that 
even highly-mature optimizers such as SQL Server
are missing equivalence rules necessary to identify common scenarios~\cite{wetune}.
It is also  inefficient
given cloud-scale volumes of complex queries, where the query optimizer quickly becomes a bottleneck.
Manual approaches, commonly used in %
many relational OLAP databases---including state-of-the-art cloud-based analytics systems like Snowflake~\cite{snowflake}, BigQuery, and NAPA--- %
 require users to manually identify common computations and create materialized views, which is error-prone, tedious and simply does not scale.
Signature-based view materialization approaches, like CloudViews ~\cite{jindal2018selecting}, use Merkle tree-like signatures for efficient identification of syntactically-identical subexpressions. However, this approach sacrifices completeness as it may miss opportunities for identifying semantically-equivalent subexpressions, as illustrated in \autoref{lst:example1}.
At the other end of the spectrum, 
verification-based approaches, such as Cosette~\cite{chu2017hottsql} and SPES~\cite{zhou2020symbolic}, formally prove the semantic equivalence of queries using automated proof assistants or SMT solvers. While these approaches are highly effective, they suffer from scalability issues. Exhaustively evaluating all pairs of subexpressions over a single day of jobs at cloud-scale would require over a trillion expensive formal verifications and more than a century of compute time!

\sloppy{In this paper, we introduce \textit{\sys} (a \textbf{G}eneral \textbf{Eq}uivalence 
\textbf{O}ptimizer})
, which addresses the aforementioned challenges. 
\sys is a general framework for efficiently identifying 
semantically-equivalent subexpressions at scale. 
It applies a series of  \textit{equivalence filters} to sets of subexpressions, enabling accelerated detection. To ensure correctness, \sys finally applies an expensive formal verifier---but only after filtering most nonequivalent subexpressions, which constitute the vast majority of the pairs. As a result, \sys produces subexpression pairs that are, with perfect precision and near-perfect recall, semantically equivalent. 

A desirable equivalence filter has two important properties: it should (i) admit virtually all of the equivalences (i.e., exhibit a high true positive rate; TPR) and (ii) reject most non-equivalences (i.e., have a high true negative rate; TNR).
\autoref{table:filter-characteristics} illustrates this for \sys's filters (detailed below), where 
the TPR is near-perfect, and the TNR steadily increases until all negatives have been eliminated.

To maximize performance, \sys arranges filters to rapidly reject   ``easy'' nonequivalent subexpression pairs, with faster filters applied first, as shown in Table \ref{table:filter-characteristics}. Slower but increasingly complex filters are then applied to identify more difficult cases. This trade-off allows \sys to achieve performance close to optimal, assuming an oracle that verifies only equivalent pairs, and is almost $200\times$ faster than verifying all subexpression pairs.

\filterTableAlternateXX

While prior work has established quick-but-low-precision heuristic-based filters---i.e., matching common table and column sets~\cite{goldstein2001optimizing}, which we refer to as \textit{schema filter (\sfm)} in Table~\ref{table:filter-characteristics}---and expensive \textit{automated verifiers} (\textit{AV} in \autoref{table:filter-characteristics}) that are slow with perfect precision, there currently exists no ``middle ground'': a way to filter non-equivalent subexpressions rapidly with high precision. \sys fills this gap by introducing two such filters.

First, \sys's \textit{vector matching filter (\vpm)} embeds subexpressions in a learned vector space and identifies likely equivalent pairs by applying an approximate nearest neighbor search (ANNS).  ANNS is a popular, high-performance technique~\cite{DBLP:conf/kdd/Qin000W21,DBLP:journals/corr/abs-2212-07588} with moderate precision.  \sys leverages the \vpm to efficiently prune moderately-difficult cases not handled by the SF, while at the same time ensuring that equivalence pairs are admitted with high recall.

Next, \sys's \textit{equivalence model filter} (\ecpm) employs a novel, high-precision, 
supervised ML model trained over a workload sample to predict semantic equivalence. 
As we detail below, the \ecpm is database- and schema-agnostic and can be easily transferred to other workloads.
As far as we are aware, \sys is the first work to present a machine-learning-accelerated framework for detecting semantic equivalence at scale.

A key challenge in training the \ecpm is the need for large amounts of labeled data. Although the cloud makes collecting query workloads much more accessible, labeling the equivalent subexpressions within the workload requires running expensive equivalence verifiers on all subexpression pairs (i.e., trillions of invocations). To reduce this cost, \sys employs a semi-supervised feedback loop (\ssfl) pipeline that iteratively improves the accuracy of the \ecpm until it matures. 
The \ssfl employs inexpensive filters (i.e., the SF and \vpm) to 
ensure approximately balanced classes in its generated training data.
This approach enables \sys to both avoid the cold start training problem and 
 fine-tune its \ecpm model as new workload data becomes available for training.

A second challenge addressed by \sys involves ensuring that its learned \ecpm model is not tied to a fixed database schema. For example, the \ecpm should be able to determine that the subexpressions shown in \autoref{lst:example1} are equivalent even if table \texttt{A}'s name was replaced with \texttt{C}.
Unlike existing instance-based ML-for-DB solutions~\cite{containrate}, GEqO uses a database and schema-agnostic approach that focuses on learning general semantic equivalence patterns. It accomplishes this during \ecpm featurization by replacing references to
database schema with symbolic correspondences. This allows \sys\ to pretrain on existing database workloads %
and apply the resulting model to %
new database workloads.

\sys is a standalone framework that can be used alongside a query optimizer to complement its ability to detect equivalent computation. Unlike adding new rewrite rules, which requires changing the core database engine code, \sys can learn any equivalence relationship in a workload, including those missed by the optimizer. We focus on subexpressions that
contain selections, projections, and joins (i.e., SPJ subexpressions) with conjunctive predicates.

Through detailed experiments, we demonstrate the efficiency and effectiveness of \sys in detecting common computations. We systematically 
evaluate its filters
and discuss their trade-offs between prediction accuracy and the overhead involved.

\vspace{1mm}
\noindent\textbf{Contributions.} The paper makes the following contributions:
\begin{itemize}
\item We introduce \sys, a scalable ML-framework for detecting semantically-equivalent subexpressions. \sys's novel \vpm and \ecpm filters quickly prune non-equivalent subexpressions, reducing the overhead of running expensive equivalence  verifiers by up to $200\times$ (\S\ref{sec:prelim-overview}). %

\item We introduce a database agnostic featurization technique that generalizes instance-specific (non)equivalent subexpression pairs into (non)equivalent patterns, making the %
\ecpm transferable to new workloads and databases (\S\ref{sec:logical-encoding}).
\item We address the challenge of requiring large volumes of labeled training data by introducing a semi-supervised feedback loop (\ssfl) to iteratively improve \sys's \ecpm filter (\S\ref{sec:labeling}).  This process is aided by drawing high-quality samples leveraging the cheaper \sfm and \vpm filters.
\item Our evaluation demonstrates the efficiency and effectiveness of \sys %
(\S\ref{sec:evaluation}).
\end{itemize}

\eat{
The rest of the paper is organized as follows. Section~\ref{sec:prelim-overview} introduces preliminaries and provides a system overview of \sys. Section~\ref{sec:feature_engineering} details the db-agnostic feature engineering in \sys. Section~\ref{sec:labeling} discusses the ML-based pre-filtering approach and the semi-supervised learning feedback loop. Section~\ref{sec:predication} describes the efficient prediction of equivalence using \sys. Section~\ref{sec:evaluation} presents experimental evaluation of \sys. Section~\ref{sec:relatedWork} discusses the related work, and finally Section~\ref{sec:conclusion} concludes this paper.}

\section{Preliminaries and Overview}\label{sec:prelim-overview}

This section defines the key concepts used in the paper and provides an %
overview of \sys.

\subsection{Problem definition}
\label{sec:prelim}

\sys assumes that a SQL query can be transformed into a tree (\ie a logical plan) $Q$ consisting of operator nodes (we use $ops(Q)$ to denote the set of all operators in $Q$).  We term each subtree rooted at node $i$ to be a \textit{subexpression} $q_i$ of $Q$.
Let $S(Q)=\{q_1,...,q_n\}$ be the set of all subexpressions induced by $Q$.
Note that $Q \in S(Q)$; the root of the logical plan is itself a (trivial) subexpression of $Q$.

\sys assumes that as a subtree in a logical query plan, subexpressions are unambiguously executable.
Let $q_i(d)$ denote the result of executing subexpression $q_i$ on some database instance $d$.  Let $D$ be the set of all database instances.
Given two subexpressions $q_i$ and $q_j$, they are
\textit{semantically equivalent}, denoted as $q_i \equiv q_j$,
if and only if 
$\forall d \in D, q_i(d)=q_j(d)$.
Note that $q_i$ and $q_j$ need not be drawn from the same query $Q$, and that this definition holds under both set and bag semantics~\cite{cohen2006equivalence}.  

An \textit{equivalence verifier} applies an automated technique (e.g., a proof assistant~\cite{de2015lean} %
or formal solver
~\cite{ebner2017metaprogramming})
to decide $q_i \equiv q_j$.    
We denote equivalence determined using an automated verifier $AV$ as $q_i \stackrel{\mathclap{\normalfont\mbox{\tiny AV}}}{\equiv} q_j$.
A verifier is correct but not complete (\ie $(q_i \stackrel{\mathclap{\normalfont\mbox{\tiny AV}}}{\equiv} q_j)\Rightarrow (q_i \equiv q_j)$ but $(q_i \equiv q_j) \centernot\Rightarrow (q_i \stackrel{\mathclap{\normalfont\mbox{\tiny AV}}}{\equiv} q_j)$) and in general run in exponential time.
Finally, given a pair of subexpressions, 
an \textit{equivalence filter} applies a model, heuristic, or similar technique to approximately decide equivalence (\ie \textit{pseudo-equivalence}).
In \sys, filters trade off speed and precision to reduce the false positives that must be checked by an equivalence verifier.
We denote pairwise pseudo-equivalence determined using a filter $f$ as $q_i \stackrel{\mathclap{\normalfont\mbox{\tiny f}}}{\approxequiv} q_j$.

Given the above, we now formally define the core problem addressed by \sys:
\begin{problem*}[Workload equivalence]
\label{problem:equivalence-set}
Given a workload $W=\{q_1, ..., q_n\}$ of subexpressions,
\sys approximates $E(W)=\{(q_i, q_j) \in W \times W \mid q_i \equiv q_j \}$, \ie the \textit{equivalence set} amongst all the pairwise combinations of subexpressions in $W$.
\end{problem*} %

There are two important special cases of the workload equivalence problem. In the first case, the workload just has a pair of subexpressions $W=\{q_i, q_j\}$. The task reduces to just detecting \textit{pairwise equivalence} ($q_i \equiv q_j$). This version of the problem is common for applications such as query rewriting or view matching. The second special case is when the input is a set of queries $\{Q^1, ... Q^m\}$. Then the workload is the enumeration of all the subexpressions of the input queries, \ie $W=\bigcup_k S(Q^k)$. This formulation is of critical importance to applications such as view recommendation, when the goal is to find common computation among a large set of queries. Although \sys can handle pairwise equivalence detection very well, it is designed more as an efficient and scalable solution for supporting general workload equivalence when the workload set $W$ is large (which includes the second special use case).

\newpage 
\subsection{{\sys} Overview}
\label{sec:overview}

\arch

The overall architecture of \sys is illustrated in \autoref{fig:arch}.   
\sys approximates computing an equivalence set by applying the series of filters $F=\langle f_1, ..., f_n\rangle$ listed in \autoref{table:filter-characteristics} to a workload of subexpressions.  Filters are applied in decreasing order of speed and increasing order of precision.
Each filter is applied to every subexpression pair in the target workload $W$ to approximate the equivalence set.
To ensure correctness (e.g., for use in a view materialization algorithm), \sys utilizes an automated verifier to eliminate false positives from the resulting equivalence set. It is important to note that if a pair is determined to be non-equivalent by a filter, it is not evaluated by subsequent filters and it is not verified (i.e., filters short-circuit).

We formalize the above process with the following two functions: %
\begin{equation}
{\sys}_{\textsc{set}}(W, F) = \{(q_i,q_j) \in W \times W \mid~
  {\sys}_{\textsc{pair}}(q_i, q_j, F)\}
\label{eq:geco}
\end{equation}

\begin{equation}
\label{geco:pair}
{\sys}_{\textsc{pair}}(q_i, q_j, F) =
  \begin{cases}
    q_i \stackrel{\mathclap{\normalfont\mbox{\tiny AV}}}{\equiv} q_j & \text{if }F = \emptyset \\
    \bot & \text{if }q_i \stackrel{\mathclap{\normalfont\mbox{\tiny $f_1$}}}{\not\approxequiv} q_j \\
    {\sys}_{\textsc{pair}}(q_i, q_j, F\setminus f_1) & \text{otherwise}\\
  \end{cases}
\end{equation}

\subsubsection{\textbf{Detecting an Equivalence Set for a Large Workload}}
\label{sec:prediction}
We now describe, given a large workload of subexpressions, how \sys applies the filters in Table~\ref{table:filter-characteristics} to efficiently narrow down the candidate equivalent subexpression pairs, before calling the expensive automated verifier (AV).

The first filter applied is the widely-used \textit{schema filter (\sfm)}. 
Subexpressions that access different sets of tables or return different numbers of columns are highly unlikely to be equivalent. Therefore, \sys groups all subexpressions in the workload based on the tables used and the number of columns returned, resulting in \textit{\sfm-groups}. From this point forward, only subexpression pairs from the same \sfm-group are considered by subsequent filters.

In the second step, for each \sfm-group, the \textit{vector matching filter (\vpm)} embeds the subexpressions in a learned vector space and identifies likely equivalent pairs by employing approximate nearest neighbor search (ANNS). It is formalized as follows:
\begin{definition}[Vector matching filter (\vpm)]
Let $e(q)$ be a function that embeds a subexpression $q$ in a vector space $\mathcal{V}$.  Let $d$ be a distance metric on $\mathcal{V}$ and $\tau$ be a threshold distance.
Given subexpressions $q_1$ and $q_2$, let
$q_1 \stackrel{\mathclap{\normalfont\mbox{\tiny \vpm}}}{\approxequiv} q_2$ when
$d(e(v_1), e(v_2)) < \tau$.  
\end{definition}
To further improve efficiency, we construct a hierarchical navigable small world (HNSW) index ~\cite{malkov2018efficient}, a common approach to applying ANNS at scale ~\cite{huang2015query}.

In the third step, \sys applies the \textit{equivalence model filter (\ecpm)}, which is a trained deep learning model, to predict whether each candidate subexpression pair from the \vpm filter are equivalent. 

Finally, \sys utilizes an \textit{automated verifier (AV)} (we leverage SPES~\cite{zhou2020symbolic}) to verify the correctness of the prediction from \ecpm.

Among the filters used in \sys, both \vpm and \ecpm are machine learning based. The \ecpm  is a deep learning model comprising multiple tree convolutions and fully connected layers. On the other hand, the \vpm utilizes the learned tree convolution from \ecpm to embed subexpressions into its metric space. %

\subsection{Equivalence Model Filter (\ecpm) Overview}
\label{sec:emf-overview}

The \ecpm is a deep learning model trained to classify equivalence. We now briefly describe its training process and the semi-supervised feedback loop (SSFL) %
to iteratively improve the model.

To train the \ecpm, \sys first featurizes~(\S\ref{sec:feature_engineering}) and labels a set of subexpression pairs as the training data. Labels are generated using the SPES automated verifier. During featurization, in addition to converting subexpressions to a fixed-length vector representation, the \ecpm applies its database-agnostic (db-agnostic) transformation~(\S\ref{sec:db-agnostic-encoding}). This transformation replaces references to specific tables and column names with symbolic correspondences between subexpression pairs, generalizing the \ecpm learning from specific examples of (non)equivalent subexpressions to \textit{patterns} of (non)equivalent subexpressions. It also ensures that the model learned on a particular workload and database is \textit{transferable} to other workloads and databases, %
allowing for user-supplied or synthetically generated initial training workloads (\S\ref{sec:synthetic-workload}).

\sys employs the \ssfl as a guardrail against regressions. %
It monitors the confidence levels of \ecpm's predictions, and  if confidence falls below a threshold (e.g., due to new or evolving workloads), %
it iteratively fine-tunes the \ecpm model through the \ssfl pipeline.

The key challenge in the \ssfl pipeline is generating high-quality samples with balanced positive and negative examples for model fine-tuning in each iteration. Even a modest workload produces an intractably large training dataset that is quadratic in the number of subexpression pairs---1000 queries with 10 subexpressions each produces a training dataset of almost 100 million pairs! This dataset is also highly imbalanced, since most subexpression pairs are unlikely to be equivalent.

To address this challenge, \sys employs the cheap \sfm and \vpm filters to efficiently identify pseudo-equivalent subexpression pairs (i.e., it computes $q_i \stackrel{\mathclap{\normalfont\mbox{\tiny SF}}}{\approxequiv} q_j \land q_i \stackrel{\mathclap{\normalfont\mbox{\tiny VMF}}}{\approxequiv} q_j$ over a workload sample).  This computation approximates \autoref{geco:pair} without the verification step (\S\ref{sec:prediction}).
Together with another set of randomly-generated, likely non-equivalent pairs, they form an approximately-balanced new sample. As before, \sys labels and applies its db-agnostic transformation to the new sample.
It then augments its training dataset with the new data and fine-tunes the \ecpm.

\rev{As previously highlighted, \sys identifies \textit{general} semantic equivalence, agnostic to the underlying database. It therefore does not consider database constraints or other instance-specific metadata. Nonetheless, extending \sys to incorporate database-specific constraints~\cite{deutsch2009fol} remains an interesting direction for future work.}

\subsection{Complexity Analysis of \sys Filters}
This subsection provides a complexity analysis (summarized in \autoref{table:filter-characteristics}) for applying each \sys filter on a workload $W$ containing $n$ subexpressions.

\vspace{-2mm}
\subsubsection*{\textbf{Schema Filter (\sfm)}}
Assuming a constant-sized schema, %
\sys groups $n$ subexpressions by the used tables and the number of returned columns in $\mathcal{O}(n)$ time.
\vspace{-1mm}
\subsubsection*{\textbf{Vector Matching Filter (\vpm)}}

Given that the HNSW  index used by the \vpm has claimed search complexity logarithmic in the number of indexed objects  ~\cite{malkov2018efficient},
\sys indexes the workload subexpressions in $\mathcal{O}(n)$ time (we assume a constant embedding size; see \S\ref{sec:evaluation}).
Next, for each vector, %
it performs a $\mathcal{O}(\log n)$ radius search for neighbors within Euclidean distance $\tau$, with  total complexity in $\mathcal{O}(n \log n)$. %

\subsubsection*{\textbf{Equivalence Model Filter (\ecpm)}}
As shown in \S\ref{sec:emf}, \sys's equivalence model contains two convolution layers followed by three fully connected layers. 
Its input is a pair of subexpressions, each with $ops(q_i)$ nodes.
We assume that there are many more subexpressions in our workload than operators in the largest tree, i.e., $\max \{ops(q_i) \mid q \in W\} \ll n$.
Total complexity is thus dominated by the matrix multiplication in the fully connected layers (i.e., $\mathcal{O}(n^3)$).\footnotemark

\footnotetext{The two convolution layers are each in $\mathcal{O}(n^2)$~\cite{DBLP:conf/nips/VaswaniSPUJGKP17}.}

\subsubsection*{\textbf{Automated Verification (AV)}}
To ensure correctness, \sys verifies pairs produced by its filters. \sys's AV leverages SPES~\cite{zhou2020symbolic}, which uses the Z3 SMT prover~\cite{de2008z3} to check equivalence. \rev{A SMT program can be transformed into an equivalent SAT formulation containing $\gamma$ symbols, which is solvable in $\mathcal{O}(2^{\Omega(\gamma)})$ time.}

\eat{
\subsection{Workload Analysis}
\RA{we did not use the templates in our recent experiment. We generated them earlier to understand the workload. The sampling is based on clustering using the GMM model, the clustering is happening at the convoluted vectors level}
\begin{figure*}[t]
\includegraphics[width=\textwidth]{Figure/QueriesTemplates_Dis_Amoeba_EQ.png}
\caption{Queries Histogram EQ Amoeba Workload  \label{fig:eq_amoeba_workload_templates}}
\end{figure*}

\begin{figure*}[t]
\includegraphics[width=\textwidth]{Figure/QueriesTemplates_Dis_WeTune_EQ.png}
\caption{Queries Histogram EQ WeTune Workload\label{fig:eq_wetune_workload_templates}}
\end{figure*}
\begin{figure*}[t]
\includegraphics[width=\textwidth]{Figure/QueriesTemplates_Dis_CNT.png}
\caption{Queries Histogram CNT Workload\label{fig:cnt_amoeba_augmented_workload_templates}}
\end{figure*}/
}

\section{Feature Engineering}
\label{sec:feature_engineering}

In this section,  we describe the features used by the \ecpm (\S\ref{sec:feature_selection}) and how these features are mechanically featurized (\S\ref{sec:query_featurizer}).

\subsection{Feature Selection}
\label{sec:feature_selection}

After conducting extensive feature analysis, we find that logical plans play the most important role in predicting equivalence, since the logical plan captures the semantics of a subexpression.
As a result, in \sys, we use the logical plans of the subexpression pairs as inputs to the \ecpm model.

We additionally considered leveraging cardinalities as an auxiliary feature.  Intuitively, since $q_i \equiv q_j \Rightarrow \lvert q_i \rvert = \lvert q_j \rvert$, this would appear to be a strongly positive signal for equivalence.  However, while our initial analysis indicates that cardinalities do improve \ecpm recall, actual subexpression cardinalities are not generally available for use as inputs to \sys and executing candidate subexpressions to determine cardinality is infeasible at scale. 
Conversely, estimated 
cardinalities are quick to compute but  yield only marginal benefit to the prediction task. Thus, %
we exclusively rely on logical plans as input features to \ecpm.

\sys canonicalizes the conjunctive predicates in selection and join operators by splitting each $n$-clause predicate into a composite containing $n$ single-clause predicates.
For example, \sys transforms a relational selection operator $\sigma_{\texttt{x} > 25 \land \texttt{y} < 10000}(R)$, 
into the composite 
$\sigma_{\texttt{x} > 25}(\sigma_{\texttt{y} < 10000}(R))$. As a result, each node in the logical plan has at most one selection or join predicate.

\subsection{\textbf{Logical Plan Featurization}}
\label{sec:query_featurizer}

\rev{Featurizing the tree structure of a logical plan is challenging since it is difficult to express an arbitrarily-shaped, variable-size tree as a fixed-size feature vector without losing the structure of the tree. To address this, we apply a \emph{tree-vector} transformation that converts an arbitrary logical plan into a fixed-length vector~\cite{mou2016convolutional}.  Our transformation is inspired by %
~\cite{MarcusNMZAKPT19}; however, we use a different encoding for each node in the %
tree. }

Specifically, we first encode each node in the logical plan as a \textit{node vector (NV)}. Each NV has the same size and format (to be described in \S\ref{sec:logical-encoding}), but the number of NVs (\ie number of nodes in the logical plan) can vary widely. 
Given a tree of NVs,  
\sys next performs a breath-first traversal of the tree and concatenates each visited NV into a
\rev{$m \times l$
matrix $M$, where $m$ is the number of nodes in a logical plan and $l$ is the size of each NV}. We finally apply the tree convolution layers of the \ecpm, to be described in \S\ref{sec:emf}, which
transforms $M$ into a vector of a fixed dimension $h$ that summarizes a subexpression. In our prototype, $h=128$ bytes. \rev{As demonstrated by prior work~\cite{MarcusNMZAKPT19,negi2021steering,marcus2020bao}, tree convolution has been proven effective at representing tree-structured SQL query plans for various ML-for-DB tasks.}

\newpage
\section{Logical Plan Encoding} \label{sec:logical-encoding}

We now detail how \sys encodes each node in a logical plan as a node vector (NV). We begin by describing an \textit{instance-based encoding}, \ie one where the encoding is specific to a workload on a particular database instance. 
Though the specifics vary, this is a common transformation and most existing  approaches are instance-based~\cite{MarcusNMZAKPT19, containrate,DBLP:journals/corr/abs-1905-06425}.
We then extend this approach to our novel \textit{db-agnostic} encoding which is oblivious to the specific workload or database. 

\workloadEncoding

\subsection{Instance-Based Encoding} \label{sec:instance-encoding}

\textbf{Notation.} Our instance-based encoding is inspired by the encoding technique described in~\cite{containrate}. 
Given a workload $W$ on a database instance, let $T_W$, $C_W$, $O_W$, and $J_W$ respectively represent the set of tables, columns, arithmetic operators (\eg $\leq$, $=$, $\geq$, $\neq$), and join types ($\bowtie$, $\leftouterjoin$, $\rightouterjoin$, $\fullouterjoin$) referenced in the workload.
Let $onehot(e, U)$ produce a one-hot encoded vector of size $\lvert U \rvert$ with entry $e \in U$ set to one, $null(x)$ indicate whether $x$ is null, and $norm(x)$ normalize $x$ over
all scalars in a workload.

\vspace{1mm}
\noindent\textbf{Encoding method.} 
\autoref{fig:workloadEncoding} illustrates the instance-based encoding process. Each NV consists of table, join and selection segments, denoted as $V_{\text{table}}$, $V_{\text{join}}$, and $V_{\text{select}}$, respectively. For a scan operator on table $t$, \sys generates the segment $V_{\text{table}} = onehot(t, T_W)$. For a selection operator with a predicate referencing a column $c$, an arithmetic operator $o$, and up to one constant value $v$ (we perform constant folding prior to encoding), \sys generates the segment $V_{\text{select}} = onehot(c, C_W) \oplus onehot(o, O_W) \oplus norm(v) \oplus null(v)$, where $\oplus$ is the concatenation operation. Finally, a join operator has a join predicate referencing a left-side column $c_l$, an arithmetic operator $o$, right-side column $c_r$, and a join type $j$. \sys generates the join segment $V_{\text{join}} = onehot(c_l, C_W) \oplus onehot(o, O_W) \oplus onehot(c_r, C_W) \oplus onehot(j, J_W)$.

As is common in ML featurization, we simply concatenate the table, join, and selection segments to form the final vector, i.e. $NV = V_{\text{table}} \oplus V_{\text{join}} \oplus V_{\text{select}}$. For a segment that does not apply to a tree node, \sys sets it to be zero, e.g. the join segment for a non-join operator is all zeros. Note that $\lvert NV \rvert = |T_W| + 3 \cdot |C_W|+ 2 \cdot |O_W| + |J_W| + 2$ (in our prototype $\lvert NV \rvert = 210$; see \S\ref{sec:evaluation}).

\begin{table}[ht]
  \begin{minipage}[t]{0.6\columnwidth}
    \centering
\begin{lstlisting}[basicstyle=\footnotesize\ttfamily]
q1: SELECT t1.c3, t2.c3 FROM t1,t2 
    WHERE t1.c1 = t2.c1 
          AND t1.c2 > t2.c2 + 10 
          AND t2.c2 > 10
q2: SELECT t1.c3, t2.c3 FROM t2,t1
    WHERE t2.c1 = t1.c1 
          AND t2.c2 + 10 < t1.c2 
          AND t2.c2 + 10 > 20 
          AND t1.c2 > 20  
\end{lstlisting}
\vspace{4pt}
\captionof{figure}{Symbolized versions of the  subexpression pairs highlighted in Figure~\ref{lst:example1}.
}
\label{fig:symbols}
  \end{minipage}%
  \hfill
  \begin{varwidth}[t]{0.37\columnwidth}
    \centering
    \small
    \begin{tabular}{ l l }
      \toprule
      \textbf{Reference} & \textbf{Symbol} \\
      \midrule
      A & t1 \\
A.joinKey & t1.c1 \\
A.val & t1.c2 \\
A.x & t1.c3 \\
B & t2 \\
B.joinKey & t2.c1 \\
B.val & t2.c2 \\
B.y & t2.c3 \\
      \bottomrule
    \end{tabular}
    \vspace{4mm}
    \caption{Symbols generated for queries in \autoref{lst:example1} under db-agnostic encoding.%
    }
    \label{table:symbols}
  \end{varwidth}%
\end{table}
\encodingConverter
\subsection{DB-Agnostic Encoding}\label{sec:db-agnostic-encoding}
Instance-based encoding makes sense for solving problems such as cardinality estimation and query optimization, where the solution targets a specific workload on a particular database instance. In contrast, the problem of learning equivalent subexpressions can be reformulated to be \textit{database agnostic}. To motivate, consider the two subexpressions highlighted in Figure~\ref{lst:example1}.  If we were to change the table and column names to those shown in Figure~\ref{fig:symbols}, the two new subexpressions remain equivalent, even though they are now for a completely different database, workload, or dataset. This observation is the basis for our db-agnostic node vector encoding technique. 
Conceptually, for each labeled training data point, we generalize the pair of subexpressions into subexpression patterns, and feed these generalized patterns into our model. As a result, we are able to transfer the learning from one workload for a database instance to a different workload on a different database.

For equivalence detection, what really matters is the tables and columns referenced in the pair of subexpressions. Further, in terms of columns, only the columns actually referenced by the join conditions, selection predicates, and projections (instead of all columns from the referenced tables) need to be considered. Moreover, the actual names of the tables and columns are unimportant.
As a result, we can convert the tables and columns in a pair of subexpressions into a generic symbolic form to derive their underlying patterns. 

\sys does this by transforming  referenced tables into a set of distinct, generic table symbols $\{t_1, ..., t_n\}$ based on an arbitrary lexicographical order (we sort alphanumerically in our implementation).  It similarly symbolizes the referenced columns  as $c_1$,..., $c_n$. \autoref{table:symbols} shows this for the example %
in Figure~\ref{fig:symbols}.

With db-agnostic encoding, we set $T'_W=\{t_1, ..., t_n\}$, where $n$ is the maximum number of symbolized table correspondences expected in any workload, and $C'_W=\{t_1.c_1,..., t_1.c_m, ... , t_n.c_1,,..., t_n.c_m\}$, where $m$ is the maximum number of symbolized column correspondences per table expected in any workload. We can set $n$ and $m$ to be large enough numbers to cover arbitrarily complex subexpressions. %
However, in general $n$ and $n\times m$ are much smaller than the total number of tables and columns in the workload, respectively. 

Following this transformation,
we now treat $T'_W$ and $C'_W$ as our new ``workload tables'' and ``workload columns'' to replace $T_W$ and $C_W$ described in \S\ref{sec:instance-encoding}.  Then to encode, for each pair of subexpressions, we first symbolize each subexpression into symbolic pattern and apply the previously-introduced instance-based encoding to produce a final db-agnostic encoding ${NV}_\alpha$ of size $\lvert {NV}_\alpha \rvert = |T'_W| + 3 \cdot |C'_W|+ 2 \cdot |O_W|+ |J_W| + 2$.
\subsubsection{Scaling to large workloads}
\label{sec:encoding-converter}
The db-agnostic encoding described above is for a pair of subexpressions. The encoding of one subexpression is different depending upon which other subexpression it is paired with during featurization. Given a large workload of $n$ subexpressions, we have to re-compute the encoding for each pair---an $O(n^2)$ computation. In contrast, in the instance-based encoding, the encoding of one subexpression stays unchanged no matter what it is paired with. In other words, we only need to compute the encoding for each subexpression once (an $O(n)$ computation). For offline training, the performance of db-agnostic encoding is not so crucial, but for online inference any improvement in the encoding process is valuable. To speed up the db-agnostic encoding process, we develop an efficient method to quickly convert an instance-based encoding to a db-agnostic encoding. With this approach,  we only incur $O(n)$ computation to produce the instance-based encoding, then apply a lightweight converter for each pair of subexpressions.

The converter takes as input the instance-based tree matrices for both subexpressions.
For each subexpression, it first projects out the submatrix $S_T = M[T]$ that corresponds to the table segment, and the submatrices $S_{C_s} = M[C_s]$, $S_{C_l}=M[C_l]$, and $S_{C_r}=M[C_r]$ that correspond to the column encoding from the selection segment, the left-side column encoding, and the right-side column encoding from the join segment. We first union the column submatrices by applying bit-wise $or$ to compute the column submatrix $S_C = S_{C_s} \lor S_{C_l} \lor S_{C_r}$. For the table submatrix $S_T$, 
we compute a vector $r$ that represents the column-wise union of the tables referenced in \textit{each} subexpression, with $r_j = \bigvee_{i} S_T[i,j]$. Next, we generate a mask $m_T$ by unioning the vectors from both subexpressions; this represents all tables referenced in either of subexpressions. We then apply this mask on $S_T$ from both subexpressions to eliminate matrix columns corresponding to unreferenced tables. The resulting submatrix is $S'_T$. We apply the same process on the column matrices to compute the mask $m_C$, and use $m_C$ to eliminate matrix columns corresponding to unreferenced table columns from $S_{C_s}$, $S_{C_l}$, and $S_{C_r}$ for each subexpression, resulting in $S'_{C_s}$, $S'_{C_l}$, and $S'_{C_r}$. Finally, for each subexpression in the pair, we replace the submatricies $S_T$, $S_{C_s}$, $S_{C_l}$, and $S_{C_r}$ with their transformed variants. The result is a pair of db-agnostic tree matrices $M_\alpha$ which eliminates references not found in either subexpression.
Figure~\ref{fig:encoding-converter} illustrates the conversion process for  table fragments in a pair of subexpressions. %

Through experiments, we find that 
applying the converter described above is $1.8\times$ faster than computing pairwise db-agnostic encodings from scratch.

\subsubsection{Tensor-based extensions}

In the previous subsection we described the db-agnostic encoding process for a single pair of subexpressions.  We now describe two tensor-based extensions.

\vspace{1mm}
\textbf{Batch pairwise encoding.}
Db-agnostic encoding can be easily extended to support batch encoding $n$ pairs.
To do so, rather than performing $n$ discrete operations on pairs of two-dimensional submatrices of size $|q_i| \times |X|$---where $|q_i|$ represents the number of tree nodes in a subexpression's logical plan and $X$ is a table or column segments of the NV---\sys represents the batch as a pair of three-dimensional tensors of size $\max(|q_i|) \times |X| \times n$.  Subexpressions with fewer than $\max(q_i)$ operators are zero-padded, which does not affect correctness.

The resulting tensor is amenable to being operated on using tensor-oriented frameworks such as PyTorch~\cite{paszke2019pytorch}.  As we describe in \autoref{sec:emf}, the \ecpm  batch-converts workloads using this approach.

\vspace{1mm}
\textbf{Generalizing from pairs to $\bm{n}$-subexpressions.}
The db-agnostic transformation we describe above is a binary operation over two subexpressions.
A second generalization involves extending it to be an $n$-ary operation over many subexpressions.  This extension only impacts the computation of the mask (e.g., in \autoref{fig:encoding-converter}, rather than $r_1 \lor r_2$ we compute $r_1 \lor ... \lor r_n$); other operations are unchanged.  This extension is also amenable to tensor execution.

In fact, this $n$-ary, tensor-based encoding is used in the \vpm filter in \sys. Recall that after applying the \sfm filter, \sys groups input workload into \sfm-groups based on tables accessed and the number of columns returned (\S\ref{sec:prediction}). 
In the \vpm, \sys then applies the $n$-ary db-agnostic transformation to all subexpressions in each \sfm-group.  It then convolves the encoded node vectors
to produce a fixed-size vector for each subexpression (\S\ref{sec:query_featurizer}).
It finally conducts an approximate nearest-neighbor search (ANNS) on the resulting vectors to identify candidate subexpression pairs that are likely to be equivalent. 
Note that all subexpressions in an \sfm-group access the same set of tables. 
Therefore, the group-based db-agnostic encodings for subexpressions from the same \sfm-group approximate their pairwise db-agnostic counterparts.

\section{Equivalence Model Filter (\ecpm)}
\label{sec:emf}

\emfArchitecture

In this section, we discuss the architecture and training process of the equivalence model filter (\ecpm).  
Recall from \S\ref{sec:emf-overview} that the \ecpm is a schema-independent deep learning model trained to classify equivalence between a pair of subexpressions.
In building the \ecpm, 
we evaluated many candidate architectures, including various supervised classifiers, logistic regression (LR)~\cite{neter1996applied}, random forests (RF)~\cite{ho1995random}, and multi-layer perceptrons (MLP)~\cite{goodfellow2016deep}. While the LR and RF models are simple to train and exhibit moderate performance, they suffered from one fundamental limitation: they do not allow incremental training and fine-tuning.  As we detail in \S\ref{sec:labeling}, the ability to incrementally fine-tune a model is critical to adapting to changing workloads and maximizing transferability.
On the other hand, MLPs are more expensive to train but support incremental training,
thus we only need to feed newly-labeled samples to fine-tune the previous model. %

As a result, we utilize the MLP model for classification in the \ecpm.
The overall architecture of the \ecpm is illustrated in \autoref{fig:emf-archiecture}.  
It comprises two tree convolution layers and three fully connected layers. 
As inputs, the \ecpm accepts a pair of instance-encoded logical plans, where each node is an instance-encoded vector of size $\lvert NV \rvert$ (see \S\ref{sec:instance-encoding}).  These plans are then transformed into their db-agnostic counterparts (i.e., vectors of size $\lvert NV_\alpha \rvert$) by applying the transformation described in \S\ref{sec:db-agnostic-encoding}.  
Next, the \ecpm applies two tree convolutions to the db-agnostic plans.  Each convolution is followed by batch normalization and parametric rectified linear unit (PReLU) activation.  The two resulting 128-byte summaries of each subexpression logical plan are then concatenated and passed through three fully connected layers for classification.

\textbf{\ecpm training and testing data.}
\label{sec:synthetic-workload}
\rev{To train the \ecpm, \sys requires a large set of labeled training data (i.e., subexpression pairs)}.
\rev{Because our db-agnostic encoding technique enables transferability between workloads and database instances,
we can initially train the \ecpm model on a high-quality \textit{synthetic} workload that contains a wide range of positively- and negatively-labeled subexpression pairs. To 
generate such data, we leverage two state-of-the-art query generation tools: 
AMOEBA~\cite{liu2022automatic} and WeTune~\cite{wetune}.}

\rev{AMOEBA employs a domain-specific fuzzing technique to generate a set of base queries $B_{Q}$. It then applies a set of semantic-preserving query rewrite rules $R$ on a given query $q_i \in B_{Q}$ and generates a set of queries $Q_i' = \{q'^1_i, ..., q'^n_i\}$ equivalent to $q_i$}.%

\rev{\sys leverages AMOEBA by applying it to produce a dataset of positively-labeled pairs $W_{+}=\bigcup_i \{(q_\alpha, q_\beta) \mid q_\alpha \in Q_i' \cup \{q_i\} \land q_\beta \in Q_i' \cup \{q_i\} \setminus q_\alpha\}$. }

\rev{To ensure a wide variety of training examples, we further leverage WeTune~\cite{wetune}%
, which is an optimizer rule generator that automatically generates a set of non-reducible and interesting rewrite rules (including rules missed by prominent commercial query optimizers).  We apply WeTune-generated rules %
to rewrite the set $B_{Q}$ of base queries produced by AMOEBA.  We then repeat the process described above to produce a WeTune-augmented training data set $W'_{+}$}.

\rev{The above process yields a diverse set of equivalent subexpression pairs.  To generate a corresponding set of non-equivalent pairs,
we group all subexpressions in $B_{Q}$ into schema-compatible groups ($\{B^1_Q, ..., B^n_Q\}$) by applying the \sfm.  Each group contains subexpressions that reference the same base tables and return the same number of columns (i.e., they are non-degenerate and would not be subsequently filtered by the \sfm).  Next, 
we generate a set of negative examples by randomly pairing the subexpressions in each group: $W_- = \{(q_\alpha, q_\beta) \mid q_\alpha, q_\beta \in B^i_Q \land (q_\alpha, q_\beta) \not\in W_+\}$.  While this process might (with low probability) yield a false negative (i.e., by negatively labeling a pair that is actually equivalent), model training is resilient to small amounts of noise in training data and we did not observe a decrease in performance.  Nonetheless, a perfect dataset could be produced by applying the automated verifier (AV) to confirm the label of each negative pair.}

Using the above, \sys finally draws a balanced set of labeled examples from $W_+$ and $W_{-}$ to produce a synthetic training dataset
that contains a variety of syntactically dissimilar and semantically equivalent logical plans.  This, along with well-known machine learning techniques such as tree convolution and dynamic pooling~\cite{mou2016convolutional} that provide resiliency to minor perturbations, enables \sys to generalize to plans of varying shapes and sizes.

\begin{figure}[t!]
\centering
\subfigure[0.47\columnwidth][Mean error by convolution layer size with two linear layers (128 and 64).]{
\includegraphics[width=0.47\columnwidth]{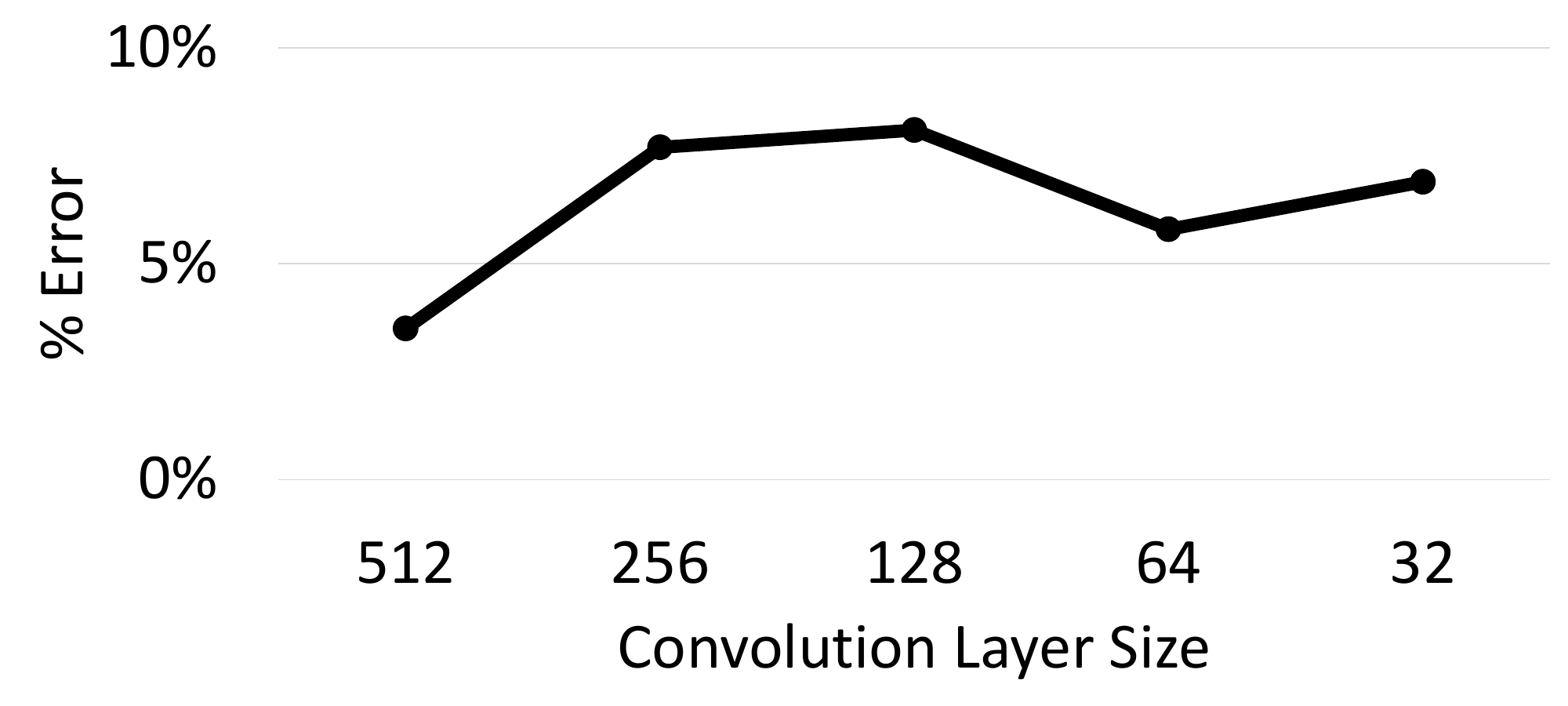}  
\label{figure:error-convolution}
}\hfill
\subfigure[0.47\columnwidth][Mean error for two convolution layers (sized 512 and 128) by linear layer size.]{
\includegraphics[width=0.47\columnwidth]{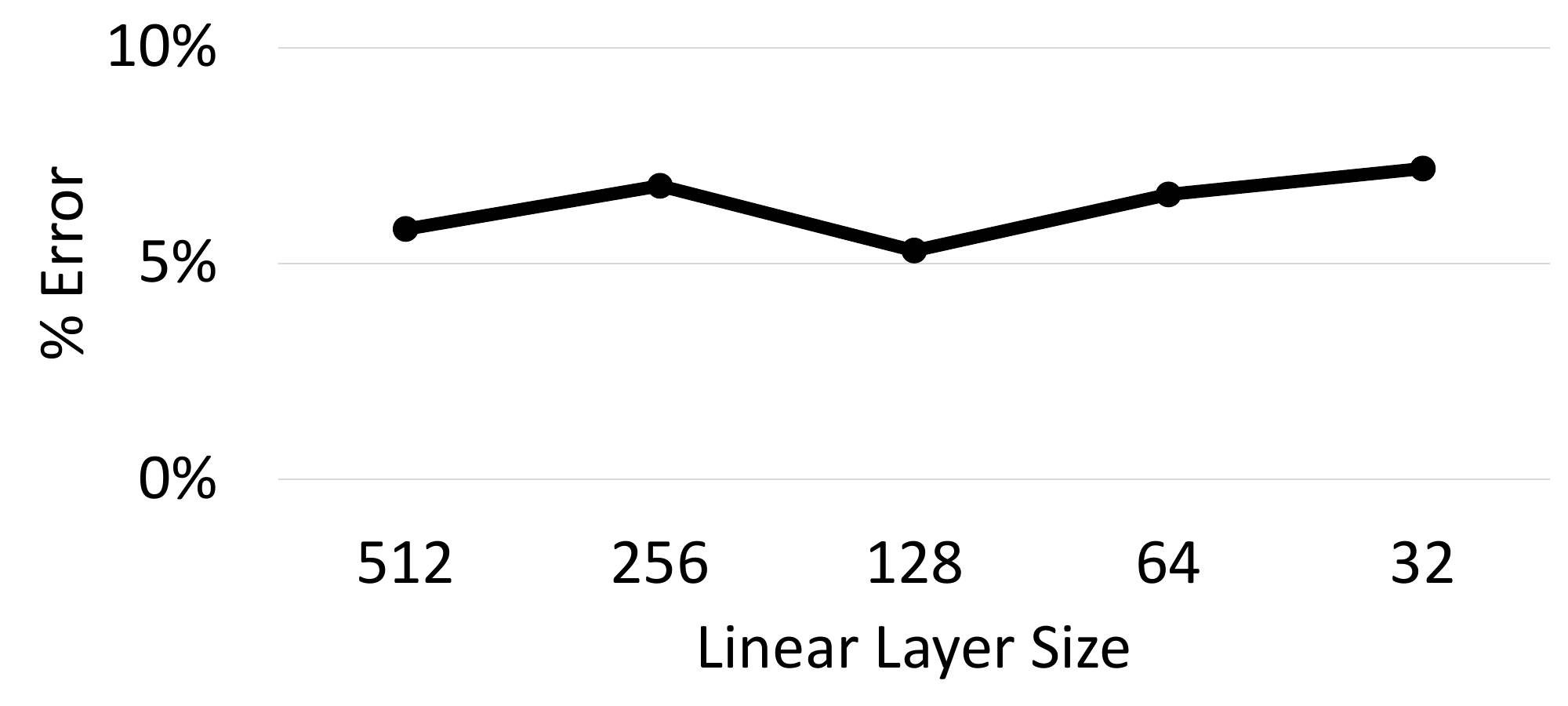}  
}
\caption{Mean error by linear and convolution layer sizes. 
\vspace{2em}
}
\label{figure:error-linear}
\end{figure}

\vspace{0.5mm}
\textbf{Hyperparameter tuning.}
To maximize \ecpm performance, we perform a search over model structure and hyperparameters.
Our search considers various network architectures (\ie between 1--5 linear and convolution layers and sizes 32--512),
activation functions, dropout, and optimizer parameters such as learning rate and decay. 
We evaluate on the synthetic dataset based on TPC-H as described in \S\ref{sec:evaluation}.

As we show in \autoref{figure:error-linear},
we find that increasing the number of convolution and hidden layers beyond two did not improve accuracy. Layer sizes have a modest impact on accuracy.
Optimizer choice and learning rate had a negligible impact on performance.

\section{Semi-Supervised Feedback Loop (\ssfl)}
\label{sec:labeling}

When applied to a new workload or 
as the distribution of (non)equivalent subexpressions in a workload drifts over time,
the performance of the previously-trained \ecpm model may suffer. To mitigate this, \sys employs a \emph{semi-supervised learning bootstrapping feedback loop (\ssfl)} inspired by Zhu et al.~\cite{zhu2009introduction}.  The \ssfl continuously monitors \ecpm performance and retrains with newly-generated training data when needed.

To accomplish this,
\sys continuously measures the confidence level of classifications made by the \ecpm. 
If this confidence level falls below
a threshold $T_h$, the \ssfl dynamically 
samples a new, balanced set of labeled samples from the current workload.
It uses this sample to fine-tune the \ecpm. 
The \ssfl iterates this process
until \ecpm performance reaches a desirable confidence level. We formalize the \ssfl confidence level as follows:

\begin{definition}[\ssfl Confidence Level]
Let $W$ be a set of queries we wish to compute 
${\sys}_{\textsc{set}}(W, F)$ over (q.v. \autoref{eq:geco}).
Let $P^{p}_{1}$ be the probability estimate that the pair $p \in W \times W$ exhibits an equivalence relationship, and $P^{p}_{0}$ the probability that $p$ exhibits a non-equivalence relationship. We compute the \ssfl confidence level $SSFL\text{-}CL$ of $W$ as:
\begin{equation*}
    SSFL\text{-}CL(W, P_0, P_1) = \frac{\sum_{p \in W \times W} \big[\max(P^{p}_{0}, P^{p}_{1}) \geq T_h\big]}{|W \times W|}
\end{equation*}
\end{definition}

\begin{algorithm}[t]
\SetAlgoNoEnd 
\SetAlgoLined
\DontPrintSemicolon 
\SetKwProg{Fn}{function}{:}{}

\KwIn{A workload $W$.}
\KwOut{A \textit{EMF} model, fine-tuned if confidence is low.}

\Fn{$SSFL(W)$}{
    $P_0 \gets \varnothing$\;
    $P_1 \gets \varnothing$\;
    \ForEach{$(q_i, q_j) \in W \times W$}{
      $\rho \gets Sigmoid\big(EMF(q_i, q_j)\big)$\;
      $P_1 \gets P_1 \cup \{\rho\}$\;
      $P_0 \gets P_0 \cup \{1-\rho\}$\;
      
    }
    
      \If{$SSFL\text{-}CL(W, P_0, P_1) \leq T_h$}{
        $S_+ \gets AV\Big(VMF\big(SF(W \times W)\big)\Big)$\;
        $S_{-} \gets sample\big((W \times W) \setminus S_+, \lvert S_+ \rvert\big)$\;
        $\textit{EMF} \gets train(\textit{EMF}, S_+ \cup S_-)$\;    
      }
    
      \Return \textit{EMF}
    }
\caption{The semi-supervised feedback loop (\textit{SSFL}).}
\label{alg:ssfl}
\end{algorithm}

Ideally, we want the semi-supervised learning feedback to only trigger a few rounds of fine-tuning before it reaches a satisfactory confidence level. In order to achieve this, we need to choose good samples with good positive and negative examples in each iteration for retraining or fine-tuning the \ecpm model. A naive sampling approach is to random sample pairs of subexpressions from the workload. However, this simple approach is like shooting in the dark and is unlikely to provide sufficient positive examples for the model to learn, since positive examples (equivalent subexpressions) are generally rare events compared to negative examples in a typical workload. The key is to make sure we find some good positive examples for training. 

As described in \S\ref{sec:prediction}, by leveraging \sfm and \vpm, we can quickly identify a set of likely equivalent subexpression pairs from a large search space, then label the pairs by actually running the equivalence verifier. We keep all the positive and negative examples. Moreover, if more negative examples are needed for a balanced sample, we can also get random sample pairs of subexpressions from the workload. As we show in \S\ref{sec:evaluation}, this \textit{filter-balanced sampling} mechanism can significantly improve the model quality with fewer labeled sample data compared to random sampling.

We formalize the \ssfl algorithm in \autoref{alg:ssfl}.  It accepts a workload $W$ and examines each pair in the cross product (line 4).  For each pair, it applies the Sigmoid function to the \ecpm output to compute the probability $\rho$ that the pair $(q_i, q_j)$ exhibits an equivalence relationship (line 5-7).  
Note that probability estimates are trivial to compute when applying the \ecpm during prediction.
Next, it computes a confidence level (line 8) and, if the model is insufficiently confident, generates a sample of likely-equivalent pairs by applying the \sfm and \vpm (line 9).  It then produces a complimentary sample of size $\lvert S_+ \rvert$ sampled randomly from non-equivalent pairs to form $S_-$(line 10).  It finally retrains or fine-tunes the \ecpm using the full sample (line 11).

\section{Experimental Evaluation}
\label{sec:evaluation}

We now present an experimental evaluation of \sys. The goals of our evaluation are as follows: to (i) compare various \ecpm models to determine their effectiveness in predicting equivalence relationships, as well as assessing their ability to transfer learning across different workloads and databases (\S\ref{sec:compare_model}); (ii) study the performance of the \vpm filter in terms of its ability to filter out ``easy''equivalence cases (\S\ref{sec:vmf_performance}); (iii) evaluate the \ssfl pipeline with the filter-based sampling mechanism (\S\ref{sec:model_eval}); (iv) examine runtimes of \vpm and \ecpm filters on CPU- and GPU-based implementations (\S\ref{sec:compute}); and (v) evaluate the impact of \sys on scaling %
state-of-the-art equivalence solvers for a large workload (\S\ref{sec:baseline_expr}).

\vspace{2mm}
\noindent \textbf{Implementation.}
We implement \sys using Python 3.10.0 and Java 18.0.2.  We manipulate subexpressions, parse and generate abstract syntax trees, and perform instance-based featurization using Calcite 1.27.0.  The \ecpm is implemented using PyTorch 1.12~\cite{paszke2019pytorch} and employs the Adam optimizer~\cite{kingma20153rd} with a learning rate of $10^{-3}$ and a weight decay of $5^{-4}$.  We train using a dropout of 50\% applied to all layers.  The \vpm is implemented using FAISS 1.7.2, where we construct a quantizer using 128-bit locality-sensitive hashes (LSH), an 128-dimension inverted index, and limit neighbor searches to a radius of $d=1$.  Finally, we set the \ssfl confidence level to $T_h=0.9$.
\BH{The GPU-based FAISS index does not support range searches, so we substituted an $k$-nearest neighbors approach that iterates until all relevant neighbors have been retrieved.}

\vspace{2mm}
\noindent\textbf{Experimental Setup.}
We conducted our experiments using a single
machine with two CPU sockets (Intel Xeon Platinum 8272CL) each with 16 physical cores (32 with hyper-threading), 264GB of main memory, and 512GB  storage device. %
Our GPU-based experiments %
are executed on a single Nvidia Tesla T4 with 16GB memory.

\vspace{2mm}
\noindent\textbf{Workloads.}
We generate a set of base subexpressions on the TPC-DS and TPC-H schema using AMOEBA augmented with rules from WeTune (\S\ref{sec:synthetic-workload}) as our workload queries. The set of TPC-DS subexpressions comprises $\sim$34k queries, while the TPC-H dataset contains $\sim$19k queries. Section \ref{sec:synthetic-workload} describes how we obtained our balanced, labeled data to train our initial model.

\subsection{\ecpm performance}\label{sec:compare_model}
We first evaluate the performance of \ecpm model %
in terms of model architecture, computational cost, and ability to transfer to unseen workloads and database schema.

\subsubsection{\textbf{Model type.}}
This experiment compares the effectiveness of
three candidate \ecpm classifiers: multi-layer perceptrons (MLP), random forests (RF), and logistic regression (LR).
We train the three variants on the TPC-H workload and measure performance on the TPC-DS dataset. 
Table~\ref{table:mlTransferLearning} summarizes the results. %
The MLP model provides superior accuracy versus the simpler models. 

Figure~\ref{fig:heatmap} shows confusion matrices for each model type, drilling down into
how the prediction aligns with the ground truth for each model. 
Since the \ecpm serves as a \textit{filter},
it should be the one that strives to simultaneously minimize the false positives (i.e., $\alpha$ error in the top right quadrant) and false negatives (i.e., $\beta$ error in the bottom left quadrant) of the prediction. Here $\beta$ error is most important---since \sys always invokes the equivalence verifier to verify the predicated equivalence, false positives from the \ecpm model do not affect the correctness of \sys, but represent wasted computation (i.e., by invoking the expensive automated verifier). 
By contrast, false negatives represent the missed equivalent queries by the \ecpm model and thus should be minimized at all costs. Clearly shown in Figure~\ref{fig:heatmap}, MLP is by far the clear winner in simultaneously minimizing the false positives and false negatives. In particular, the false negatives for MLP is kept around 0.1\%, which is orders of magnitude smaller than the other two models.  Due to the superiority of the MLP architecture to detect equivalence, all subsequent experiments in this section utilize this  model.

\CMTransferLearning

\subsubsection{\textbf{Computational Cost.}}
 We next analyze the training, prediction, and space costs of the \ecpm trained using the architecture described in \S\ref{sec:emf} averaged over five runs.
We train the \ecpm using $\sim$47k subexpression pairs drawn from the TPC-H dataset. 
On average, a training run with 20 epochs takes approximately 40 minutes. 
The size of the model when
serialized to disk is approximately 2.3MB, including all the learned parameters.
\ecpm prediction time is 0.00319s per pair of subexpressions averaged over $\sim$70k random TPC-DS subexpression pairs.

\subsubsection{\textbf{Transfer Learning}}
\label{sec:transferleaning}

\mlTransferLearningFinal

We now discuss the ability of the \ecpm to transfer to unseen datasets.
First, note that the results shown in Table~\ref{table:mlTransferLearning} and Figure~\ref{fig:heatmap} already illustrate this ability, where the \ecpm is trained on the TPC-H workload and tested on %
TPC-DS workload. %

\mlTransferLearningAltFinal

Next, we generate five additional datasets ranging from approximately 1k to 50k on a random schema using the method described in \S\ref{sec:emf}.  We then evaluate the \ecpm using the TPC-H-trained model and report model performance in \autoref{table:mlTransferLearningAlt}.  The high performance on additional unseen datasets reinforces \ecpm's ability to easily adapt to new, unseen workloads.

\eat{
\RA{\textbf{IMPORTANT} The result of RF in Table~\ref{table:mlTransferLearning} is after using the recent changes in the convolution code. The previous results of RF was: 95.05\% accuracy and 94.07\% (I can still reproduce the results) F1 score, should we report the previous results or the recent results? }
}

\subsection{\vpm performance}
\label{sec:vmf_performance}
In Table ~\ref{table:vmf}, we study the performance of the \vpm filter, which filters out ``easy'' equivalence cases before \sys applies the \ecpm.
As in \S\ref{sec:compare_model}, we evaluate the \vpm by applying it to the TPC-DS workload. %
We observe that the \vpm is able to substantially reduce the search space and serves as an excellent filter prior to invoking the \ecpm. 

\VMFperformance

\subsection{\ssfl performance} 
\label{sec:model_eval}

\eqfeedbackloopTPCDSTPCHSep

In this experiment, we evaluate the semi-supervised feedback loop (\ssfl) in \sys. To do so, we iteratively train on additional labeled samples to fine-tune the \ecpm model. We compare our filter-based sampling method (\S\ref{sec:labeling}) against random sampling. 

For this experiment, we start with a scenario where the workload changes with new equivalent and non-equivalent patterns that the model has never seen before. We expect an initial model with low quality that improves with subsequent \ssfl iterations. To model this, we first create a degenerate TPC-H dataset by omitting all queries that contain joins.  We then train an initial model on the degenerate dataset and test on the TPC-DS workload.

Figure~\ref{fig:EQCMFeedbackLoopAcc} shows the accuracy and F1 score of the variants as they are exposed to additional labeled samples. Since the initial model has only been exposed to limited forms of  
equivalent and non-equivalent patterns, it does not perform well on the new workload which contains lots of subexpressions with joins. In each iteration of the feedback loop, we draw 512 labeled samples, using either filter-based or random sampling, from the new workload to help improve the model. 

With random sampling, performance does not improve meaningfully.  Due to the non-equivalence of most subexpressions in the workload, identifying positive examples through random sampling is nearly impossible. 
As a result, the accuracy and F1 score remain extremely low. 

By contrast, the filter-based sampling is more intelligent in selecting balanced samples that contain both positive and negative examples. This leads to significant improvements in both accuracy and F1 score. It takes only $\sim$4k samples to improve model accuracy and F1 score to 90\%.

\eqfeedbackloopPer
\GECOFeedbackLoopDrilldown

We next measure \ssfl execution time at various batch sizes. Figure~\ref{fig:EQCMFeedbackLoop} shows the result. Each bar in the figure shows the end-to-end \ssfl runtime, including both the time for sampling and training. Obviously, the filter-based sampling is more expensive than random sampling since it needs to do extra work to identify likely equivalent subexpression pairs (e.g., by executing the \sfm and \vpm filters and verifying). 
As we see in the figure, as more samples are trained over, 
the difference between the two reduces from 6.9$\times$ to less than 2$\times$. At the same time, it's worth recalling that the filter-based sampling requires many fewer iterations to achieve a satisfactory model accuracy and F1 score.
Additionally, the \ssfl process may be performed out-of-band with model prediction and the improved model may be substituted after training, mitigating performance impact on the prediction path.
Once a model has stabilized,
the \ssfl will no longer be active, and no further overhead will be incurred.

Finally, Figure~\ref{fig:EQCMFeedbackLoopdrilldown} shows the breakdown of the time for the feedback loop with filter-based sampling. As can be seen, the time spent in featurization, sampling, and verification is modest and does not substantially increase with batch size.  
On the other hand, the increase in training time is more dramatic and it quickly dominates \ssfl runtime.

\subsection{\vpm \& \ecpm compute performance}
\label{sec:compute}

In the previous sections we evaluated the performance of the \vpm and \ecpm in terms of their ability to identify equivalences and eliminate non-equivalences.  In this section, we further examine the runtimes of each filter.
To do so, we execute each filter on increasingly large subsets of the TPC-DS dataset.  We reduce confounds by disabling all other filters and compare performance using a CPU- and GPU-based implementation.

\begin{figure}[t]
\centering
\includegraphics[trim={0 0 0 19cm},clip,width=0.36\columnwidth]{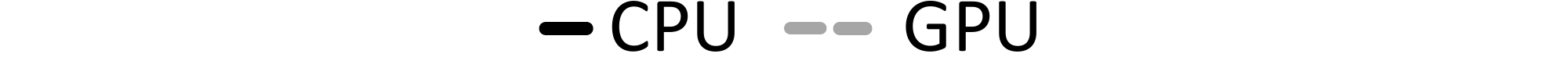}  
\subfigure[\vpm Runtime]{
  \includegraphics[trim={0 4cm 0 0},clip,width=0.5\columnwidth]{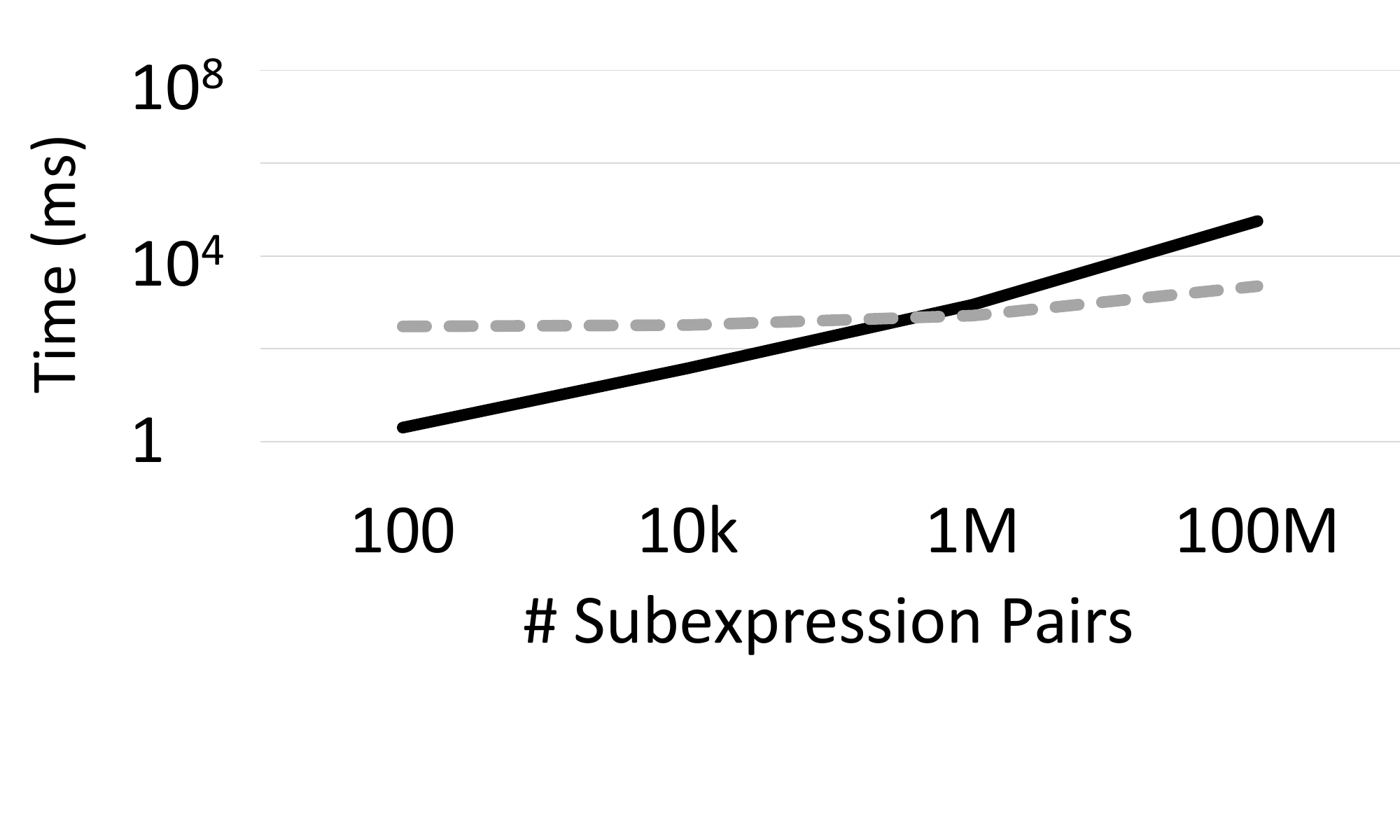}
  \label{fig:filter-runtimes-vmf}
}%
\subfigure[\ecpm Runtime]{
  \includegraphics[trim={0 4cm 0 0},clip,width=0.5\columnwidth]{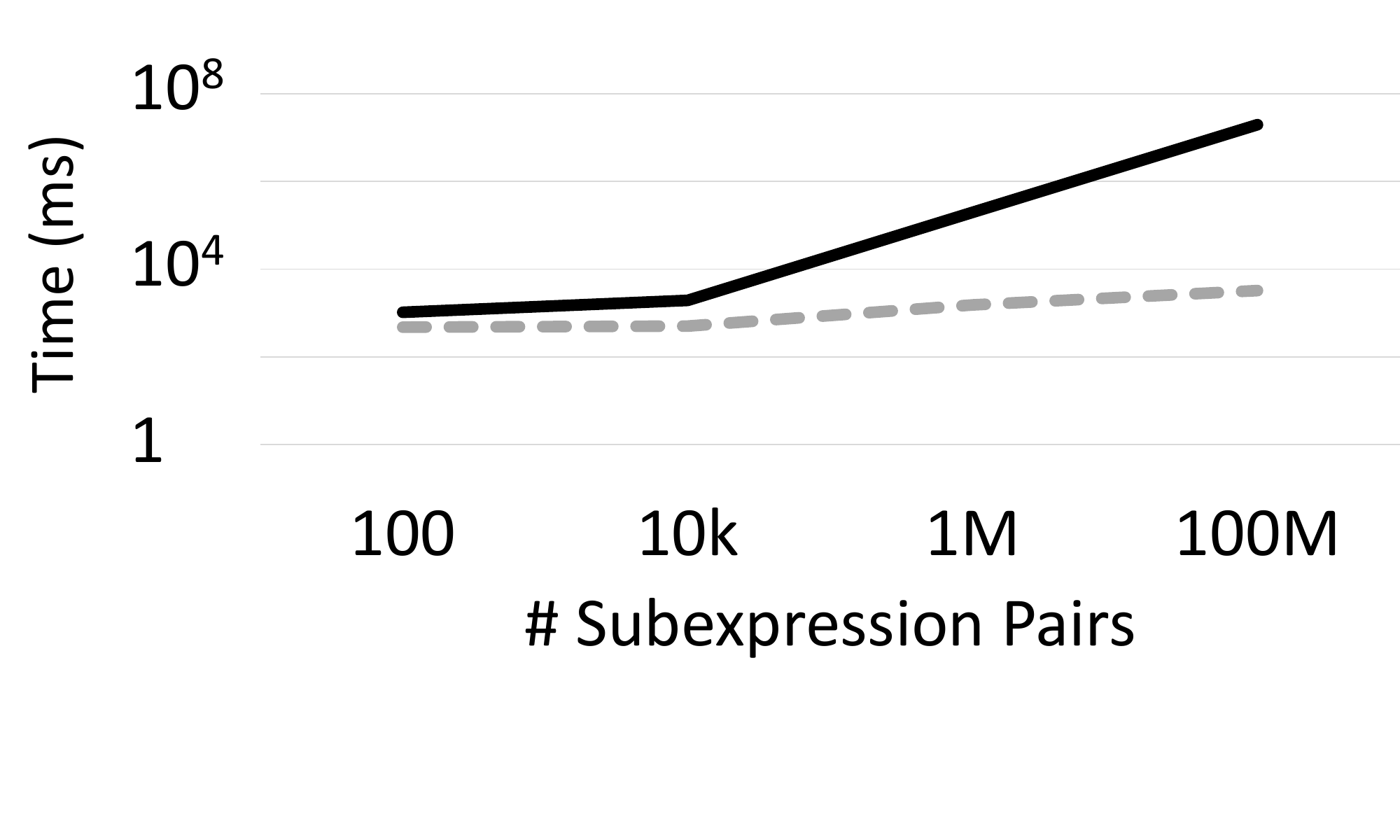}
  \label{fig:filter-runtimes-emf}
}
\caption{Total runtimes of the \vpm and \ecpm filters on varying number of subexpression pairs (log-log scale).}
\label{fig:filter-runtimes}
\end{figure}

In \autoref{fig:filter-runtimes-vmf}, we observe that the CPU-based \vpm exhibits excellent performance for smaller numbers of subexpression pairs, whereas the GPU-based variant surpasses it for ${\geq} {\sim}1$ million pairs due to a decrease in the proportion of data transfer I/O overheads in the overall runtime. In contrast, in \autoref{fig:filter-runtimes-emf}, the \ecpm consistently shows superior performance with GPU-based execution, although the CPU variant performs well at lower numbers of pairs. These results highlight the flexibility of \sys in targeting and adapting to heterogeneous hardware, providing a distinct advantage over other heuristic- and optimizer-based techniques

\subsection{End-to-End \sys performance} 
\label{sec:baseline_expr}

We now evaluate \sys performance in detecting equivalent subexpression pairs in various workloads. 
To do so, we randomly create a series of forty ${\sim}50k$ pair datasets generated on the TPC-DS schema and unseen by the \sys model.
We verify that each dataset contains approximately 8, 16, 32, 64, or 128 equivalent pairs.
We ensure that we have at least five (or more) datasets at each equivalence count.\footnotemark~  We then execute \sys on each dataset by executing ${\sys}_{\textsc{set}}(W, \{SF, VMF, EMF\})$ defined in \S\ref{sec:overview} along with the baselines described below.  For this experiment, we assume that the equivalences admitted by the AV constitute ground truth and \sys has executed its \ssfl and reached a confidence level above its minimum threshold ($T_h \ge 0.9$).

\footnotetext{A small number of datasets had additional (up to 6.25\%) equivalences as a byproduct of the randomized selection process.}

We compare \sys against three baselines: (i) the SPES query equivalence solver~\cite{zhou2020symbolic}; (ii) signature-based equivalence detection based on~\cite{jindal2018computation}, which compares signatures computed on each subexpression's  abstract syntax tree (AST); 
and (iii) an optimizer-based equivalence detection technique that leverages the Calcite %
optimizer to check whether two subexpressions are equivalent.

\GECOEndtoEnd

As shown in \autoref{subfig:e2e-tpr}, we observe that \sys identifies nearly all the semantic equivalences in the dataset (with a true positive rate averaging 88\% across all datasets and equivalence rates), whereas the Calcite and signature-based techniques average far fewer.  Unsurprisingly, SPES correctly verifies all equivalences.  

Next, \autoref{subfig:e2e-spes} shows the runtimes for each method.  
SPES's runtime is more than $200\times$ more expensive than the other methods.
\autoref{subfig:e2e-runtime} omits SPES and illustrates that
the Calcite and signature-based methods have approximately constant runtimes across all datasets, whereas \sys 
exhibits a curve that is similar at low numbers of equivalences and gradually rises for datasets with more equivalences.  %
These plots demonstrate that \sys is able to detect equivalences at an accuracy level near that of SPES but at a runtime similar to the heuristic-based techniques.

Finally, we observe that while \autoref{subfig:e2e-runtime} suggests a gradually rising runtime for \sys, this occurs because \textit{it detects more equivalences}.  \autoref{subfig:e2e-velocity} plots the runtime per equivalence detected and demonstrates that \sys spends approximately the same amount of time as Calcite and the signature-based method per equivalence.  For scaling reasons, we do not show SPES values in this plot, which ranged from 13.8 to 118.2 seconds per identified equivalence.%

\vspace{2mm}

\subsection{\rev{\sys Filters Ablation Study}}

\ablation

\rev{
Next we explore the relative contribution of each \sys filter. We 
execute ${\sys}_{\textsc{set}}(W, F)$ (see  \S\ref{sec:overview}), by varying $F$ to be some combination of the filters available in \sys
(i.e., the nonempty power set of $\{\vpm, \ecpm, \sfm\}$)
and $W$ to be each of the 32-equivalence datasets described in \S\ref{sec:baseline_expr}.
We report mean runtime over evaluated workloads, including verification time of the filtered pairs.
}

\rev{
\autoref{fig:ablation} shows the result of this experiment.  
We observe that \sys achieves best performance only when applying \textit{all} filters; no other combination minimizes the total runtime.  This implies that \sys's filters are complimentary to each other, and not redundantly filtering the same sets of subexpression pairs.
}

\reuse
\subsection{\rev{A Case Study on Result Caching}}
\rev{In our final experiment, we evaluate how \sys can be utilized in a result caching or materialization application, where results of queries are cached under a storage budget, to save computation for future semantically-equivalent queries.  
Using the workload from \S\ref{sec:evaluation} on the 100GB TPC-DS dataset, we obtain approximately ${\sim}23$k unique expressions after excluding those that produce empty results.  Our experiment assumes no updates. }

\rev{When executing with unlimited storage budget, the result cache using \sys could materialize the first occurrence of each equivalent expression (there are 5,277 equivalence classes in the workload), resulting in a total of ${\sim}2$GB storage space (in this workload, the expressions are computationally expensive but return small results), which we use as the \textit{upper-bound} for our storage budget. We then vary the storage budget for the cache, and simulate a caching policy that materializes the most expensive queries (leveraging past runtime statistics). Figure~\ref{fig:resue} shows a reduction of up to 61.5\% in the total workload execution time (running on a modern commercial database system), with 10\% of storage budget. With 100\% storage budget, a total of 96.2\% computation reduction is achieved.}

\section{Related work}\label{sec:relatedWork}

\noindent\textbf{Materialized Views and Query Rewriting.} 
As one of the most widely used approaches for computation reuse, materialized views are supported in many analytics engines. However, most systems---even including some modern cloud-based analytics engines like Snowflake~\cite{snowflake}, BigQuery~\cite{bigquery}, and NAPA~\cite{napa}---still require manual identification of common computation and creation of views. To automate view materialization, many view selection algorithms have been proposed~\cite{agrawal2000automated, alibaba,jindal2018computation,automv} to choose views that maximize computation reuse for a workload. 

\rev{Efficient and effective detection of overlapping computation is crucial in optimally selecting views to materialize. 
Some classical view selection methods heavily depend on the query optimizer to identify equivalences.
They consider factors such as  
resource constraints 
when selecting which views to materialize.
CloudViews~\cite{jindal2018selecting} employs Merkle tree-like signatures to quickly detect equivalent subexpressions. The ML-based view selection algorithm in~\cite{alibaba} utilizes a SQL equivalence verifier, called EQUITAS~\cite{zhou2019automated}, to detect equivalent subexpressions (we discuss verifiers in detail below). In terms of efficiency and scalability, the signature-based approach is clearly the best; however, it is the least effective since it only admits syntactic equivalence. By contrast, equivalence verifiers are superior in detecting semantic equivalence, but are computationally expensive. The optimizer's ability to detect semantic equivalence is bound by its rewrite rules. Interestingly, the work in~\cite{wetune} found that even a mature optimizer like the one in SQL Server could still miss some rewrite rules. In addition, repeatedly invoking the optimizer to check equivalence at cloud scale could easily turn the optimizer into a bottleneck. Compared to all these existing approaches, \sys is designed to be efficient and scalable, %
achieving effectiveness close to that of a verifier}. 

To utilize materialized views, view matching algorithms (e.g.,~\cite{goldstein2001optimizing}) match a query against previously-materialized views to determine whether the query can be rewritten into an improved variant by leveraging the views at runtime. In fact, query rewrite is generally an important query optimization step applied in many settings inside or outside the optimizer~\cite{cascadeopt, volcanoopt}. Most optimizers continue to rely on rewrite rules to identify equivalence and transform the original query into a semantically equivalent alternative. \sys can be used to learn equivalence relationships present in the given workload and complement these existing rewrite rules.

\vspace{2mm}
\noindent\textbf{Query Equivalence Verification.}
Verification of SQL query equivalence has been a long-standing topic of research in database theory~\cite{abiteboul1995foundations}. Several practical verifiers have been proposed~\cite{chu2017hottsql, chu2018axiomatic, zhou2019automated, zhou2020symbolic,tate2009equality}. Cosette~\cite{chu2017hottsql} and its extended version UDP~\cite{chu2018axiomatic} transform SQL queries into algebraic expressions and then utilize the Coq proof assistant~\cite{paulin2011introduction} to compare the two resultant algebraic expressions. However, these two approaches are computationally expensive due to the large number of normalized algebraic representations.  %
Recently, EQUITAS~\cite{zhou2019automated} and its extension SPES~\cite{zhou2020symbolic} address this limitation by efficiently deriving symbolic representation of SQL queries and use satisfiability modulo theories (SMT) to determine their equivalence under set and bag semantics. \rev{Approaches such as Peggy~\cite{tate2009equality}  leverage equality saturation, where the optimizer enumerates equivalent expressions for a given input expression based on predefined rules and collects them in a compact graph representation. %
While saturating every subexpression in a workload is not scalable, this technique could be leveraged by \sys as an alternative equivalence verifier. %
}

Since \sys is a general framework, it can plug in any of the above equivalence verifiers, or even new customized ones, to verify the predication from the \ecpm filter.

\section{Conclusion and Future Work} %
\label{sec:conclusion}
In this paper, we presented \sys, a portable lightweight ML-based framework for efficiently identifying semantically equivalent subexpressions 
at scale. We introduced \vpm and \ecpm 
filters to fill in the gap between a simple but very coarse schema-based filter and the accurate but very expensive equivalence verifier. We trained a deep-learning-based model to efficiently predict equivalence relationship between a pair of subexpressions. The db-agnostic featurization allows the learning from one workload and database to be transferrable to another. We also introduced an end-to-end semi-supervised learning feedback loop  with clever sampling to circumvent the expensive data labeling process. Our experimental evaluation demonstrates that \sys is  up to $200\times$ faster than verifiers on TPC-DS subexpressions.

\subsection{\rev{Extension to Complex Subexpressions}}

\rev{%
Even though a query workload might contain non-SPJ queries, \sys can still detect the equivalence of the SPJ subexpressions in the workload. Therefore, it represents a significant step forward in building an end-to-end framework for efficiently identifying semantically equivalent computations at scale. }
\rev{
Nevertheless, We plan to extend \sys to support complex subexpressions beyond SPJ (e.g., unions, aggregation, and complex predicates), using a similar approach to ~\cite{containrate}.
We next briefly sketch potential encoding extensions.}

\rev{
\textbf{\textit{OR and IN operators.}} We convert the WHERE clause with the OR operator to DNF, considering each conjunctive as a separate query and introduce unions. Each conjunctive branch is encoded as described in Section ~\ref{sec:feature_engineering}.  However, this approach encounters scalability issues due to the exponential growth in the number of clauses and the redundant encoding across union branches. The $IN$ clause can be considered as a shorthand representation for multiple $OR$ conditions. Unnesting the clause and introducing $OR$ conditions still pose scalability issues, which we intend to investigate.}

\rev{
\textbf{\textit{Union and except operators}.} We add a one-hot vector %
indicating union or except operators. }

\rev{
\textbf{\textit{Group By and aggregation operators}}. We add a new group-by segment that contains a one-hot vector for each of the group-by columns 
and %
an aggregate segment [AGG,COL], which is a concatenation of AGG and COL one-hot vectors of supported aggregation functions and columns. 
}

\rev{In addition to addressing encoding scalability issues with certain operators, we plan to assess the effectiveness of the current \ecpm model on complex queries and determine if any enhancement to the current  architecture (\eg augmenting the number of convolution layers) is necessary.}

\subsection{\rev{Extension to Query Containment}}
\rev{A second important future direction
involves using \sys to scalably detect semantic containment,
which is crucial for some view selection algorithms~\cite{ahmed2020automated,chirkova2012materialized}.}

\rev{We think the \sys framework should be applicable to semantic containment. The \ecpm model can be directly extended to classify containment. We conduct a preliminary experiment to demonstrate this
by training a new containment model over
TPC-H subexpressions with one-way joins and up to three predicates.  This model achieved $\sim$98\% accuracy on a test TPC-DS workload of similar complexity. 
As we increased the complexity of the workload (e.g., with additional joins),
the accuracy dropped to ${\sim}78$\%. We believe these results are promising since detecting containment is strictly harder than equivalence. 
}

\rev{In the prediction pipeline, the \sfm filter is adaptable to support containment. For instance, for a given pair, the table set of one of the subexpressions should be a subset of the other subexpression's table set, same condition applies on the projected columns. %
However, the distance metric used in the \vpm filter is not as easily adpatable, and we leave this as future work.
In terms of automated verification of semantic containment, this problem is well-studied under set semantics~\cite{abiteboul1995foundations}, but far less understood under bag semantics (e.g., the class of unions of conjunctive queries is undecidable under bag semantics~\cite{ioannidis1995containment, jayram2006containment}). We direct readers to the survey ~\cite{halevy2001answering}, which describes several practical containment checking algorithms in the context of rewriting queries using views, among which the algorithm in~\cite{goldstein2001optimizing} is a popular one that has been adopted by SQL Server and Calcite~\cite{calcite} optimizers.}

\bibliographystyle{ACM-Reference-Format}
\bibliography{bibliography}


\begin{thebibliography}{56}


\ifx \showCODEN    \undefined \def \showCODEN     #1{\unskip}     \fi
\ifx \showDOI      \undefined \def \showDOI       #1{#1}\fi
\ifx \showISBNx    \undefined \def \showISBNx     #1{\unskip}     \fi
\ifx \showISBNxiii \undefined \def \showISBNxiii  #1{\unskip}     \fi
\ifx \showISSN     \undefined \def \showISSN      #1{\unskip}     \fi
\ifx \showLCCN     \undefined \def \showLCCN      #1{\unskip}     \fi
\ifx \shownote     \undefined \def \shownote      #1{#1}          \fi
\ifx \showarticletitle \undefined \def \showarticletitle #1{#1}   \fi
\ifx \showURL      \undefined \def \showURL       {\relax}        \fi
\providecommand\bibfield[2]{#2}
\providecommand\bibinfo[2]{#2}
\providecommand\natexlab[1]{#1}
\providecommand\showeprint[2][]{arXiv:#2}

\bibitem[Abiteboul et~al\mbox{.}(1995)]%
        {abiteboul1995foundations}
\bibfield{author}{\bibinfo{person}{Serge Abiteboul}, \bibinfo{person}{Richard Hull}, {and} \bibinfo{person}{Victor Vianu}.} \bibinfo{year}{1995}\natexlab{}.
\newblock \bibinfo{booktitle}{\emph{Foundations of Databases}}. Vol.~\bibinfo{volume}{8}.
\newblock \bibinfo{publisher}{Addison-Wesley Reading}.
\newblock


\bibitem[{Agiwal, Ankur and Lai, Kevin and Manoharan, Gokul Nath Babu and Roy, Indrajit and Sankaranarayanan, Jagan and Zhang, Hao and Zou, Tao and Chen, Min and Chen, Jim and Dai, Ming and others}(2021)]%
        {napa}
\bibfield{author}{\bibinfo{person}{{Agiwal, Ankur and Lai, Kevin and Manoharan, Gokul Nath Babu and Roy, Indrajit and Sankaranarayanan, Jagan and Zhang, Hao and Zou, Tao and Chen, Min and Chen, Jim and Dai, Ming and others}}.} \bibinfo{year}{2021}\natexlab{}.
\newblock \showarticletitle{Napa: Powering Scalable Data Warehousing with Robust Query Performance at Google}.
\newblock  \bibinfo{volume}{14}, \bibinfo{number}{12} (\bibinfo{year}{2021}).
\newblock


\bibitem[Agrawal et~al\mbox{.}(2000)]%
        {agrawal2000automated}
\bibfield{author}{\bibinfo{person}{Sanjay Agrawal}, \bibinfo{person}{Surajit Chaudhuri}, {and} \bibinfo{person}{Vivek~R Narasayya}.} \bibinfo{year}{2000}\natexlab{}.
\newblock \showarticletitle{Automated selection of materialized views and indexes in SQL databases}. In \bibinfo{booktitle}{\emph{VLDB}}, Vol.~\bibinfo{volume}{2000}. \bibinfo{pages}{496--505}.
\newblock


\bibitem[Ahmed et~al\mbox{.}(2020)]%
        {ahmed2020automated}
\bibfield{author}{\bibinfo{person}{Rafi Ahmed}, \bibinfo{person}{Randall Bello}, \bibinfo{person}{Andrew Witkowski}, {and} \bibinfo{person}{Praveen Kumar}.} \bibinfo{year}{2020}\natexlab{}.
\newblock \showarticletitle{Automated generation of materialized views in oracle}. In \bibinfo{booktitle}{\emph{VLDB}}, Vol.~\bibinfo{volume}{13}. \bibinfo{pages}{3046--3058}.
\newblock


\bibitem[{Amazon}(2023)]%
        {automv}
\bibfield{author}{\bibinfo{person}{{Amazon}}.} \bibinfo{year}{2023}\natexlab{}.
\newblock \bibinfo{title}{{Amazon Redshift: Automated materialized views}}.
\newblock \bibinfo{howpublished}{\url{https://docs.aws.amazon.com/redshift/latest/dg/materialized-view-auto-mv.html}}.
\newblock
\newblock
\shownote{Accessed: 2023}.


\bibitem[{{Apache Software Foundation}}(2023a)]%
        {calcite}
\bibfield{author}{\bibinfo{person}{{{Apache Software Foundation}}}.} \bibinfo{year}{2023}\natexlab{a}.
\newblock \bibinfo{title}{{Apache Calcite: The foundation for your next high-performance database}}.
\newblock \bibinfo{howpublished}{\url{https://calcite.apache.org}}.
\newblock


\bibitem[{{Apache Software Foundation}}(2023b)]%
        {spark}
\bibfield{author}{\bibinfo{person}{{{Apache Software Foundation}}}.} \bibinfo{year}{2023}\natexlab{b}.
\newblock \bibinfo{title}{{Apache Spark: Unified Engine for large-scale data analytics}}.
\newblock \bibinfo{howpublished}{\url{https://spark.apache.org}}.
\newblock


\bibitem[Chirkova and Yang(2012)]%
        {chirkova2012materialized}
\bibfield{author}{\bibinfo{person}{Rada Chirkova} {and} \bibinfo{person}{Jun Yang}.} \bibinfo{year}{2012}\natexlab{}.
\newblock \showarticletitle{Materialized views}.
\newblock \bibinfo{journal}{\emph{Foundations and Trends{\textregistered} in Databases}} \bibinfo{volume}{4}, \bibinfo{number}{4} (\bibinfo{year}{2012}), \bibinfo{pages}{295--405}.
\newblock


\bibitem[Chu et~al\mbox{.}(2018)]%
        {chu2018axiomatic}
\bibfield{author}{\bibinfo{person}{Shumo Chu}, \bibinfo{person}{Brendan Murphy}, \bibinfo{person}{Jared Roesch}, \bibinfo{person}{Alvin Cheung}, {and} \bibinfo{person}{Dan Suciu}.} \bibinfo{year}{2018}\natexlab{}.
\newblock \showarticletitle{Axiomatic foundations and algorithms for deciding semantic equivalences of SQL queries}. In \bibinfo{booktitle}{\emph{VLDB}}, Vol.~\bibinfo{volume}{11}. \bibinfo{pages}{1482–1495}.
\newblock


\bibitem[Chu et~al\mbox{.}(2017)]%
        {chu2017hottsql}
\bibfield{author}{\bibinfo{person}{Shumo Chu}, \bibinfo{person}{Konstantin Weitz}, \bibinfo{person}{Alvin Cheung}, {and} \bibinfo{person}{Dan Suciu}.} \bibinfo{year}{2017}\natexlab{}.
\newblock \showarticletitle{HoTTSQL: Proving query rewrites with univalent SQL semantics}. In \bibinfo{booktitle}{\emph{{SIGPLAN}}}, Vol.~\bibinfo{volume}{52}. \bibinfo{pages}{510--524}.
\newblock


\bibitem[Cohen(2006)]%
        {cohen2006equivalence}
\bibfield{author}{\bibinfo{person}{Sara Cohen}.} \bibinfo{year}{2006}\natexlab{}.
\newblock \showarticletitle{Equivalence of queries combining set and bag-set semantics}. In \bibinfo{booktitle}{\emph{{PODS}}}. \bibinfo{pages}{70--79}.
\newblock


\bibitem[{Dageville, Benoit and Cruanes, Thierry and Zukowski, Marcin and Antonov, Vadim and Avanes, Artin and Bock, Jon and Claybaugh, Jonathan and Engovatov, Daniel and Hentschel, Martin and Huang, Jiansheng and others}(2016)]%
        {snowflake}
\bibfield{author}{\bibinfo{person}{{Dageville, Benoit and Cruanes, Thierry and Zukowski, Marcin and Antonov, Vadim and Avanes, Artin and Bock, Jon and Claybaugh, Jonathan and Engovatov, Daniel and Hentschel, Martin and Huang, Jiansheng and others}}.} \bibinfo{year}{2016}\natexlab{}.
\newblock \showarticletitle{The Snowflake Elastic Data Warehouse}. In \bibinfo{booktitle}{\emph{SIGMOD}}. \bibinfo{pages}{215–226}.
\newblock


\bibitem[De~Moura and Bj{\o}rner(2008)]%
        {de2008z3}
\bibfield{author}{\bibinfo{person}{Leonardo De~Moura} {and} \bibinfo{person}{Nikolaj Bj{\o}rner}.} \bibinfo{year}{2008}\natexlab{}.
\newblock \showarticletitle{Z3: An efficient SMT solver}. In \bibinfo{booktitle}{\emph{{TACAS}}}. \bibinfo{pages}{337--340}.
\newblock


\bibitem[de~Moura et~al\mbox{.}(2015)]%
        {de2015lean}
\bibfield{author}{\bibinfo{person}{Leonardo de Moura}, \bibinfo{person}{Soonho Kong}, \bibinfo{person}{Jeremy Avigad}, \bibinfo{person}{Floris Van~Doorn}, {and} \bibinfo{person}{Jakob von Raumer}.} \bibinfo{year}{2015}\natexlab{}.
\newblock \showarticletitle{The Lean theorem prover (system description)}. In \bibinfo{booktitle}{\emph{{CADE-25}}}. \bibinfo{pages}{378--388}.
\newblock


\bibitem[Deutsch(2018)]%
        {deutsch2009fol}
\bibfield{author}{\bibinfo{person}{Alin Deutsch}.} \bibinfo{year}{2018}\natexlab{}.
\newblock \showarticletitle{{FOL} Modeling of Integrity Constraints (Dependencies)}.
\newblock In \bibinfo{booktitle}{\emph{Encyclopedia of Database Systems, Second Edition}}.
\newblock


\bibitem[Dong et~al\mbox{.}(2023)]%
        {DBLP:journals/corr/abs-2212-07588}
\bibfield{author}{\bibinfo{person}{Yuyang Dong}, \bibinfo{person}{Chuan Xiao}, \bibinfo{person}{Takuma Nozawa}, \bibinfo{person}{Masafumi Enomoto}, {and} \bibinfo{person}{Masafumi Oyamada}.} \bibinfo{year}{2023}\natexlab{}.
\newblock \showarticletitle{{DeepJoin}: Joinable Table Discovery with Pre-trained Language Models}. In \bibinfo{booktitle}{\emph{VLDB}}, Vol.~\bibinfo{volume}{16}. \bibinfo{pages}{2458–2470}.
\newblock


\bibitem[Ebner et~al\mbox{.}(2017)]%
        {ebner2017metaprogramming}
\bibfield{author}{\bibinfo{person}{Gabriel Ebner}, \bibinfo{person}{Sebastian Ullrich}, \bibinfo{person}{Jared Roesch}, \bibinfo{person}{Jeremy Avigad}, {and} \bibinfo{person}{Leonardo de Moura}.} \bibinfo{year}{2017}\natexlab{}.
\newblock \showarticletitle{A metaprogramming framework for formal verification}. In \bibinfo{booktitle}{\emph{{ICFP}}}, Vol.~\bibinfo{volume}{1}. \bibinfo{pages}{1--29}.
\newblock


\bibitem[Goldstein and Larson(2001)]%
        {goldstein2001optimizing}
\bibfield{author}{\bibinfo{person}{Jonathan Goldstein} {and} \bibinfo{person}{Per-{\AA}ke Larson}.} \bibinfo{year}{2001}\natexlab{}.
\newblock \showarticletitle{Optimizing queries using materialized views: a practical, scalable solution}. In \bibinfo{booktitle}{\emph{{SIGMOD}}}, Vol.~\bibinfo{volume}{30}. \bibinfo{pages}{331--342}.
\newblock


\bibitem[Goodfellow et~al\mbox{.}(2016)]%
        {goodfellow2016deep}
\bibfield{author}{\bibinfo{person}{Ian Goodfellow}, \bibinfo{person}{Yoshua Bengio}, {and} \bibinfo{person}{Aaron Courville}.} \bibinfo{year}{2016}\natexlab{}.
\newblock \bibinfo{booktitle}{\emph{Deep learning}}.
\newblock \bibinfo{publisher}{MIT press}.
\newblock


\bibitem[Google(2023)]%
        {bigquery}
\bibfield{author}{\bibinfo{person}{Google}.} \bibinfo{year}{2023}\natexlab{}.
\newblock \bibinfo{title}{{BigQuery}}.
\newblock \bibinfo{howpublished}{\url{https://cloud.google.com/bigquery}}.
\newblock


\bibitem[Graefe(1995)]%
        {cascadeopt}
\bibfield{author}{\bibinfo{person}{Goetz Graefe}.} \bibinfo{year}{1995}\natexlab{}.
\newblock \showarticletitle{The Cascades Framework for Query Optimization}.
\newblock \bibinfo{journal}{\emph{IEEE Data Eng. Bull.}}  \bibinfo{volume}{18} (\bibinfo{year}{1995}), \bibinfo{pages}{19--29}.
\newblock


\bibitem[Graefe and McKenna(1993)]%
        {volcanoopt}
\bibfield{author}{\bibinfo{person}{Goetz Graefe} {and} \bibinfo{person}{William~J. McKenna}.} \bibinfo{year}{1993}\natexlab{}.
\newblock \showarticletitle{The Volcano Optimizer Generator: Extensibility and Efficient Search}. In \bibinfo{booktitle}{\emph{ICDE}}. \bibinfo{pages}{209–218}.
\newblock


\bibitem[Gupta et~al\mbox{.}(2015)]%
        {redshift}
\bibfield{author}{\bibinfo{person}{Anurag Gupta}, \bibinfo{person}{Deepak Agarwal}, \bibinfo{person}{Derek Tan}, \bibinfo{person}{Jakub Kulesza}, \bibinfo{person}{Rahul Pathak}, \bibinfo{person}{Stefano Stefani}, {and} \bibinfo{person}{Vidhya Srinivasan}.} \bibinfo{year}{2015}\natexlab{}.
\newblock \showarticletitle{Amazon Redshift and the Case for Simpler Data Warehouses}. In \bibinfo{booktitle}{\emph{{SIGMOD}}}. \bibinfo{pages}{1917–1923}.
\newblock


\bibitem[Halevy(2001)]%
        {halevy2001answering}
\bibfield{author}{\bibinfo{person}{Alon~Y Halevy}.} \bibinfo{year}{2001}\natexlab{}.
\newblock \showarticletitle{Answering queries using views: A survey}.
\newblock \bibinfo{journal}{\emph{The VLDB Journal}} \bibinfo{volume}{10}, \bibinfo{number}{4} (\bibinfo{year}{2001}), \bibinfo{pages}{270--294}.
\newblock


\bibitem[Hayek and Shmueli(2020)]%
        {containrate}
\bibfield{author}{\bibinfo{person}{Rojeh Hayek} {and} \bibinfo{person}{Oded Shmueli}.} \bibinfo{year}{2020}\natexlab{}.
\newblock \showarticletitle{Improved Cardinality Estimation by Learning Queries Containment Rates}. In \bibinfo{booktitle}{\emph{{EDBT}}}. \bibinfo{pages}{157--168}.
\newblock


\bibitem[Ho(1995)]%
        {ho1995random}
\bibfield{author}{\bibinfo{person}{Tin~Kam Ho}.} \bibinfo{year}{1995}\natexlab{}.
\newblock \showarticletitle{Random decision forests}. In \bibinfo{booktitle}{\emph{{ICDAR}}}, Vol.~\bibinfo{volume}{1}. \bibinfo{pages}{278--282}.
\newblock


\bibitem[Huang et~al\mbox{.}(2015)]%
        {huang2015query}
\bibfield{author}{\bibinfo{person}{Qiang Huang}, \bibinfo{person}{Jianlin Feng}, \bibinfo{person}{Yikai Zhang}, \bibinfo{person}{Qiong Fang}, {and} \bibinfo{person}{Wilfred Ng}.} \bibinfo{year}{2015}\natexlab{}.
\newblock \showarticletitle{Query-aware locality-sensitive hashing for approximate nearest neighbor search}. In \bibinfo{booktitle}{\emph{{VLDB}}}, Vol.~\bibinfo{volume}{9}. \bibinfo{pages}{1--12}.
\newblock


\bibitem[Ioannidis and Ramakrishnan(1995)]%
        {ioannidis1995containment}
\bibfield{author}{\bibinfo{person}{Yannis~E Ioannidis} {and} \bibinfo{person}{Raghu Ramakrishnan}.} \bibinfo{year}{1995}\natexlab{}.
\newblock \showarticletitle{Containment of conjunctive queries: Beyond relations as sets}.
\newblock \bibinfo{journal}{\emph{TODS}} \bibinfo{volume}{20}, \bibinfo{number}{3} (\bibinfo{year}{1995}), \bibinfo{pages}{288--324}.
\newblock


\bibitem[Jayram et~al\mbox{.}(2006)]%
        {jayram2006containment}
\bibfield{author}{\bibinfo{person}{TS Jayram}, \bibinfo{person}{Phokion~G Kolaitis}, {and} \bibinfo{person}{Erik Vee}.} \bibinfo{year}{2006}\natexlab{}.
\newblock \showarticletitle{The containment problem for real conjunctive queries with inequalities}. In \bibinfo{booktitle}{\emph{PODS}}. \bibinfo{pages}{80--89}.
\newblock


\bibitem[Jindal et~al\mbox{.}(2018a)]%
        {jindal2018selecting}
\bibfield{author}{\bibinfo{person}{Alekh Jindal}, \bibinfo{person}{Konstantinos Karanasos}, \bibinfo{person}{Sriram Rao}, {and} \bibinfo{person}{Hiren Patel}.} \bibinfo{year}{2018}\natexlab{a}.
\newblock \showarticletitle{Selecting subexpressions to materialize at datacenter scale}. In \bibinfo{booktitle}{\emph{{VLDB}}}, Vol.~\bibinfo{volume}{11}. \bibinfo{pages}{800--812}.
\newblock


\bibitem[Jindal et~al\mbox{.}(2021)]%
        {jindal2021production}
\bibfield{author}{\bibinfo{person}{Alekh Jindal}, \bibinfo{person}{Shi Qiao}, \bibinfo{person}{Hiren Patel}, \bibinfo{person}{Abhishek Roy}, \bibinfo{person}{Jyoti Leeka}, {and} \bibinfo{person}{Brandon Haynes}.} \bibinfo{year}{2021}\natexlab{}.
\newblock \showarticletitle{Production Experiences from Computation Reuse at Microsoft}. In \bibinfo{booktitle}{\emph{EDBT}}. \bibinfo{pages}{623--634}.
\newblock


\bibitem[Jindal et~al\mbox{.}(2018b)]%
        {jindal2018computation}
\bibfield{author}{\bibinfo{person}{Alekh Jindal}, \bibinfo{person}{Shi Qiao}, \bibinfo{person}{Hiren Patel}, \bibinfo{person}{Zhicheng Yin}, \bibinfo{person}{Jieming Di}, \bibinfo{person}{Malay Bag}, \bibinfo{person}{Marc Friedman}, \bibinfo{person}{Yifung Lin}, \bibinfo{person}{Konstantinos Karanasos}, {and} \bibinfo{person}{Sriram Rao}.} \bibinfo{year}{2018}\natexlab{b}.
\newblock \showarticletitle{Computation reuse in analytics job service at Microsoft}. In \bibinfo{booktitle}{\emph{{ICDM}}}. \bibinfo{pages}{191--203}.
\newblock


\bibitem[Kingma and Ba(2015)]%
        {kingma20153rd}
\bibfield{author}{\bibinfo{person}{Diederik~P. Kingma} {and} \bibinfo{person}{Jimmy Ba}.} \bibinfo{year}{2015}\natexlab{}.
\newblock \showarticletitle{Adam: {A} Method for Stochastic Optimization}. In \bibinfo{booktitle}{\emph{{ICLR}}}.
\newblock


\bibitem[Liu et~al\mbox{.}(2022)]%
        {liu2022automatic}
\bibfield{author}{\bibinfo{person}{Xinyu Liu}, \bibinfo{person}{Qi Zhou}, \bibinfo{person}{Joy Arulraj}, {and} \bibinfo{person}{Alessandro Orso}.} \bibinfo{year}{2022}\natexlab{}.
\newblock \showarticletitle{Automatic Detection of Performance Bugs in Database Systems using Equivalent Queries}. In \bibinfo{booktitle}{\emph{{ICSE}}}. \bibinfo{pages}{225--236}.
\newblock


\bibitem[Malkov and Yashunin(2018)]%
        {malkov2018efficient}
\bibfield{author}{\bibinfo{person}{Yu~A Malkov} {and} \bibinfo{person}{Dmitry~A Yashunin}.} \bibinfo{year}{2018}\natexlab{}.
\newblock \showarticletitle{Efficient and robust approximate nearest neighbor search using hierarchical navigable small world graphs}. In \bibinfo{booktitle}{\emph{{PAMI}}}, Vol.~\bibinfo{volume}{42}. \bibinfo{pages}{824--836}.
\newblock


\bibitem[Marcus et~al\mbox{.}(2021)]%
        {marcus2020bao}
\bibfield{author}{\bibinfo{person}{Ryan Marcus}, \bibinfo{person}{Parimarjan Negi}, \bibinfo{person}{Hongzi Mao}, \bibinfo{person}{Nesime Tatbul}, \bibinfo{person}{Mohammad Alizadeh}, {and} \bibinfo{person}{Tim Kraska}.} \bibinfo{year}{2021}\natexlab{}.
\newblock \showarticletitle{Bao: Making learned query optimization practical}. In \bibinfo{booktitle}{\emph{SIGMOD}}. \bibinfo{pages}{1275--1288}.
\newblock


\bibitem[Marcus et~al\mbox{.}(2019)]%
        {MarcusNMZAKPT19}
\bibfield{author}{\bibinfo{person}{Ryan~C. Marcus}, \bibinfo{person}{Parimarjan Negi}, \bibinfo{person}{Hongzi Mao}, \bibinfo{person}{Chi Zhang}, \bibinfo{person}{Mohammad Alizadeh}, \bibinfo{person}{Tim Kraska}, \bibinfo{person}{Olga Papaemmanouil}, {and} \bibinfo{person}{Nesime Tatbul}.} \bibinfo{year}{2019}\natexlab{}.
\newblock \showarticletitle{Neo: {A} Learned Query Optimizer}. In \bibinfo{booktitle}{\emph{{VLDB}}}, Vol.~\bibinfo{volume}{12}. \bibinfo{pages}{1705--1718}.
\newblock


\bibitem[{Microsoft}(2023)]%
        {synapse}
\bibfield{author}{\bibinfo{person}{{Microsoft}}.} \bibinfo{year}{2023}\natexlab{}.
\newblock \bibinfo{title}{{Azure Synapse Analytics}}.
\newblock \bibinfo{howpublished}{\url{https://azure.microsoft.com/en-us/services/synapse-analytics}}.
\newblock


\bibitem[Mou et~al\mbox{.}(2016)]%
        {mou2016convolutional}
\bibfield{author}{\bibinfo{person}{Lili Mou}, \bibinfo{person}{Ge Li}, \bibinfo{person}{Lu Zhang}, \bibinfo{person}{Tao Wang}, {and} \bibinfo{person}{Zhi Jin}.} \bibinfo{year}{2016}\natexlab{}.
\newblock \showarticletitle{Convolutional neural networks over tree structures for programming language processing}. In \bibinfo{booktitle}{\emph{AAAI}}, Vol.~\bibinfo{volume}{30}.
\newblock


\bibitem[Negi et~al\mbox{.}(2021)]%
        {negi2021steering}
\bibfield{author}{\bibinfo{person}{Parimarjan Negi}, \bibinfo{person}{Matteo Interlandi}, \bibinfo{person}{Ryan Marcus}, \bibinfo{person}{Mohammad Alizadeh}, \bibinfo{person}{Tim Kraska}, \bibinfo{person}{Marc Friedman}, {and} \bibinfo{person}{Alekh Jindal}.} \bibinfo{year}{2021}\natexlab{}.
\newblock \showarticletitle{Steering Query Optimizers: A Practical Take on Big Data Workloads}. In \bibinfo{booktitle}{\emph{{ICMD}}}. \bibinfo{pages}{2557--2569}.
\newblock


\bibitem[Neter et~al\mbox{.}(1996)]%
        {neter1996applied}
\bibfield{author}{\bibinfo{person}{John Neter}, \bibinfo{person}{Michael~H Kutner}, \bibinfo{person}{Christopher~J Nachtsheim}, {and} \bibinfo{person}{William Wasserman}.} \bibinfo{year}{1996}\natexlab{}.
\newblock \showarticletitle{Applied linear statistical models}.
\newblock  (\bibinfo{year}{1996}).
\newblock


\bibitem[Ortiz et~al\mbox{.}(2019)]%
        {DBLP:journals/corr/abs-1905-06425}
\bibfield{author}{\bibinfo{person}{Jennifer Ortiz}, \bibinfo{person}{Magdalena Balazinska}, \bibinfo{person}{Johannes Gehrke}, {and} \bibinfo{person}{S.~Sathiya Keerthi}.} \bibinfo{year}{2019}\natexlab{}.
\newblock \showarticletitle{An Empirical Analysis of Deep Learning for Cardinality Estimation}.
\newblock \bibinfo{journal}{\emph{CoRR}}  \bibinfo{volume}{abs/1905.06425} (\bibinfo{year}{2019}).
\newblock


\bibitem[Paszke et~al\mbox{.}(2019)]%
        {paszke2019pytorch}
\bibfield{author}{\bibinfo{person}{Adam Paszke}, \bibinfo{person}{Sam Gross}, \bibinfo{person}{Francisco Massa}, \bibinfo{person}{Adam Lerer}, \bibinfo{person}{James Bradbury}, \bibinfo{person}{Gregory Chanan}, \bibinfo{person}{Trevor Killeen}, \bibinfo{person}{Zeming Lin}, \bibinfo{person}{Natalia Gimelshein}, \bibinfo{person}{Luca Antiga}, {et~al\mbox{.}}} \bibinfo{year}{2019}\natexlab{}.
\newblock \showarticletitle{Pytorch: An imperative style, high-performance deep learning library}.
\newblock \bibinfo{journal}{\emph{{NeurIPS}}}  \bibinfo{volume}{32} (\bibinfo{year}{2019}).
\newblock


\bibitem[Paulin-Mohring(2011)]%
        {paulin2011introduction}
\bibfield{author}{\bibinfo{person}{Christine Paulin-Mohring}.} \bibinfo{year}{2011}\natexlab{}.
\newblock \showarticletitle{Introduction to the Coq proof-assistant for practical software verification}. In \bibinfo{booktitle}{\emph{{LASER}}}. \bibinfo{pages}{45--95}.
\newblock


\bibitem[Qin et~al\mbox{.}(2021)]%
        {DBLP:conf/kdd/Qin000W21}
\bibfield{author}{\bibinfo{person}{Jianbin Qin}, \bibinfo{person}{Wei Wang}, \bibinfo{person}{Chuan Xiao}, \bibinfo{person}{Ying Zhang}, {and} \bibinfo{person}{Yaoshu Wang}.} \bibinfo{year}{2021}\natexlab{}.
\newblock \showarticletitle{High-Dimensional Similarity Query Processing for Data Science}. In \bibinfo{booktitle}{\emph{{SIGKDD}}}. \bibinfo{pages}{4062--4063}.
\newblock


\bibitem[Sellis(1988)]%
        {mqo88}
\bibfield{author}{\bibinfo{person}{Timos~K. Sellis}.} \bibinfo{year}{1988}\natexlab{}.
\newblock \showarticletitle{Multiple-Query Optimization}. In \bibinfo{booktitle}{\emph{{TODS}}}, Vol.~\bibinfo{volume}{13}. \bibinfo{pages}{23–52}.
\newblock


\bibitem[Tate et~al\mbox{.}(2009)]%
        {tate2009equality}
\bibfield{author}{\bibinfo{person}{Ross Tate}, \bibinfo{person}{Michael Stepp}, \bibinfo{person}{Zachary Tatlock}, {and} \bibinfo{person}{Sorin Lerner}.} \bibinfo{year}{2009}\natexlab{}.
\newblock \showarticletitle{Equality saturation: a new approach to optimization}. In \bibinfo{booktitle}{\emph{POPL}}. \bibinfo{pages}{264--276}.
\newblock


\bibitem[Tu et~al\mbox{.}(2022)]%
        {mqo22}
\bibfield{author}{\bibinfo{person}{Yicheng Tu}, \bibinfo{person}{Mehrad Eslami}, \bibinfo{person}{Zichen Xu}, {and} \bibinfo{person}{Hadi Charkhgard}.} \bibinfo{year}{2022}\natexlab{}.
\newblock \showarticletitle{Multi-Query Optimization Revisited: A Full-Query Algebraic Method}. In \bibinfo{booktitle}{\emph{{Big Data}}}. \bibinfo{pages}{252--261}.
\newblock


\bibitem[Vaswani et~al\mbox{.}(2017)]%
        {DBLP:conf/nips/VaswaniSPUJGKP17}
\bibfield{author}{\bibinfo{person}{Ashish Vaswani}, \bibinfo{person}{Noam Shazeer}, \bibinfo{person}{Niki Parmar}, \bibinfo{person}{Jakob Uszkoreit}, \bibinfo{person}{Llion Jones}, \bibinfo{person}{Aidan~N. Gomez}, \bibinfo{person}{Lukasz Kaiser}, {and} \bibinfo{person}{Illia Polosukhin}.} \bibinfo{year}{2017}\natexlab{}.
\newblock \showarticletitle{Attention is All you Need}. In \bibinfo{booktitle}{\emph{{NIPS}}}. \bibinfo{pages}{5998--6008}.
\newblock


\bibitem[Wang et~al\mbox{.}(2022)]%
        {wetune}
\bibfield{author}{\bibinfo{person}{Zhaoguo Wang}, \bibinfo{person}{Zhou Zhou}, \bibinfo{person}{Yicun Yang}, \bibinfo{person}{Haoran Ding}, \bibinfo{person}{Gansen Hu}, \bibinfo{person}{Ding Ding}, \bibinfo{person}{Chuzhe Tang}, \bibinfo{person}{Haibo Chen}, {and} \bibinfo{person}{Jinyang Li}.} \bibinfo{year}{2022}\natexlab{}.
\newblock \showarticletitle{WeTune: Automatic Discovery and Verification of Query Rewrite Rules}. In \bibinfo{booktitle}{\emph{SIGMOD}}. \bibinfo{pages}{94–107}.
\newblock


\bibitem[Yuan et~al\mbox{.}(2020)]%
        {alibaba}
\bibfield{author}{\bibinfo{person}{Haitao Yuan}, \bibinfo{person}{Guoliang Li}, \bibinfo{person}{Ling Feng}, \bibinfo{person}{Ji Sun}, {and} \bibinfo{person}{Yue Han}.} \bibinfo{year}{2020}\natexlab{}.
\newblock \showarticletitle{Automatic view generation with deep learning and reinforcement learning}. In \bibinfo{booktitle}{\emph{{ICDE}}}. \bibinfo{pages}{1501--1512}.
\newblock


\bibitem[Zhou et~al\mbox{.}(2012)]%
        {scope}
\bibfield{author}{\bibinfo{person}{Jingren Zhou}, \bibinfo{person}{Nicolas Bruno}, \bibinfo{person}{Ming-Chuan Wu}, \bibinfo{person}{Per-Ake Larson}, \bibinfo{person}{Ronnie Chaiken}, {and} \bibinfo{person}{Darren Shakib}.} \bibinfo{year}{2012}\natexlab{}.
\newblock \showarticletitle{SCOPE: Parallel Databases Meet MapReduce}. In \bibinfo{booktitle}{\emph{{VLDB}}}, Vol.~\bibinfo{volume}{21}. \bibinfo{pages}{611–636}.
\newblock


\bibitem[Zhou et~al\mbox{.}(2019)]%
        {zhou2019automated}
\bibfield{author}{\bibinfo{person}{Qi Zhou}, \bibinfo{person}{Joy Arulraj}, \bibinfo{person}{Shamkant Navathe}, \bibinfo{person}{William Harris}, {and} \bibinfo{person}{Dong Xu}.} \bibinfo{year}{2019}\natexlab{}.
\newblock \showarticletitle{Automated verification of query equivalence using satisfiability modulo theories}.
\newblock \bibinfo{journal}{\emph{{VLDB}}} \bibinfo{volume}{12}, \bibinfo{number}{11}, \bibinfo{pages}{1276--1288}.
\newblock


\bibitem[Zhou et~al\mbox{.}(2022)]%
        {zhou2020symbolic}
\bibfield{author}{\bibinfo{person}{Qi Zhou}, \bibinfo{person}{Joy Arulraj}, \bibinfo{person}{Shamkant~B. Navathe}, \bibinfo{person}{William Harris}, {and} \bibinfo{person}{Jinpeng Wu}.} \bibinfo{year}{2022}\natexlab{}.
\newblock \showarticletitle{SPES: A Symbolic Approach to Proving Query Equivalence Under Bag Semantics}. In \bibinfo{booktitle}{\emph{ICDE}}. \bibinfo{pages}{2735--2748}.
\newblock


\bibitem[Zhu and Goldberg(2009)]%
        {zhu2009introduction}
\bibfield{author}{\bibinfo{person}{Xiaojin Zhu} {and} \bibinfo{person}{Andrew~B Goldberg}.} \bibinfo{year}{2009}\natexlab{}.
\newblock \showarticletitle{Introduction to semi-supervised learning}.
\newblock \bibinfo{journal}{\emph{Synthesis lectures on artificial intelligence and machine learning}} \bibinfo{volume}{3}, \bibinfo{number}{1} (\bibinfo{year}{2009}), \bibinfo{pages}{1--130}.
\newblock


\bibitem[Zhu et~al\mbox{.}(2021)]%
        {zhu2021kea}
\bibfield{author}{\bibinfo{person}{Yiwen Zhu}, \bibinfo{person}{Subru Krishnan}, \bibinfo{person}{Konstantinos Karanasos}, \bibinfo{person}{Isha Tarte}, \bibinfo{person}{Conor Power}, \bibinfo{person}{Abhishek Modi}, \bibinfo{person}{Manoj Kumar}, \bibinfo{person}{Deli Zhang}, \bibinfo{person}{Kartheek Muthyala}, \bibinfo{person}{Nick Jurgens}, {et~al\mbox{.}}} \bibinfo{year}{2021}\natexlab{}.
\newblock \showarticletitle{KEA: Tuning an Exabyte-Scale Data Infrastructure}. In \bibinfo{booktitle}{\emph{SIGMOD}}. \bibinfo{pages}{2667--2680}.
\newblock


\end{thebibliography}
\appendix

\end{document}